\documentclass{aa}  
\usepackage{graphicx}
\usepackage{txfonts}
\usepackage{longtable}
\usepackage{float}
\usepackage{cleveref}
\usepackage{color}

\DeclareUnicodeCharacter{2212}{-}

\begin{document} 

%


   \title{Physical properties of Brightest Cluster Galaxies up to redshift 1.80 based on HST data}

   \author{A. Chu
          \inst{1}
          \and
          F. Durret
          \inst{1}
          \and
          I. M\'arquez
          \inst{2}
          }

   \institute{Sorbonne Universit\'e, CNRS, UMR 7095, Institut d’Astrophysique de Paris, 98bis Bd Arago, 75014, Paris, France \\
              \email{aline.chu@iap.fr}
         \and
             Instituto de Astrof\'isica de Andaluc\'ia, CSIC, Glorieta de la Astronom\'ia s/n, 18008, Granada, Spain
             }

   \date{Accepted, Received}

%
 
  \abstract
   {Brightest cluster galaxies (BCGs) have grown by accreting numerous smaller galaxies and can be used as tracers of cluster formation and evolution in the cosmic web. However, there is still a controversy on the main epoch of formation of BCGs, 
  since some authors believe they have already formed before redshift z=2, while others still find them to evolve at more recent epochs. }
   {We aim to analyse the physical properties of a large sample of BCGs covering a wide redshift range  up to z=1.8 and analysed in a homogeneous way, to see if their characteristics vary with redshift. As a first step, we also present a new tool to define for each cluster which galaxy is the BCG.}
   {For a sample of 137 clusters with HST images in the optical and/or infrared, we analyse the BCG properties by applying GALFIT with one or two S\'ersic components. For each BCG, we thus compute the S\'ersic index, effective radius, major axis position angle, surface brightness. We then search for correlations of these quantities with redshift. }
   {We find that the BCGs follow the Kormendy relation (between the effective radius and the mean surface brightness), with a slope that remains constant with redshift, but with a variation with redshift of the ordinate at the origin. Although the trends are faint, we find that both the absolute magnitudes and effective radii tend to become respectively brighter and bigger with decreasing redshift. On the other hand, we find no significant correlation of the mean surface brightnesses or S\'ersic indices with redshift. The major axes of the cluster elongations and of the BCGs agree within 30 degrees for 73\% of our clusters at redshift z $\leq$ 0.9. }
   {Our results agree with the BCGs being mainly formed before redshift z=2. The alignment of the major axes of BCGs with their clusters agree with the general idea that BCGs form at the same time as clusters by accreting matter along the filaments of the cosmic web. }


\maketitle
%

%


\section{Introduction}

Galaxy clusters are the largest and most massive gravitationally bound structures observed in the Universe. As so, they constitute perfect probes to test cosmological models and help us better understand the history of the Universe, as they will constrain the limits of observed physical parameters through time, such as mass or brightness, in numerical simulations \citep{Kravtsov_2012}. The $\Lambda$CDM model proposes a hierarchical evolution scenario starting from small fluctuations that assemble together via the gravitational force, and grow to form bigger and bigger structures. As a result, galaxy clusters are the latest and most massive structures to have formed.

Clusters are believed to be located at the intersection of cosmic filaments, and to form by merging with other smaller clusters or groups of galaxies, and by constantly accreting gas and galaxies that preferentially move along cosmic filaments and end up falling towards the center of the gravitational potential well, which often coincides with the peak of the X-ray emission \citep[see][and references therein]{De_Propris_2020}. Generally, the brightest galaxy of the cluster, the Brightest Cluster Galaxy (BCG hereafter) lies at the center of the cluster. It is usually a supermassive elliptical galaxy, that is formed and grows by mergers with other galaxies, and can be up to two magnitudes brighter than the second brightest galaxy. This property makes BCGs easily recognisable. BCGs have often been referred to as cD galaxies, i.e. supergiant ellipticals with a large and diffuse halo of stars. Since their properties are closely linked to those of their host cluster \citep{lauer2014brightest}, they can be extremely useful to trace how galaxy clusters have formed and evolved. BCGs tend to be aligned along the major axis of the cluster, which also hints at the close link between the BCG and its host cluster \citep{donahue2015,durret2016,west2017ten,De_Propris_2020}. This alignment suggests that the accretion of galaxies may be done along a preferential axis, with galaxies falling into clusters along cosmic filaments.

Most of the stars in today's BCGs were already formed at redshift $z \geq 2$ \citep{thomas10.1111/j.1365-2966.2010.16427.x}. BCGs, especially the most massive ones, can present an extended halo made of stars that were stripped from their host galaxy during mergers, and form the Intra Cluster Light (ICL). When measuring photometric properties of galaxies, some parameters such as the BCG major axis can be difficult to measure accurately as the separation between the ICL and the external envelope of the BCG is not clear. However, the ICL is a very faint component, and as we are observing bright galaxies, the ICL should not strongly affect our study, so the ICL will not be considered in this paper. 

The evolution of BCG properties with redshift is of interest to study cluster formation and evolution, but this topic remains quite controversial.
Some authors report no evolution of the sizes of the BCGs with redshift \citep[see][and references therein]{bai2014inside, Stott_2011}: \citet{Stott_2011} found that there was no significant evolution of the sizes or shapes of the BCGs between redshift 0.25 and 1.3; whereas \citet{Ascaso_2010} found that, although the shapes show little change, they have grown by a factor of 2 in the last 6 Gyrs. \citet{Bernardi_2009} found a 70\% increase of the sizes of BCGs since z = 0.25, and an increase of a factor 2 since z = 0.5.

\cite{bai2014inside} found that while the inner region of the galaxies doesn't grow much, the light dispersed around the BCG forms the outer component, resulting in a shallow outer luminosity profile. This is an indication of an inside-out growth of BCGs: the inner component forms first and then stops growing while the outer component develops. \citet{edwards10.1093/mnras/stz2706} gave more evidence to justify this inside-out growth of BCGs by showing that the stars in the ICL are younger and less metal-rich than those in the cores of the BCGs. They also showed that the most extended BCGs tend to be close to the X-ray center. This last statement is supported by \citet{lauer2014brightest}, who added that the inner component would have already been formed before the cluster, while the outer component, the envelope of the BCG, is formed and grows later. In fact, numerical simulations with AGN suppressed cooling flows, showed  that about 80\% of the stars are already formed at redshift z $\approx$ 3 in the BCG progenitors that merge together to form today's BCGs \citep{De_Lucia_2007}. \citet{Cooke_2019} found that BCG progenitors in the COSMOS field have an active star formation phase before z = 2.25, followed by a phase of dry and wet mergers until z = 1.25 that leads to more star formation and increases the stellar mass of the progenitors, after which the stellar mass of progenitors mainly grows through dry mergers, and half of the stellar mass is formed at z~=~0.5. Similarly, \citet{Cerulo_2019} did not find significant stellar mass growth between z~=~0.35 and z~=~0.05, suggesting that most of the BCGs stellar masses were formed by z~=~0.35. 
\citet{Durret_2019} observed a possible variation with redshift of the effective radius of the outer S\'ersic component of BCGs for 38 BCGs in the redshift range 0.2 $\leq$ z $\leq$ 0.9, agreeing with a scenario in which BCGs at these redshifts mostly grow by accreting smaller galaxies. 

Several conflicts also arise on the growth of the stellar masses of the BCGs: \citet{Collins_1998, Collins_2009, Stott_2010} found little to no evolution. On the other hand, other studies found a strong evolution in the stellar masses of BCGs since redshift z~=~2 \citep{Aragon_Salamanca_1998, Lidman_2012, Lin_2013, bellstedt10.1093/mnras/stw1184, Zhang_2016}.

In the present paper, we will try to characterize how the properties of BCGs have evolved since z = 1.80, based on HST data to have the best possible spatial resolution, which is particularly necessary at high redshift.
When dealing with a large amount of data, identifying individually the BCG of a cluster to build a sample can be a long task to do. This is why we present here a method based on several photometric properties of the BCGs, that will allow to detect BCGs automatically. We analyze a sample of 137 galaxy clusters, covering the redshift range 0.187 $\leq$ z $\leq$ 1.80, and including various types of BCGs (star forming BCGs, SF BCGs hereafter, interacting BCGs, hosts of possible AGNs, supercluster members, etc.).

\begin{figure}[ht]
    \centering
    \includegraphics[width=\linewidth]{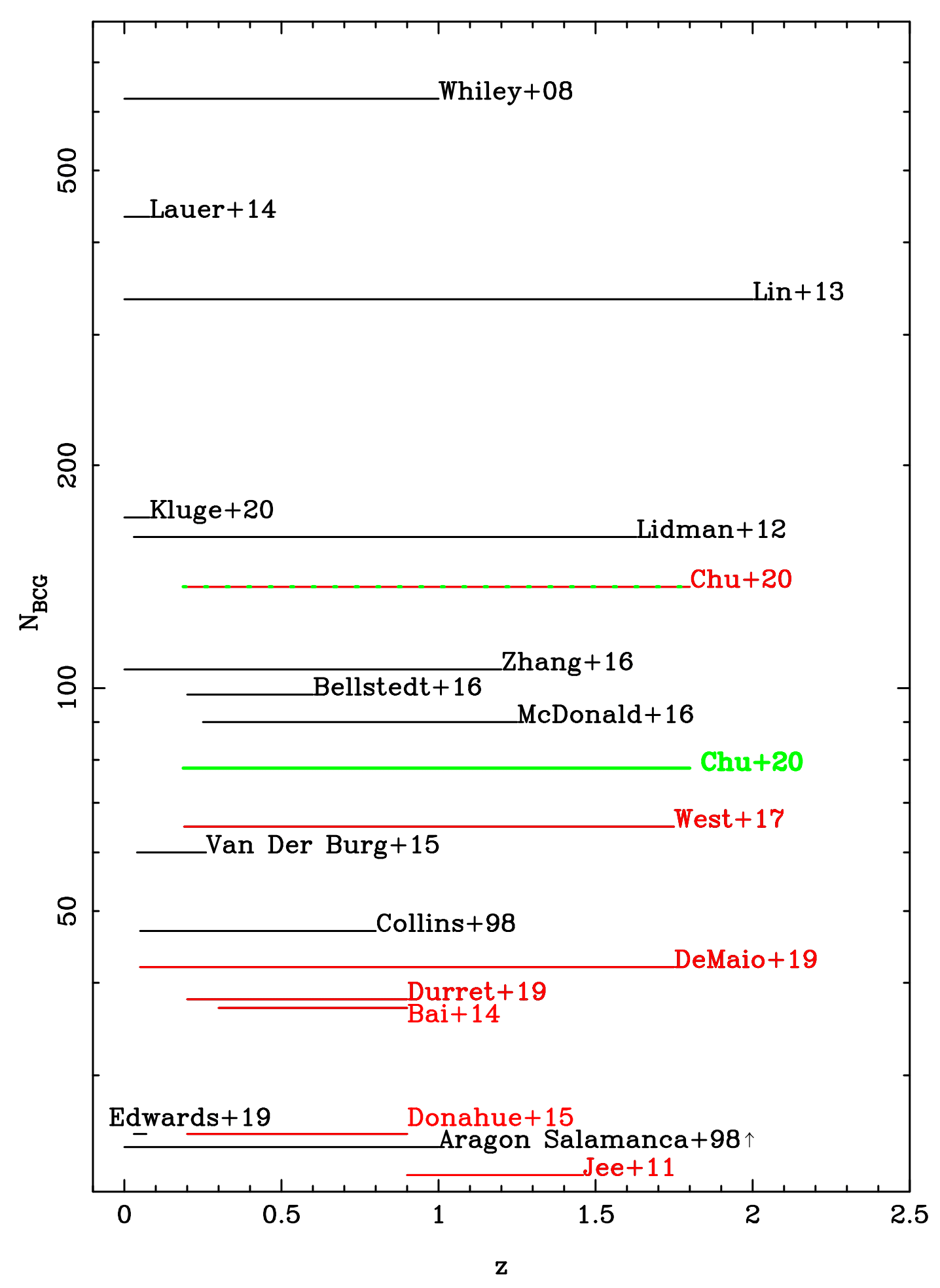}
    \caption{Comparison of the various samples of BCGs found in the literature, considering the redshift range, the number of galaxies analysed and the type of data used. Only samples with at least 20 objects are represented here.
     \citet{Cerulo_2019} with a sample of 74275 BCGs is not represented here for better lisibility. The samples represented in black use ground-based telescope data, space-based data excluding HST, or a mix of ground-based and space-based data, while those in red only use HST data. Our initial sample is represented by the red and green dotted line, and our final BCG sample in green (see \Cref{section:luminosity_profiles}).}
    \label{fig:comp_study}
\end{figure}

The present paper (in green in \Cref{fig:comp_study}, the figure will be described in more detail in the next section) will allow to cover a large redshift range with one of the largest samples observed with HST (indicated in red). This will enable us to obtain more significant statistics on the evolution of BCG properties. 

The paper is organized as follows. We will describe the data in \Cref{section:sample}, the method to detect automatically the BCGs in \Cref{section:detection}, and the modelization of their luminosity profiles in \Cref{section:luminosity_profiles}. The results obtained as well as a short study of the link between the BCG masses, distance between the BCG and the X-ray center of the cluster, and physical properties are given in \Cref{section:results}. A final discussion and conclusions are presented in \Cref{section:conclusion}.

Throughout this paper, we assume H$_{0}$ = 70 km s$^{-1}$ Mpc$^{-1}$, $\Omega_{M}$ = 0.3 and $\Omega_{\Lambda}$ = 0.7. The scales and physical distances are computed using the astropy.coordinates package\footnote{https://docs.astropy.org/en/stable/coordinates/}. Unless specified, all magnitudes are given in the AB system.


\section{Sample and data}
\label{section:sample}

\subsection{The sample}

\begin{figure}[ht]
    \centering
    \includegraphics[width=\linewidth]{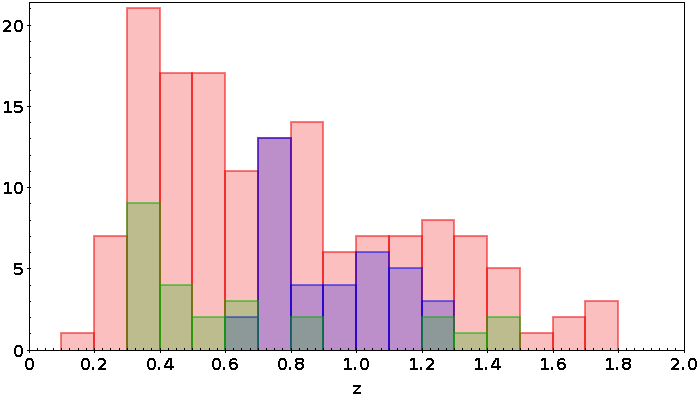}
    \caption{Histogram of the redshifts of the 149 BCGs in our sample. The red histogram shows all the BCGs studied while the blue histogram (37 BCGs) shows those observed in too blue rest frame filters compared to the 4000 \AA\ break (see \Cref{section:detection}). The green histogram (25 BCGs) shows all BCGs with an important inner component (see  \Cref{section:luminosity_profiles}).}
    \label{fig:hist_redshift}
\end{figure}{}

The sample studied in this paper consists of 137 galaxy clusters with HST imaging taken from  \citet{jee2011scaling, Postman_2012, bai2014inside, donahue2015, west2017ten, DeMaio_2019, Durret_2019, Sazonova_2020} . We also add five more distant clusters at z $\geq$ 0.8, as well as the cluster Abell~2813 at z = 0.29. Among them, we identify twelve clusters which present in their center two BCGs similar in magnitude and size (see Section~\ref{section:detection}). As a result, our final BCG sample contains 149 BCGs (the number of BCGs is not equal to the number of clusters studied because of these clusters with two BCGs).
This sample is a good representation of the most massive BCGs in the range 0.1 $\leq$ z $\leq$ 1.80. The redshift distribution is shown on \Cref{fig:hist_redshift}. 

All of these clusters have data available from the Hubble Space Telescope, obtained with the Advanced Camera for Surveys (ACS) in optical bands and/or the Wide Field Camera 3 (WFC3) in infrared bands, resulting in good quality images. This allows us to perform accurate photometry with relatively good precision, and to treat all the BCGs in a homogeneous way. Contrary to other studies such as \citet{bai2014inside}, we do not exclude in our study clusters with bright nearby objects that may hinder our measurements near the BCG area. 
We also identify in our sample two clusters host of blue BCGs (negative rest frame blue-red color), with active star forming regions inside the BCGs. These two BCGs will be described more in details in \Cref{section:detection}.

Our sample covers a large range in redshift, which will enable us to trace the history of cluster formation through time. 

\Cref{fig:comp_study}\footnote{By increasing number of BCGs: \citet{bai2014inside,  Durret_2019,  DeMaio_2019, vanderburg, west2017ten, McDonald_2016, bellstedt10.1093/mnras/stw1184, Zhang_2016, Lidman_2012, Kluge_2020, Lin_2013, lauer2014brightest,Whiley_2008, Cerulo_2019}.} shows the comparison of the sample sizes and redshift ranges between different studies done on BCGs. \citet{Cerulo_2019} studied a sample 74275 BCGs from the SDSS in the redshift range 0.05 $\leq$ z $\leq$ 0.35, and is not represented on this figure for better lisibility.
Most large studies, especially those with HST data (represented in red), were done exclusively on local BCGs (z $\leq$ 0.1) \citep[][]{lauer2014brightest,Cerulo_2019}, while farther clusters and BCGs were limited to relatively small samples (N $\leq$ 45) \citep{bai2014inside, DeMaio_2019, Durret_2019} and/or used ground-based data (represented in black). Our sample contains more clusters and BCGs at high redshifts (z $\geq$ 0.7) than that of \citet{Lidman_2012} \citep[33 in][73 in this paper]{Lidman_2012}. With the present study, we therefore almost double the previous samples and cover a larger range in redshift. This enables us to obtain more significant statistics on the evolution of BCG properties. 

\citet{Lidman_2012}, \citet{Lin_2013}, \citet{west2017ten} and \citet{De_Propris_2020} mainly focus on the alignment of the BCGs with their host cluster and on the evolution of the BCG stellar masses. Our work constitutes a deeper analysis since we also study the luminosity profiles of the BCGs.

\subsection{Retrieving data and cluster information}

We retrieve all the FITS images from the Hubble Legacy Archive (HLA\footnote{https://hla.stsci.edu/}). We look for combined or mosaic images according to what is available and download stacked images directly from the HLA. To avoid handling such heavy files, we first crop these images, define the new center on the cluster coordinates found in NED, and create a new image 1.2 Mpc wide. Linear scales (arcsec/Mpc) are determined from the cluster redshifts in the literature \citep[from][or found in NED for the five other clusters we added]{jee2011scaling, Postman_2012, bai2014inside, donahue2015,west2017ten,DeMaio_2019,Durret_2019,Sazonova_2020}. Cluster information can be found in \Cref{tab:BCGs_coord}. It is necessary to add the keywords 'GAIN' and 'RDNOISE' in the header of the FITS images, which will be used later by SExtractor or GALFIT. As the images are in units of electrons/s, we set the GAIN to 1 and multiply the images by the total exposure time (EXPTIME) to get back to units in electrons. 
Single exposure images were summed with AstroDrizzle to get the final combined images. We also retrieve the associated weight maps (wht fits) obtained applying the option IVM (Inverse Variance Map) of AstroDrizzle.


\section{Procedure for the detection of the BCG}
\label{section:detection}

The definition of the BCG that we use throughout this paper is the following: the BCG is the brightest galaxy of the cluster that lies close to the cluster center, defined as the center of the cluster member galaxy distribution. Generally, the cluster center is defined as the X-ray center of the cluster, as X-rays trace the mass distribution better. However, it is difficult to obtain X-ray data, particularly at high redshifts. If a large sample is considered, most probably a good fraction of the clusters do not have X-ray data available. Moreover, it was shown in several studies \citep{Patel_2006,Hashimoto_2014,De_Propris_2020} that BCGs are often displaced from the X-ray center. For these reasons and anticipating for future works with much larger samples, we use a definition which is independent from X-rays and only relies on optical and infrared photometric data. We define the center as that of the spatial distribution of cluster galaxies \citep[as in][]{Kluge_2020}. As a matter of fact, X-ray coordinates are only available for 68 out of the 137 clusters in our sample (see \Cref{table:mass_coordX}). These X-ray positions will only be used to study whether or not the BCG properties correlate with their position relatively to the X-ray center (see \Cref{section:results}). 

\subsection{Method for detecting red BCGs}

\begin{figure*}[t]
    \centering
    \includegraphics[width=\linewidth]{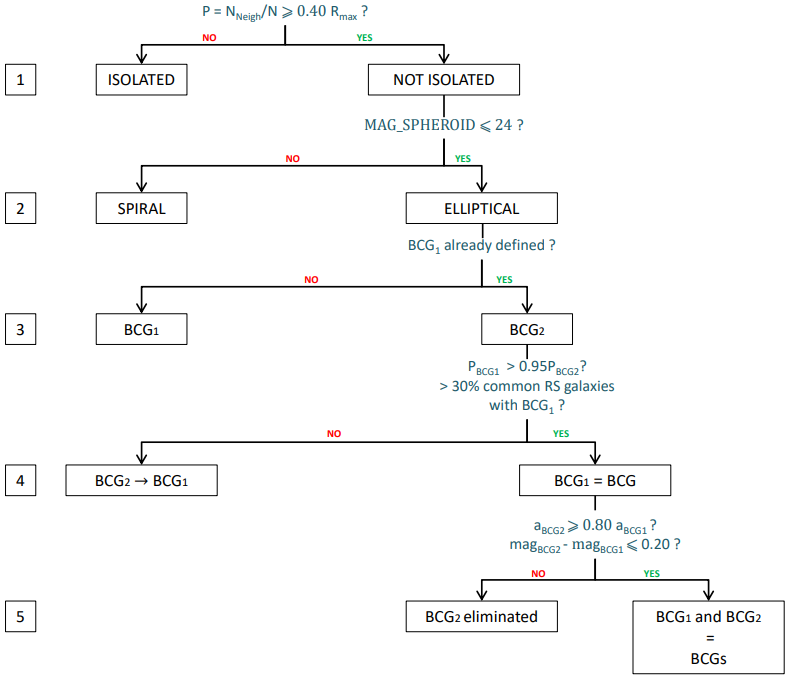}
    \caption{Flowchart showing how BCGs are automatically selected, in the case of a rich cluster (see description of the method for our definition of a rich cluster). Values may change for less rich clusters (see \Cref{subsubsection:spiral_isolated}).}
    \label{fig:schema_detec}
\end{figure*}{}

The method applied to select automatically red cluster BCGs is schematically summarized in \Cref{fig:schema_detec}, and will be described in detail below. The efficiency of the method will be discussed in \Cref{detect_results}. Blue BCGs will be mentioned in \Cref{subestion:blueBCGs}.

\subsubsection{Rejection of foreground sources}

In order to differentiate the BCG from other objects in the field, we need to identify which objects are part of the cluster and which are not. 
We describe below step by step our method to detect the BCGs among all the contaminations (stars, foreground and background galaxies, artifacts) in our images.

Measurements with SExtractor \citep[refer to][for more details on the following mentioned parameters]{SEx1996A&AS..117..393B} were done using two different deblendings (parameters DEBLEND$\_$MINCOUNT = 0.01 and DEBLEND$\_$MINCOUNT = 0.02). The smallest deblending parameter, i.e, the finest deblending, is sufficient to separate two nearby galaxies without fragmenting excessively spiral galaxies in the foreground, and provides the most accurate measurements. However, BCGs can present a very diffuse and luminous halo which may be associated with intracluster light (ICL). We noticed that, in presence of nearby bright sources in the region of the BCG, SExtractor would detect only those foreground sources and process the BCG halo as a very luminous background. We therefore decided to run SExtractor in parallel with a coarser deblending to take this into account. The two catalogs obtained with two different deblending parameters are then matched: we keep the values obtained with the finer deblending, and add all new objects detected using a coarser deblending. 

We compute the magnitude at which our catalogue is complete at 80\%, m$_{80\%}$. For this, we plot the histogram in apparent magnitudes and fit the distribution up to the magnitude at which the distribution drops.
By dividing the number of detected sources by the total number of sources that is expected to be detected in a magnitude bin (given by the fit), we compute the completeness of the catalogue at each bin. We can then determine m$_{80\%}$, and make a cut in apparent magnitude to reject all galaxies with m $\geq$ m$_{80\%}$ + 2, as the photometry will not be accurate for these faintest objects.

Our procedure to reject the various contaminations is as follows. First, we query in NED for all the sources in the region of the cluster and reject all those with a spectroscopic redshift that differs by more than 0.15 from the cluster redshift (|z - z$_{spec}$| $>$ 0.15). We identify bad detections by their magnitude values which get returned as MAG = 99.99 by SExtractor. All point sources or unresolved compact galaxies are eliminated using the parameter CLASS$\_$STAR $\geq$ 0.95 in SExtractor, which requires a PSF model to be fed into SExtractor, created with PSFex \citep{PSFex2011ASPC..442..435B}.
Most foreground galaxies can be identified by their excessively bright absolute magnitude when computed from their MAG\_AUTO magnitude and assuming they are at the cluster redshift. We thus exclude all sources with an absolute magnitude MAG\_ABS $\leq -26$.

\begin{figure}[t]
    \centering
    \includegraphics[width=0.7 \linewidth]{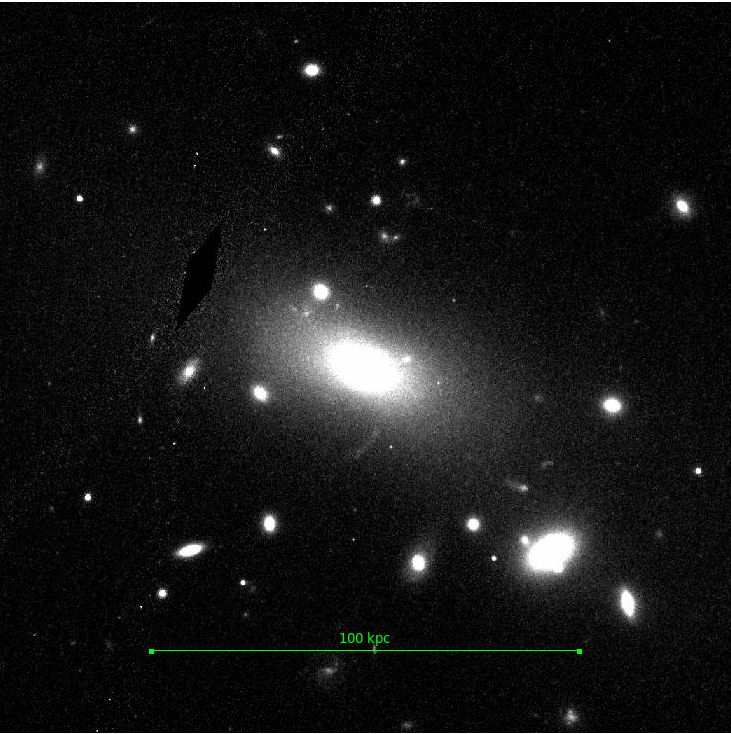}
    \caption{The BCG in RX J2129+0005 at redshift z = 0.234 has an extended and luminous halo, which makes it difficult to accurately estimate the a/b axis ratio. In this case, the major axis has most likely been overestimated, as the diffuse light is extended along this axis. The image was taken with the F775W ACS filter.}
    \label{fig:RX21290005_BCG}
\end{figure}

We can identify edge-on spiral galaxies, which appear very elongated. They can be filtered by making cuts in elongation (defined in SExtractor as the ratio of the galaxy's major to minor axis). 
As will be explained in \Cref{subsubsection:red_sequence}, we consider two different filters.
We define two different cuts depending on the filter we are looking at: in the bluest filter, we apply the criterion ELONGATION $\leq$ 2.3, and in the reddest filter, ELONGATION $\leq$ 2.6. The latter limit may seem quite high to filter efficiently all edge-on spiral galaxies. However, because of deblending issues, measuring with precision the lengths of the major and minor axes of the sources can be difficult, and will sometimes lead to a very elongated object. A very bright and elongated halo around the BCG, which can be linked to the ICL, will possibly return a high a/b axis ratio. This is the case for the BCG in RX~J2129+0005, which has the highest a/b elongation (in the F606W filter) measured in our sample, reaching a/b=2.57 (see \Cref{fig:RX21290005_BCG}). As the reddest filter is more sensitive to the ICL, we prefer to define a limit that is not too strict on this filter. It will not eliminate all edge-on galaxies (for that, we should lower the limit), but we cannot take the risk of filtering out any of the BCGs we are looking for. That's why we define a different, stricter limit on the bluest filter, as it will be more sensitive to the blue stellar population present in the disk of spiral galaxies, and less to the ICL.

{}

\subsubsection{Selection of red cluster galaxies}
\label{subsubsection:red_sequence}
Early-type galaxies in clusters are usually easily recognizable by their red  colors, since they are mostly red elliptical galaxies, without star formation. While blue spiral galaxies also exist inside the cluster, they are a minority, and red elliptical galaxies draw a red sequence in a color-magnitude diagram, which has a low dispersion. We thus apply a filter in color in order to only keep the red galaxies that form the red sequence.

To extract all the red early-type galaxies in a cluster at redshift z, we model their color using a spectral energy distribution (SED) template from \citet{bruzual2003stellar}. The model is similar to the one used by \citet{hennig2017galaxy}: a single period of star formation beginning at redshift z$_f$ = 5, with a Chabrier IMF and solar metallicity, that decreases exponentially with $\tau = 0.5$~Gyr. However, the model differs from \citet{hennig2017galaxy} on the star formation redshift: we chose a higher z$_f$ to better model clusters at higher redshifts, as \citet{hennig2017galaxy} limit their study to redshift z = 1.1. We reject all blue galaxies (blue-red $\leq$ 0) and all galaxies whose measured (blue-red) color differs by more than 0.60 magnitude from the model. While the red sequence of a cluster presents a rather narrow color-magnitude relation, and therefore very little dispersion, this large limit of 0.60 was fixed in order to take into account photometric uncertainties due to deblending issues, redshift uncertainties, or simply to the accuracy of the model used (Charlot, private communication). 
The color is computed considering a fixed aperture of 35 kpc in diameter (parameter MAG\_APER), which is large enough to contain all of the galaxy's light. All magnitudes are K-corrected (K-correction values are taken from the EZGAL BC03 computed model) and we also take into account galactic extinction. Dust maps were taken from \citet{schlegel1998maps} and reddening values for the ACS and WFC3 bandpasses were taken from \citet{Schlafly_2011}, considering a reddening law R$_{V}$ = 3.1. The color computed will depend on the filters available and on the redshift of the cluster. The colors computed for each cluster can be found in \Cref{tab:BCGs_coord}.

\begin{figure}[ht]
    \centering
    \includegraphics[width=\linewidth]{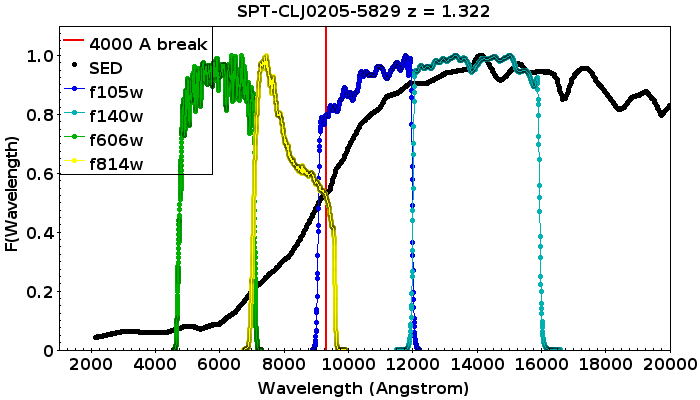}
    \caption{SED of an elliptical galaxy from the CFHTLS (black solid line) redshifted at the cluster's redshift (SPT-CL J0295-5829, z = 1.322). The filter transmissions are normalized to 1 for better visualisation, and the break at 4000~\AA\ is marked as a red vertical line for reference. In this case, the chosen (blue-red) rest frame color is F814W-F105W, and the filter we choose for the final step (modelisation of the luminosity profile with GALFIT) is F140W (see \Cref{section:luminosity_profiles}).}
    \label{fig:SED_filters}
\end{figure}

The rest frame (blue-red) color to compute is defined as the color based on two magnitudes with the smallest wavelength difference that bracket the 4000~\AA\ break at the cluster redshift. Depending on what filters are available for each cluster, the selected filters will differ. An example is given in \Cref{fig:SED_filters}, for a cluster at z = 1.322; in this case, the filters bracketing the 4000 \AA\ break are F814W and F105W \footnote{The colors F606W - F625W and F775W - F814W were excluded as the two filters are really close to each other.}. 
In the cases where the two optimal filters are not available, the color used for tracing the red sequence galaxies at different redshifts may not be efficient; for instance, the use of the  color (F606W - F140W) would not enable us to optimize the selection, as a galaxy at higher redshift (z = 1.65 for example) than the cluster redshift (here z = 1.322) would have the same  color and will not be filtered out. 

\subsubsection{Rejection of spiral and isolated galaxies}
\label{subsubsection:spiral_isolated}

The cut in colors is an important step that allows to remove most of the spiral galaxies and to maximize the number of ellipticals in our catalogues. However, a few foreground galaxies may still remain, and we describe here the method used to remove them.

The algorithm described hereafter will be applied to every single galaxy, from the brightest to the faintest, until the cluster BCG is found. We refer the reader to the sketch shown in \Cref{fig:schema_detec}. The procedure includes the following steps: 

\begin{figure}[ht]
    \centering
    \includegraphics[width=\linewidth]{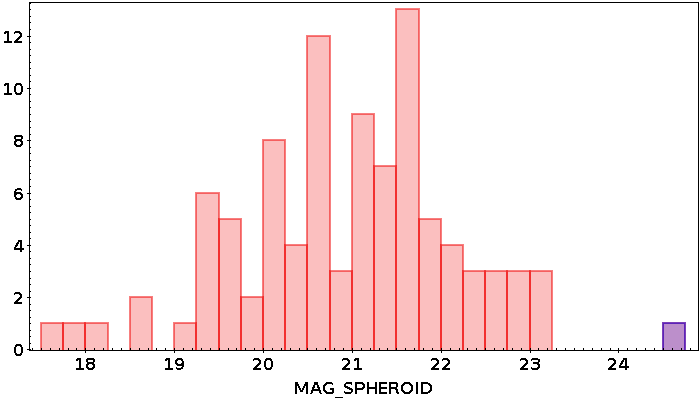}
    \caption{Histogram of the modeled bulge magnitudes (parameter MAG\_SPHEROID) returned by SExtractor. All BCGs are shown in red, the blue point represents a spiral galaxy close to the cluster ClG J1604+4304 (see Step 2).}
    \label{fig:hist_mag_spheroid}
\end{figure}

\begin{itemize}
    \item Step 1: for each cluster, we sort the catalog from the brightest to the faintest galaxies, and going down their brightnesses, we exclude galaxies that are too isolated from the rest (explained hereafter). We define the BCG as the brightest elliptical galaxy at the center of the galaxy density distribution. The method to calculate the center is as follows:
    
    We compute N$_{neigh}$, the number of cluster members, i.e. red sequence galaxies, found in a fixed aperture of 200 kpc radius centered on each galaxy in the final catalogue, and note N$_{max}$ the maximum number computed.
    If N is the total number of red sequence galaxies whose colors fall within 0.60 magnitude from the model, and N$_{neigh}$ is the number of neighbours of a given galaxy in an aperture of 200 kpc, we consider that a galaxy is isolated and unlikely to be the BCG if the ratio P = N$_{neigh}$/N is smaller than 40$\%$ of P$_{max}$ = N$_{max}$/N.
    
    Considering that we cropped our images to cover a projected area of 1.2x1.2 Mpc$^{2}$, the aperture of diameter 400 kpc taken here represents one third of the side of the images. This is small enough to detect high density areas on the image, and big enough to work on clusters with a high spatial extent. After several trials adopting different values, the value of 200 kpc radius is the one that works best. Less than 200 kpc becomes too small for extended clusters, while a higher radius makes it difficult to detect the smaller density fluctuations, as the covered area becomes large. The limit defined at R$_{lim}$ = 0.40 R$_{max}$ was also determined after several tests. This condition allows to take into account the cluster richness and spatial extent, as well as the possible offset of the BCG relatively to the cluster center. 

    \item Step 2: the next step consists in filtering out the last spiral galaxies that remain among the potential BCG candidates. We run SExtractor to model the potential BCG with a bulge and a disk component. We find that spiral galaxies have a very faint bulge (parameter MAG\_SPHEROID): as can be seen on \Cref{fig:hist_mag_spheroid}, a spiral galaxy (shown in blue) near the cluster ClG J1604+4304 prevented us from successfully detecting the BCG. We see a gap in magnitude between the spiral galaxy and the other BCGs (which are not all pure ellipticals). This enables us to define a new cut in magnitude to remove these remaining spirals: MAG\_SPHEROID $\leq$ 24.

    \item Step 3: if a galaxy complies with these conditions, i.e. not being isolated and not being a spiral, we keep it as BCG$_{1}$ if no other BCG candidate was found before, and as BCG$_{2}$ otherwise. We don't proceed to the next step until a BCG$_{2}$ is defined.
    
    \item Step 4: we check if there are more red sequence members in the same aperture for 
    BCG$_{2}$ than the number defined in Step 1 for BCG$_{1}$, by comparing their P$_{Neigh}$ ratios (defined in Step 1). We will note them P$_{BCG1}$ and P$_{BCG2}$. If P$_{BCG1}$ $\leq$ 0.95 P$_{BCG2}$, and if less than 30\% of N$_{Neigh, BCG2}$ are in common with BCG$_{1}$, BCG$_{1}$ is eliminated and we define BCG$_{2}$ as the new BCG$_{1}$. 
    
    The factor of 95\% ensures that the overdensity in which BCG$_{2}$ resides is significantly richer than the one in which BCG$_{1}$ is. The second criterion on the number of common galaxies to BCG$_{1}$ and BCG$_{2}$ is to make sure that we are not replacing a BCG that is not at the very center of the cluster by another galaxy that is closer. This criterion is necessary to avoid eliminating BCGs that are a little offset from the center of the cluster, where the density is higher.
    It allows to check that the two galaxies are not in the same area in the sky, i.e, we check that BCG$_{2}$ does not belong to the same clump (overdensity) as BCG$_{1}$, or that the two galaxies do not belong to the same cluster. 

    \item Step 5: this step is taken only if BCG$_{2}$ is defined, otherwise we repeat the previous steps until it is found. If BCG$_{1}$ and BCG$_{2}$ are similar in sizes (ratio of the major axes a$_{BCG2}$/a$_{BCG1}$ $\geq$ 0.80) and brightnesses (magnitude difference mag$_{BCG2}$ - mag$_{BCG1}$ $\leq$ 0.2), we keep both BCG$_{1}$ and BCG$_{2}$ as the BCGs of the cluster. Otherwise, BCG$_{2}$ is eliminated and BCG$_{1}$ is defined as the BCG.
\end{itemize}

The values above do not always work for poor clusters, i.e., when the number of cluster members is low, or when the density of red sequence galaxies is low. There is no problem when all the cluster members are concentrated in the same area (with a size comparable to the previously defined aperture), but if the members are dispersed over the sky and cover a large area, an aperture of 200~kpc radius becomes too small to detect density fluctuations on the sky. We thus differentiate these clusters by their number of red sequence members, N, and by the previously defined parameter P$_{max}$ (see Step 1). We separate the poor clusters with P$_{max} \leq$ 0.25 and N $\leq$ 100 (very extended cluster with no important density clumps), and P$_{max} \leq$ 0.5 and N $\leq$ 40 (low number of red sequence galaxies, extended spatial distribution). We were not able to correctly determine the BCGs for these clusters by defining a 200-kpc radius aperture, so for these poorer clusters, we consider a bigger aperture of 500 kpc radius. To take into account the bigger aperture, we also modify the second criterion in Step 4. We check that the two BCGs candidates, BCG$_{1}$ and BCG$_{2}$, have less than 50\% galaxies in common in the same aperture, to guarantee that they are not both residing in the same cluster.

\subsubsection{Results for detected red BCGs}
\label{detect_results}

\begin{figure*}[ht!]
  \centering
  \begin{tabular}{@{}c@{}}
    \includegraphics[width=.32\linewidth]{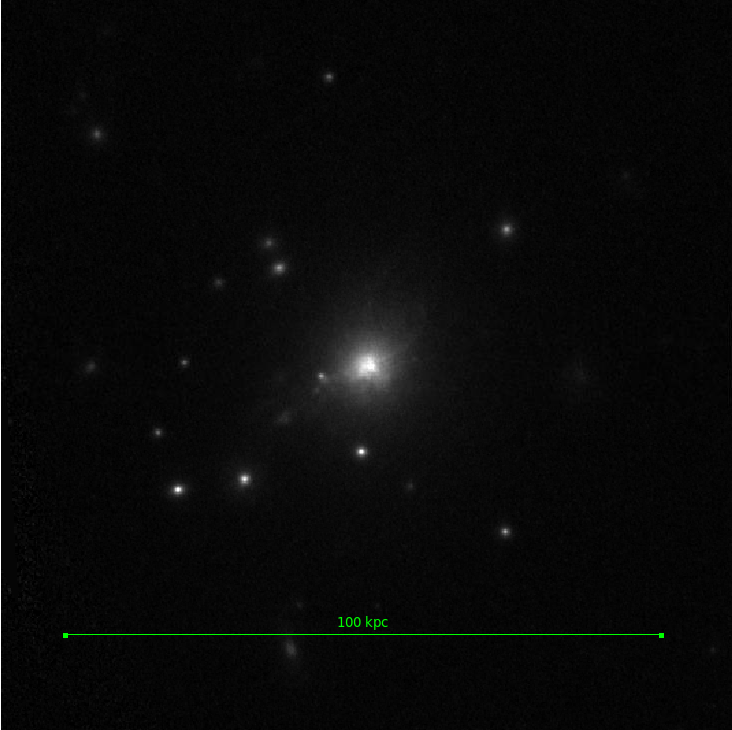}
  \end{tabular}
  \begin{tabular}{@{}c@{}}
    \includegraphics[width=.32\linewidth]{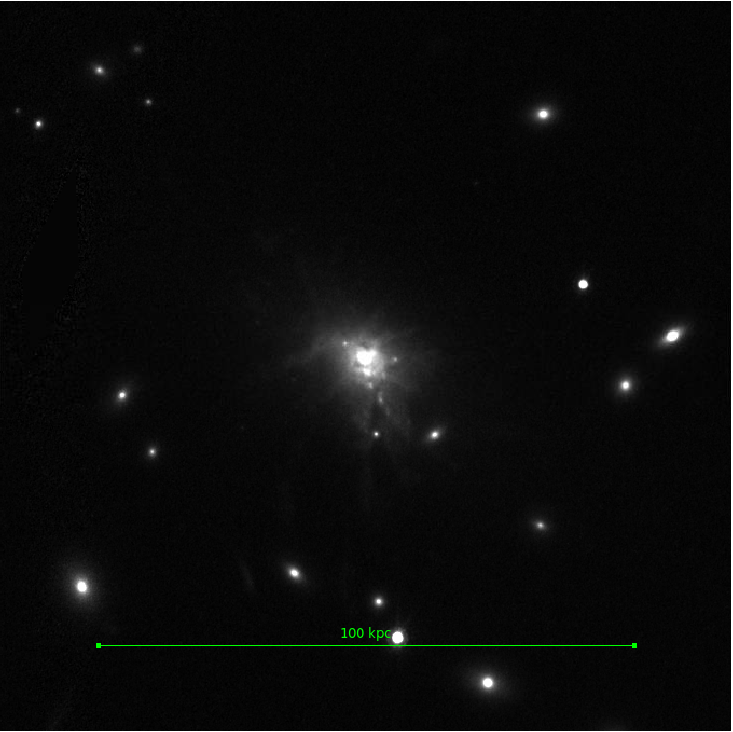}
  \end{tabular}
  \begin{tabular}{@{}c@{}}
    \includegraphics[width=.32\linewidth]{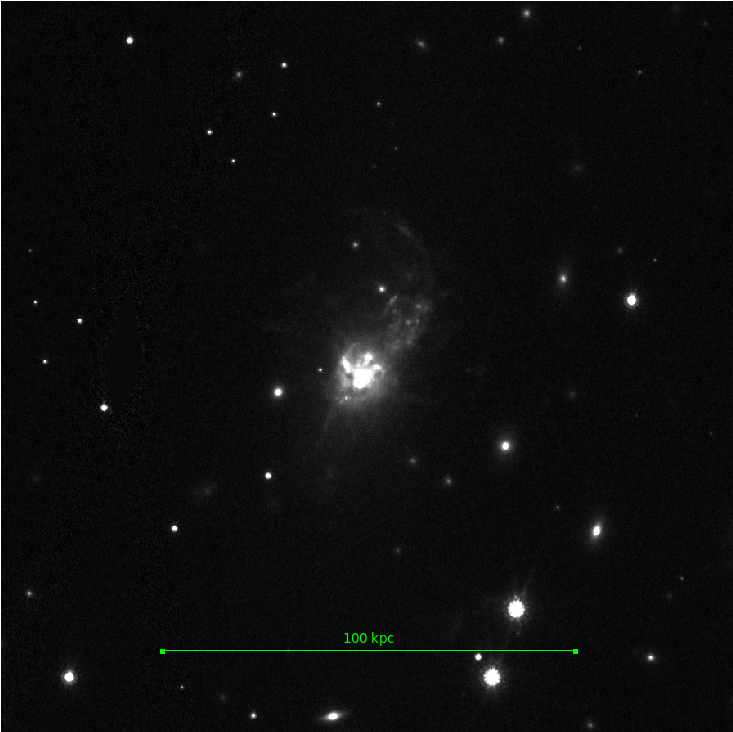}
  \end{tabular}
  \caption{From left to right: MACS J0329-0211 (z = 0.45): an example of a red star forming BCG, RX J1532+3020 (z = 0.3615),and MACS J1932-2635 (z = 0.352): the only two blue star forming BCGs in our sample.} 
    \label{fig:SF_BCGs}
\end{figure*}

Among the 137 clusters in our sample, 50 clusters only had one filter available, and were thus excluded from this procedure. For these 50 clusters without available colors, we checked visually the images to determine the BCG, and checked with X-ray maps or other studies before adding them to the final sample.

In order to assess the efficiency and accuracy of our detection method, we checked each detection visually and compared it with other studies and with any X-ray map we could find. We compared the X-ray map to the position of the detected BCG to make sure that it is not too far from the X-ray peak (but not necessarily located at the peak, in a radius of about 200 kpc). 

During this verification, we found that our detection differs from that of \citet{Durret_2019} and \citet{bai2014inside} for the BCGs in MACS-J0717.5+3745 and SpARCS-J0224 respectively. MACS-J0717.5+3745 presents a very complex structure as is it undergoing multiple mergers \citep[see][and references therein]{Limousin_2016, Ellien_2019}. \citet{Durret_2019} define the BCG as the one in the southern structure, whereas we detect a brighter galaxy in the northern structure, that we hence define as the BCG. We choose to keep our detection as it lies near the X-ray peak in the northern structure and is surrounded by galaxies at the cluster's redshift. We find, by checking visually, that the BCG in SpARCS-J0224 defined in \citet{bai2014inside} is a spiral galaxy. We thus choose to keep our detection, which is an elliptical galaxy located just south of their detection.

A few star forming BCGs can be found in our sample. We find red BCGs with a very high SFR: we give the example of MACS J0329.6-0211 at z = 0.45. This BCG has an almost starburst level of UV continuum and star formation \citep{donahue2015}. 
Images of this BCG in the UV continuum and  H$\alpha$-[NII] lines are given by \citet{Fogarty15}, illustrating the distribution of star formation throughout the galaxy. The high star formation rate of about 40 M$_{\odot}$ yr$^{-1}$ was confirmed by \citet{Fogarty17}, based on Herschel data.  \citet{Green16} also indicate that this galaxy hosts an AGN, and is quite blue ($blue-red=-0.71$), with strong emission lines and a rather high X-ray luminosity of $11.85\times 10^{44}$ erg~s$^{-1}$.

Overall, all the red BCGs, even the non pure elliptical BCGs, or those with not optimized colors because of the lack of available filters, were successfully detected with our method.
We successfully detect 97\% of the BCGs in our sample, and all the red BCGs are found. The method is effective to detect red BCGs presenting different morphologies and characteristics (mergers, star forming, traces of dust in the core, disturbed). 

It is to note, though, that this method may be less reliable for poorer clusters (as defined in the previous subsection). As we were conducting several tests, trying different values of apertures in which we computed the number of red sequence galaxies, or the threshold below which galaxies are considered as isolated, we found that the detection efficiency for poorer clusters was more sensitive to these parameters. As this method relies on the density of red sequence galaxies in a small aperture, BCGs that are a little offset from the density peak (which is more difficult to calculate for poor clusters with an extended spatial distribution) could be eliminated, and rejected as being isolated from the other red sequence galaxies.

It may also be important to note that, in the presence of more than one cluster, i.e. two clusters interacting with each other, or in the case of superclusters, only the brightest galaxy of one substructure will be detected. For MACS-J0717.5+3745 for example, which we already mentioned, the BCG of the northern clump being the brightest, it is the one that is detected by our algorithm.

\subsection{Finding blue BCGs}
\label{subestion:blueBCGs}

Out of the 98 BCGs (87 clusters, 11 clusters with two BCGs) in our sample which we tried to detect, we find two peculiar BCGs with blue colors.

BCGs are in majority quiescent galaxies, and their dominant stellar population is typically red and old. As they grow by undergoing mergers through time, all their gas is consumed, and we expect the star formation to be quenched or suppressed. However, we do observe, both today and in the distant universe, BCGs with intense UV emitting filaments or knots, hinting at active star formation. 
\citet{Cerulo_2019} found that 9\% of his sample of massive BCGs in the redshift range 0.05 $\leq$ z $\leq$ 0.35 from the SDSS and WISE surveys have blue  colors (which they define as galaxies with colors $2 \sigma$ bluer than the median  color of the cluster red sequence), and are star forming. What we will refer to from now on as star forming BCGs (SF BCGs) have only been observed in cool core clusters so far. Their morphology can be quite different from that of a simple elliptical galaxy, as was stated before. These galaxies can appear disturbed, with a complex structure showing a possible recent or ongoing merger. Such examples of SF BCGs show that BCGs 
are not all simple ellipticals.

Two BCGs, RX J1532+3020 (z = 0.3615) and MACS J1932-2635 (z = 0.352), were not correctly detected as they are cool core BCGs with an extremely active star forming center, so they were eliminated because of their blue colors. These two BCGs were identified by eye and added manually, after checking and confirming with other studies. Their images are shown in \Cref{fig:SF_BCGs}.

These two BCGs are the only blue BCGs in our sample (blue meaning a negative rest frame blue-red color), out of the 98 BCGs for which we compute a color. While comparing with other studies, we find a few other BCGs that are star forming, but are still red. 

RX J1532+3020 is one of the most extreme cool core galaxy clusters observed today, as well as one of the most massive. An intensive study by \citet{Hlavacek_Larrondo_2013} shows the existence of a western and an eastern cavity, which are used to quantify the AGN feedback at the center of the galaxy. These authors estimated that this feedback would release at least 10$^{45}$ erg s$^{-1}$, which would prevent the Intra Cluster Medium (ICM) from cooling, and would then allow to solve the cooling flow problem in cool core clusters. The BCG of this cluster is a radio loud galaxy that presents in its central regions UV filaments and knots, as well as traces of dust, hinting at recent star formation, with a SFR of at least 100~M$_\odot$ yr$^{-1}$ \citep{castignani2020molecular}. A strong and broad Lyman $\alpha$ emission and stellar UV continuum, and no other emission lines, have been observed by \citet{Donahue_2016}. CO with a large reservoir of molecular gas as well as a high level of excitation were also detected by \citet{castignani2020molecular}.

MACS J1932-2635 is another cool core cluster with a huge reservoir of cold gas in the core, of mass (1.9$\pm$ 0.3) $\times$ 10$^{10}$ M$_{\odot}$, which makes it one of the largest reservoirs observed today, in which \citet{Fogarty_2019} detected CO emission as well as UV knots and H$\alpha$ filaments around the BCG. They  measured a SFR of 250 M$_{\odot}$ yr$^{-1}$ and also observed an elongated tail that extends to the northwest, with traces of cold dust in the tail, which they suspect might be caused by a recent AGN outburst.

In order to detect these blue BCGs, we would have to relax the condition on the color. However, this condition is necessary in order to remove most of the spiral galaxies, and we find that allowing galaxies with blue colors will make the method much less reliable, as the red sequence will be ill-defined. Our method is thus only reliable to detect red BCGs, even if they are not pure ellipticals (star forming or merging galaxies for example).


\section{Luminosity profiles}
\label{section:luminosity_profiles}

We fit 2D analytical models on sources with GALFIT \citep{peng2002detailed}. Once the BCG is defined, we run SExtractor one last time to return model fit parameters in the available filter closest to the F606W rest frame at redshift z (see \Cref{fig:SED_filters}), which is at a wavelength above the 4000~\AA\ break and thus is in the spectral region where we will get the highest flux. The chosen filters can be found in \Cref{tab:BCGs_coord}. We note that there are 37 BCGs out of the 149 for which HST data are not available in the F606W rest frame or redder. The reddest filter is either bluer than the 4000~\AA\ break or contains it, which means that we are not only looking at the oldest, reddest star population, but at the youngest  bluest stars as well. These BCGs are marked by blue squares in plots. The redshift distribution of all our BCGs is plotted in \Cref{fig:hist_redshift}, the blue histogram represents the clusters with filters which are bluer than the 4000 \AA\ break. These clusters observed in too blue filters are mainly between redshifts 0.7 and 1.2.
 
\subsection{Masking}

\begin{figure*}[ht!]
  \centering
  \begin{tabular}{@{}c@{}}
    \includegraphics[width=.32\linewidth]{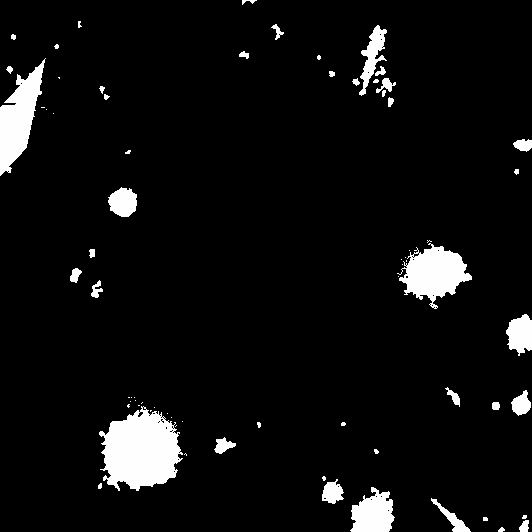}
  \end{tabular}
  \begin{tabular}{@{}c@{}}
    \includegraphics[width=.32\linewidth]{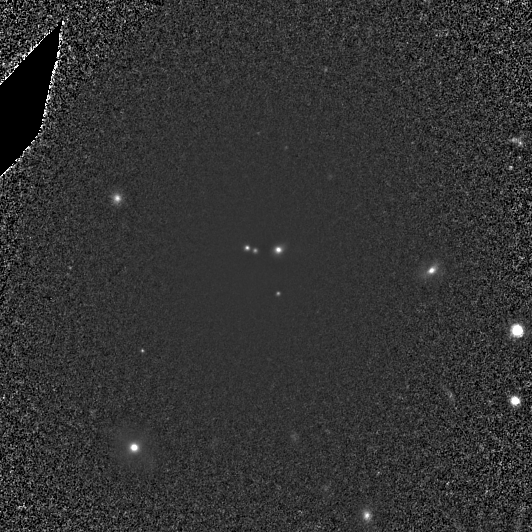}
  \end{tabular}
  \begin{tabular}{@{}c@{}}
    \includegraphics[width=.32\linewidth]{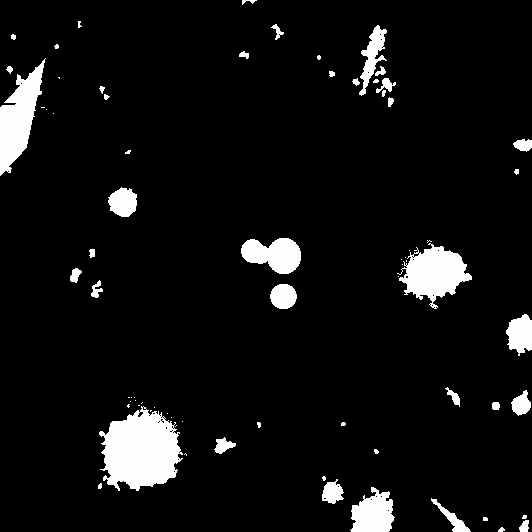}
  \end{tabular}
  \caption{Example of the BCG in the cluster Abell 2261, at z = 0.224. (Left) The segmentation map returned by SExtractor, with the BCG unmasked. The  pixels with a value of 1 are masked, those with a value of 0 are unmasked. (Middle) Sharp divided image, in which four knots in the core appear. These knots were drowned in the light of the BCG and are now visible. (Right) Final mask (including the central objects).}
\label{fig:1036412_sd}
\end{figure*}

We first need to mask all the neighbouring sources. We take the SEGMENTATION map returned by SExtractor, and unmask the BCG (which is identified by an identification number), and also mask any blank region on the image. Because of deblending issues, it is more than likely that other objects, projected on the BCG, need to be masked.

We use sharp divided images to detect any neighbouring objects that pollute the signal. Sharp divided (SD) images \citep[see e.g.][]{marquez1999near,marquez2003detection} are obtained by dividing the images by the median filtered corresponding images. This brings out all the small neighbouring sources that may have been hidden by the luminous halo of the BCG. We run SExtractor (again) on this SD image, and mask all the objects that are farther than 0.5 arcsec from the BCG coordinates (an example is given in \Cref{fig:1036412_sd}), which is the minimum distance required to avoid masking the BCG center. As can be seen on \Cref{fig:1036412_sd}, the sources masked based on the SD image detection seem larger on the final mask than on the SD images, as the SD image does not show the true sizes of the objects. We apply a factor of 6 to the minor and major axes of the sources detected by SExtractor on the SD image to create our final mask. This factor allows to include all the luminosity of the sources and to mask them efficiently. If necessary, we identify by eye and draw the regions to be masked ourselves in SAOImage DS9 and create a new mask.

\subsection{PSF model}

To obtain a successful model of the galaxy profiles that also works for the inner regions, an accurate description of the PSF is needed. While the PSF we used for the photometry may have been sufficient to distinguish stars from galaxies, GALFIT requires the PSF to meet a number of criteria: a very high SNR, a flat and zero background (if not, any pattern in the background will appear on the model image when convolved with the PSF); it should match the image (diffraction rings and spikes, speckle pattern, etc.) and be correctly centered (see GALFIT Technical FAQ). 

We first substract from the images the sky background, which is determined by masking all sources and blank areas on the cluster image, using the routine calc\_background with a 3$\sigma$ clipping method.
We then use PSFex, and make a selective sample of the stars that will go into making the PSF. We select all point-like sources with FLAGS = 0, MAG\_AUTO $\leq$ 21, ELONGATION $\leq$ 1.1, CLASS\_STAR $\geq$ 0.98, SNR $\geq$ 20 and an isophotal print ISOAREA\_IMAGE $\geq$ 20 pixels.

Since we work on HST observations that cover a small field of view, there may not be many bright stars in the field of the cluster that we could use to compute a PSF. We tried to take several faint stars and stack them to increase the SNR of the PSF. However, we find that this often results in a PSF with an uneven background, which stands out during the model fitting returned by GALFIT, and this usually ends up being a bad fit (too large effective radius, large uncertainties). Since we are working on space observations, the PSF does not vary much, and though it may vary with time, the variations should not be significant  \citep[see][]{martinet2017faint}. This means that we can replace the PSF for a given filter by another one in the same filter with a better SNR. Higher SNR PSFs return better fits. 

Modeled and theoretical PSFs are available for ACS/WFC and WFC3/IR. However, according to GALFIT Technical FAQ\footnote{https://users.obs.carnegiescience.edu/peng/work/galfit/TFAQ.html} (which we refer the reader to), the profiles obtained with models may not be realistic for space-based images, so we prefer to use observed PSFs. 
\subsection{Profile fitting}

We use GALFIT to fit two different models to our BCGs: a single S\'ersic component or two S\'ersic components, to allow different contributions from the inner and outer parts of the galaxies. We also tried to apply other models or combinations of models including a de Vaucouleurs profile, but they always provided worse results (i.e. they gave a worse $\chi^{2}$, and about 30\% of the BCGs were not well fitted with one or two de Vaucouleurs profiles), so we will only discuss the results with S\'ersic fittings.

One needs to give GALFIT an estimate of all the initial parameters: the effective surface brightness or total magnitude, the effective radius (the radius at which half of the total light of the galaxy is contained), the elongation or the position angle (PA) of the BCG. These initial guesses are taken from the SExtractor catalogs: MU\_EFF\_MODEL or MAG\_AUTO, FLUX\_RADIUS, ELONGATION, and THETA\_IMAGE. We don't have an estimate of the BCG S\'ersic index, so we start from the value corresponding to the de Vaucouleurs profile: n = 4. If the fitting does not converge, we try different S\'ersic indices in the range 0.5 to 10. For the second S\'ersic component that accounts for the inner part, the following parameters are considered: MU\_EFF\_SPHEROID, SPHEROID\_REFF\_IMAGE, SPHEROID\_SERSICN and SPHEROID\_THETA\_IMAGE. The suffix SPHEROID refers to the bulb component when SExtractor tries to model a disk and a bulb to a galaxy. We consider an elongation (minor to major axis ratio, b/a) of 0.90 for the inner part, as an initial guess. The region to fit is a box that is 2.5 r$_{Kron}$ wide (cf. GALFIT FAQ), r$_{Kron}$ being the Kron radius returned by SExtractor. This is large enough to contain all the light from the BCG as well as some sky background, and is a good compromise to obtain good fits of our galaxies.

We first run GALFIT to fit the BCGs with one single S\'ersic component. If it does not manage to converge with a S\'ersic index of n = 4, we try different values between 0.5 and 10 until it converges to a meaningful fit, and reject any fit with returned effective radius larger than half the size of the fitting region, which is to say R$_e \leq$ 2.5 r$_{Kron}$/2 pixels. We then use the output parameters as initial guesses to fit the outer part of the galaxy and add another S\'ersic component to fit the inner part of the galaxy. If it does not converge towards meaningful values, we increment the S\'ersic index until it manages to fit the BCG. For pairs of BCGs (two brightest cluster galaxies with similar sizes and magnitudes), we fit both of them simultaneously.

\subsection{Choice of the best fit model}

The quality of the fit can be estimated from the reduced $\chi^{2}$ ($\chi^{2}_{\nu}$), which should be close to 1. From our results, we notice that $\chi^{2}_{\nu} > $ 1.2 or $\chi^{2}_{\nu} < $ 0.8 often indicate a bad fit.
This happens when the model used to fit the BCG is not adapted, or when the initial parameters given are bad estimates. In this case, GALFIT may also not have converged and/or crashed.

To decide if a second component is really necessary to fit the BCG, or if one component gives equally good results, we use the F-test \citep[][]{simard2011ApJS..196...11S,margalef2016MNRAS.461.2728M}. 

As stated in \citet{bai2014inside}, as the background noise is not gaussian, the meaning of $\chi^{2}_{\nu}$ is not as significant, and when comparing two models, a $\chi^{2}_{\nu}$ closer to unity does not necessarily mean that it is a better fit. So we prefer to use a F-test.

The F-test states that if the P-value, determined from the F-value and the number of degrees of freedom, is lower than a probability P$_0$, then you can reject the null hypothesis and consider that the second model gives a significantly better result than the simpler one. The F-value is defined as the ratio of the reduced $\chi^{2}$, $\chi^{2}_{\nu}$, of both models. GALFIT returns the $\chi^{2}$ as well as the $\chi^{2}_{\nu}$ of the fit, but instead of directly considering the output $\chi^{2}_{\nu}$ computed by GALFIT, we compute $\chi^{2}_{\nu}$ as:
\begin{equation}
    \chi^{2}_{\nu} = \frac{\chi^{2}}{n_{dof}}
\end{equation}
\noindent
with n$_{dof}$ the number of degrees of freedom, which is here defined as the number of resolution elements, n$_{res}$, minus the number of free parameters in the model, n$_{free}$. n$_{res}$ can be calculated as follows \citep[see][]{simard2011ApJS..196...11S,margalef2016MNRAS.461.2728M}:
\begin{equation}
    n_{res} = \frac{n_{pixels}}{\pi \theta^{2}}
\end{equation}
\noindent
where $n_{pixels}$ is the number of unmasked pixels used for the fitting, and $\theta$ is the full width at half maximum (FWHM) of the given PSF, in units of pixels. $n_{dof}$ is then:
\begin{equation}
    n_{dof} = n_{res} - n_{free} - 1
\end{equation}
The P-value is then calculated with the routine f.cdf from scipy.stats in python. We set $P_0 = 0.32$, which represents a $1\sigma$ threshold value (Margalef private communication). 

\begin{figure}[t]
    \centering
    \includegraphics[width=\linewidth]{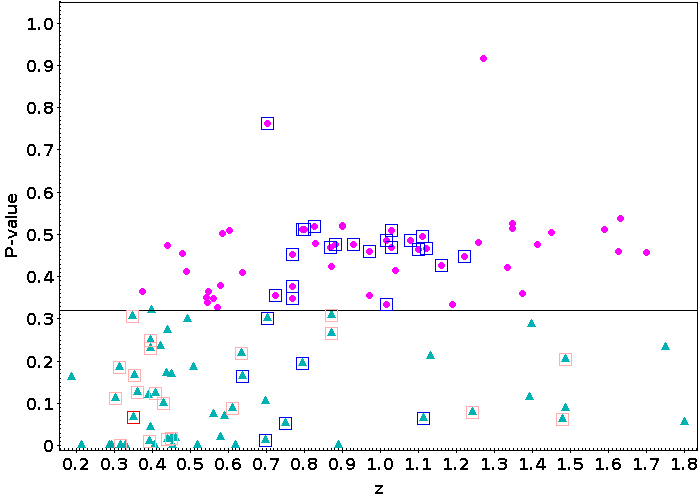}
    \caption{Distribution of P-values as a function of redshift considering a model with one S\'ersic component (magenta), and a model with two S\'ersic components (cyan). BCGs which could not be fitted by either model are not included.}
    \label{fig:redshift_Pvalue}
\end{figure}

\begin{figure}[t]
    \centering
    \includegraphics[width=\linewidth]{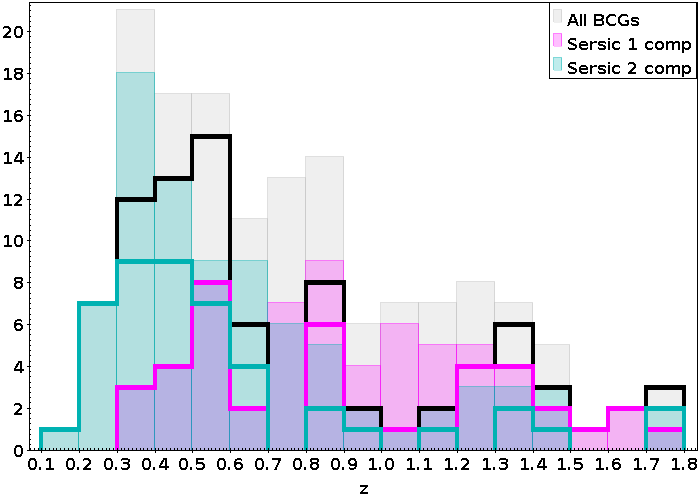}
    \caption{Distribution of redshifts for each model: a single S\'ersic component (magenta) and two S\'ersic components (cyan). In grey is represented the overall distribution. The semi-filled histograms represent the initial sample, and the unfilled histograms only contain BCGs with appropriate data.}
    \label{fig:hist_Pvalue}
\end{figure}

We show the distribution of P-values as a function of redshift in \Cref{fig:redshift_Pvalue}. A P-value $\leq$ P$_{0}$ means that we need a second component to correctly model the BCG light distribution. On this plot are not represented BCGs which could not be fitted by either model: 9 BCGs could only be fitted with a single component, 22 could only be fitted with two components, 2 could only be fitted by fixing the S\'ersic index n = 4 (de Vaucouleurs profile), and 2 BCGs could not be fitted or returned a poor fit for either of the models.
On \Cref{fig:hist_Pvalue}, it appears that BCGs that need a second component to obtain a good fit tend to be at lower redshifts (peak at z = 0.3), while the distribution for those that were well fitted with a single component is flatter. We also find BCGs with a model with two S\'ersic profiles at higher redshifts (14 BCGs at z $\geq$ 1.0). If the chosen model depended on the distance, we would have expected not to have two component BCGs at higher z, which is not the case. 

We must remember that part of our sample (37 BCGs) is studied in a too blue rest frame filter, and for these clusters we are not looking at the same star population. Without taking into account those observed in too blue filters, we find that 55 out of 72 BCGs (76\%) at redshift z $\leq$ 0.8 need a second component, while the trend is reversed at z $>$ 0.8, as 23 out of 38 BCGs (61\%) can be well modeled with only one component. We also notice that the BCGs observed with too blue filters can in majority (62\%) be modeled with only one S\'ersic.

We can also wonder if the existence of these two distinct populations (BCGs with two components at low z, and BCGs with a single component throughout redshift), with a limit around redshift z = 0.8, may be due to the fact that BCGs at higher redshifts will be less resolved than their lower redshift counterparts. 

To test this hypothesis, we bring a sample of 44 BCGs at redshifts z $\leq$ 1.0 to a common physical scale at redshift z = 1.2. We smooth the images with a gaussian and repeat the previous steps. The $\sigma_{gauss}$ of the gaussian to apply is calculated as :

$\sigma_{gauss}$ = sqrt($\sigma_{z = 1.2}^{2} - \sigma_{z, cluster}^{2})$
\noindent
with $\sigma_{z, cluster}$ computed from the FWHM of the image we want to degrade, and $\sigma_{z = 1.2}$ the $\sigma$ at the reference redshift z = 1.2, which was computed as:

$\sigma_{z = 1.2}$ = $\sigma_{z, cluster}$ * $\frac{pixscale_{z = 1.2}}{pixscale_{z, cluster}}$

Ouf of the 44 BCGs at z $\leq$ 1.0 on which we did this test, 30 returned similar results as those obtained with the original (unsmoothed) images. We also find that a few BCGs (seven) which were better fitted with two S\'ersic components can be modeled just as well, according to the F-test, with only one S\'ersic after smoothing the images. Surprisingly, the opposite also happened for seven other BCGs: among these seven BCGs, four could not initially be fitted with two components while the other three are close to the P-value limit, P$_{lim}$ = 0.32.
As 68\% of the tested BCGs showed no significant difference, we can confirm that the lack of resolution for the farthest BCGs does not cause the absence of an inner component for BCGs at higher redshifts. 

\subsection{BCGs observed in too blue filters}

In all that follows, when considering together the results from BCGs better fit with one or two S\'ersic components, we consider the values obtained for the outer S\'ersic component (R$_{e,out} \geq$ R$_{e,in}$).

As stated before, we have 37 BCGs observed in too blue filters (relatively to the 4000 \AA\ break). We must determine if they can be taken into account in our final study. For this, we run a test on 40 clusters with filters available on the blue side of the 4000 \AA\ break as well as appropriate red filters, to check if the returned parameters vary depending on the filters chosen.
We find that the absolute magnitude and mean effective surface brightness become fainter as the filter gets bluer. 
However, the dispersion is too big to simply correct for the offset to bring the BCGs observed in too blue filters to the appropriate red ones. This could be due to the fact that the filters on the blue side of the break do not always fall on the same spectral region on the SED (as the SED varies with redshift, and not all clusters were observed with the same filters). 
The effective radii can have their sizes halved when observed with too blue filters. 
As for the S\'ersic indices, we find that the BCGs that need a second component tend to have S\'ersic indices in bluer filters consistent with those measured in the appropriate red filters. The BCGs which could be fitted with only one S\'ersic have indices that vary without any clear pattern. These observations show that we cannot directly consider together the measurements obtained looking at different parts of the SED. Therefore, we chose to exclude the BCGs observed in too blue filters in what follows.

\begin{figure}[h!]
\centering
\includegraphics[width=\linewidth]{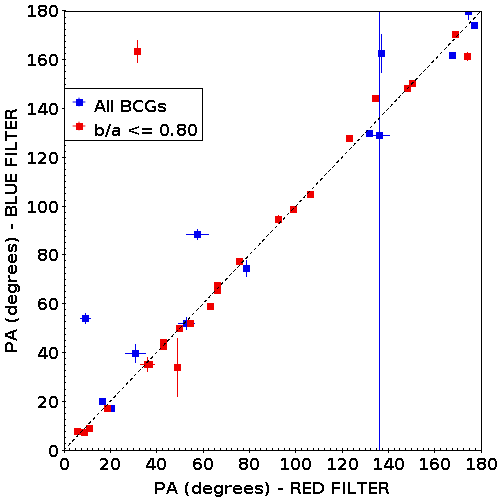}
\caption{Difference between the PA measured on the appropriate red filter (x-axis), and the one measured in a too blue filter (y-axis). In red are BCGs with elongations b/a $\leq$ 0.80.}
\label{fig:alignment_red_blue}
\end{figure}

We do however find that the position angles (PA) of the BCGs are not affected and remain consistent regardless of the filter chosen (see \Cref{fig:alignment_red_blue}). The PAs of these BCGs will thus be kept. Only one point presents a big difference between the two values (PA$_{red}$ - PA$_{blue}$ $>$ 120 degrees). We found that the ICL associated with this BCG is more extended in the reddest filter \citep[][also show that the ICL tends to be more extended in redder filters]{Ellien_2019}. The other BCGs with a significant difference between the values measured in the two different filters are circular in shape (b/a $>$ 0.80), so the PAs are ill-defined, which also explains the huge error bars. 

\section{Results}
\label{section:results}

As explained above, we tried fitting the BCGs with one or two S\'ersic profiles. In the following, the values plotted are those from the best model determined using the F-test (see previous section). The resulting parameters are summarized in \Cref{tab:GALFIT_out} and \Cref{tab:GALFIT_inn}. 

Two BCGs were not properly fitted by either model, but were correctly fitted by fixing the S\'ersic index n = 4 (corresponding to a de Vaucouleurs profile). We thus kept the parameters obtained with this fit. Two other BCGs were not correctly fitted by either model, and are thus excluded, bringing our sample size to 147 BCGs.

We summarize the total numbers of galaxies that were better fit with each model:
\begin{itemize}
    \item S\'ersic (1 component): 63 BCGs
    \item S\'ersic + S\'ersic (2 components): 84 BCGs
\end{itemize}

Without taking into account the BCGs observed in too blue filters we have: 
\begin{itemize}
    \item S\'ersic (1 component): 40 BCGs
    \item S\'ersic + S\'ersic (2 components): 70 BCGs
\end{itemize}

In all the plots shown in this paper, the BCGs better fitted with two S\'ersic components will be represented with triangles, and those fitted with only one component with diamonds.

\begin{figure}[h]
    \centering
    \includegraphics[width=\linewidth]{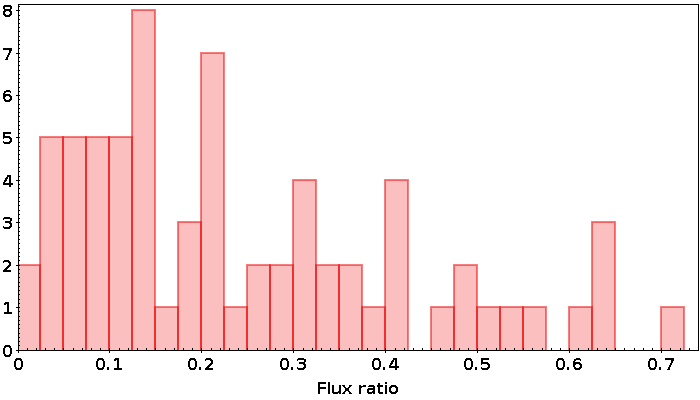}
    \caption{Histogram of the ratio of the flux of the inner component to the total flux of the BCG, for clusters better fit with two components. Clusters observed in too blue filters are excluded in this plot.}
    \label{fig:ratio_flux_inner}
\end{figure}{}

Before drawing conclusions, we need to know if we can consider together the results from BCGs better fit with one and with two S\'ersic components. 
In principle, the two subsamples can be put together if, for the two components, we assume that the outer component contains most of the light of the BCG and that the outer profile represents well enough the overall luminosity of the galaxy. The more important the contribution of the inner component to the total luminosity of the BCG is, the less accurate this statement will be. 
If an inner component is required to model the BCG, then the resulting outer profile obtained when fitting two components may not be comparable to a profile obtained with only one component. 
    
We show the histogram of the ratio of the inner component to total fluxes for the 70 BCGs requiring two S\'ersic components (see \Cref{fig:ratio_flux_inner}), and find that 24 BCGs present a very important inner component, which can contribute up to 30\% of the total luminosity of the galaxy. 
If we choose to ignore these 24 BCGs, no obvious difference can be seen in the overall relations observed in the following. However, we prefer to exclude them, as the outer profile may not be comparable to the profile obtained with a single component modeling most of the light of the galaxy. 

After excluding the galaxies with a very bright inner component and those observed in too blue filters, we obtain the final numbers:
\begin{itemize}
    \item S\'ersic (1 component): 40 BCGs
    \item S\'ersic + S\'ersic (2 components): 46 BCGs
\end{itemize}

In the following plots, BCGs observed in too blue filters will be marked with blue squares, and those fit with two S\'ersic profiles and with an important inner component will be marked with light pink squares. We will also identify the blue SF BCGs (cf. \Cref{section:detection}) with red squares and pairs of BCGs with black triangles.

\subsection{Evolution with redshift}

In order to study the evolution of the BCGs, we consider the dependence of the derived parameters as a function of redshift. 

\begin{figure*}[ht!]
  \centering
  \begin{tabular}{@{}c@{}}
    \includegraphics[width=.49\linewidth]{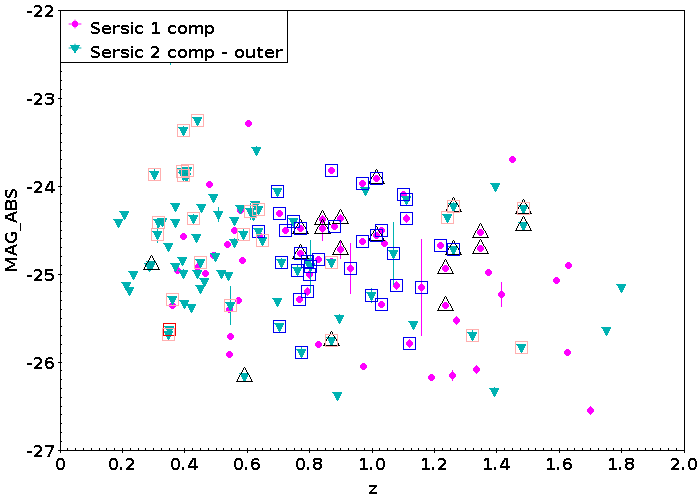} 
  \end{tabular}
  \begin{tabular}{@{}c@{}}
    \includegraphics[width=.49\linewidth]{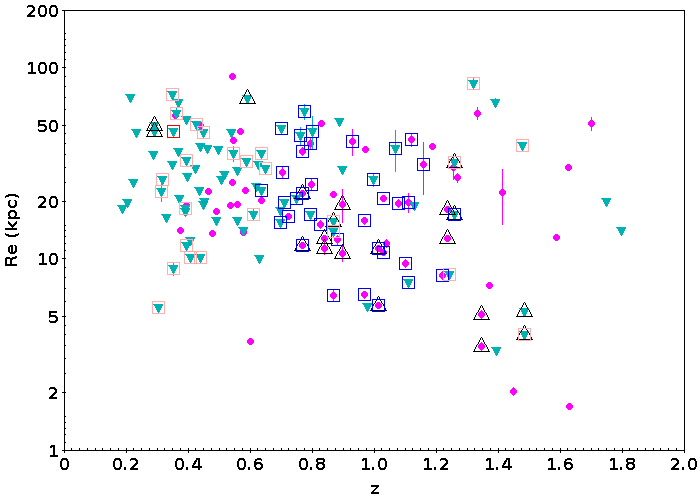}
  \end{tabular}
  \caption{(Left) Absolute magnitude and (Right) effective radius in logarithmic scale as a function of redshift. Cyan triangles are BCGs fit by two S\'ersic components (the outer component is considered), while the magenta points are BCGs fit with only one S\'ersic component. Blue squares represent the BCGs observed in too blue filters, red squares are SF BCGs (see \Cref{section:detection}), black triangles are pairs of BCGs, and the light pink squares are BCGs with an important inner component contribution.}
  \label{fig:evol_redshift_MAG_ABS_Re}
\end{figure*}

The absolute magnitudes of the BCGs, computed from the total apparent magnitudes (see \Cref{tab:BCGs_coord} for the filters considered) calculated by GALFIT, despite the very big dispersion (4 magnitudes thoughout redshift) tend to become brighter with redshift (see \Cref{fig:evol_redshift_MAG_ABS_Re}, left). The trend is faint, and can be quantified with a coefficient correlation R = $-0.29$ and a p-value p = 0.006563 (calculated from the coefficient R and the number of data points N \footnote{https://www.socscistatistics.com/pvalues/pearsondistribution.aspx}). By taking a significance level of $\alpha$ = 0.05, we show that we can reject the null hypothesis (p $< \alpha$) and conclude that the trend is significant.

There is a moderate trend for BCGs to grow with time (\Cref{fig:evol_redshift_MAG_ABS_Re}, right), as those with the smallest effective radii are at higher redshifts (z $\geq$ 1.2). The trend in logarithmic scale is quantified by a correlation coefficient R = $-0.40$, and with a p-value of p = 0.000142. BCGs observed in too blue filters generally have smaller effective radii than the others at a given redshift. Those with an important inner component contribution do not appear to occupy a special place in these relations.

\begin{figure*}[ht!]
\centering
    \includegraphics[width=0.49\linewidth]{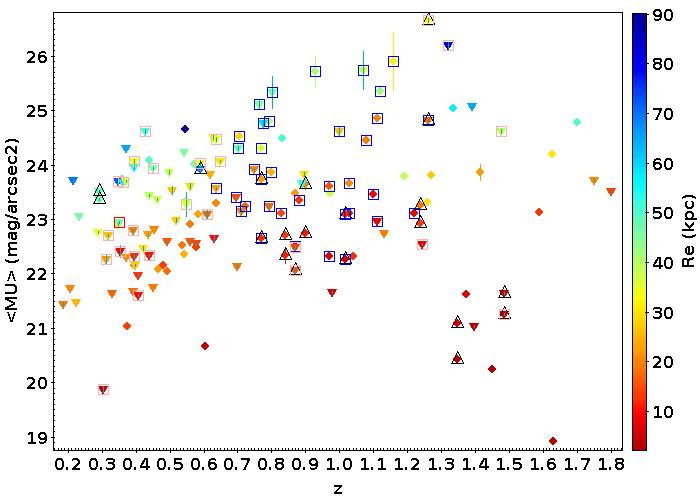}
\caption{Mean effective surface brightness as a function of redshift. See \Cref{fig:evol_redshift_MAG_ABS_Re} for color code. Additional information on the effective radii of the BCGs is shown on the right of the figure.}
\label{fig:redshift_MU_aux}
\end{figure*}

The mean effective surface brightness (\Cref{fig:redshift_MU_aux}) shows no significant evolution as a function of redshift (R $<<$ 0.1), with a very large dispersion (it spans a range of 6 magnitudes at z $\geq$ 1.25). Seven BCGs at higher redshifts (z $\geq$ 1.4) can be seen among the galaxies with the brightest mean effective surface brightnesses ($<\mu> \leq$ 22 mag arcsec$^{-2}$). We confirm that nothing peculiar was observed with these BCGs. Those observed in too blue filters and those with an important inner component contribution do not occupy a specific place in the relation.

The vertical gradient in color in \Cref{fig:redshift_MU_aux} shows that the large dispersion is also linked to the effective radius. As we go towards the biggest BCGs (increasing effective radii), the relation is shifted towards the fainter mean effective surface brightnesses. This is to be linked with the Kormendy relation, which will be shown in \Cref{subsection:kormendy}.

\begin{figure*}[ht!]
  \centering
  \begin{tabular}{@{}c@{}}
    \includegraphics[width=.49\linewidth]{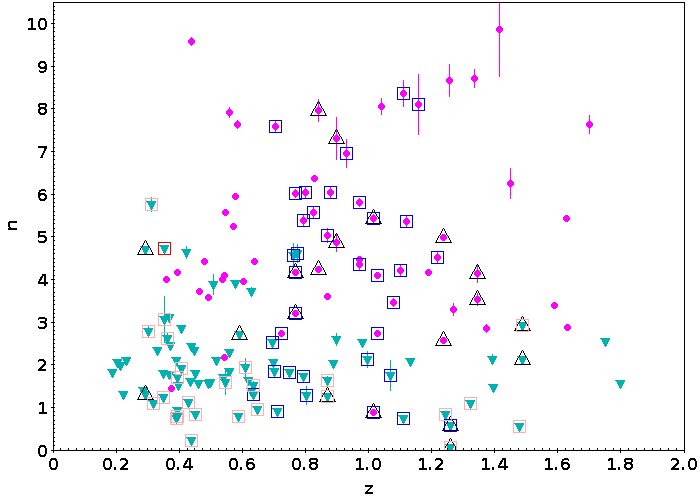}
  \end{tabular}
  \begin{tabular}{@{}c@{}}
    \includegraphics[width=.49\linewidth]{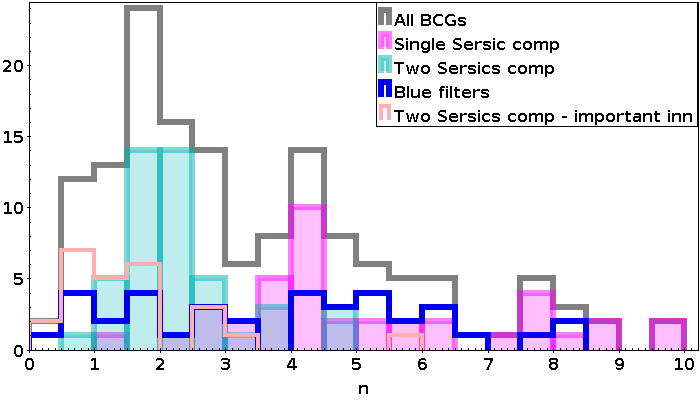}
  \end{tabular}
  \caption{(Left) S\'ersic indices as a function of redshift, see \Cref{fig:evol_redshift_MAG_ABS_Re} for color code. (Right) Distribution of the S\'ersic indices. All BCGs are represented in the grey histogram. The magenta and cyan histograms represent the distributions obtained with one and two components respectively, only for the BCGs with appropriate data. The blue and lighter pink histograms represent BCGs observed in too blue filters and with an important inner component, respectively.} 
\label{fig:redshift_n}
\end{figure*}

Finally, there is no correlation between the S\'ersic index and redshift (\Cref{fig:redshift_n}, left). However, on the right panel, we notice two different populations: the BCGs that were modeled with only one component generally have high S\'ersic indices with a strong peak at n$_{1comp,mean}$ = 4 (without considering the BCGs observed in too blue filters), while the BCGs that were better modeled with two S\'ersic components with lower S\'ersic indices show a peak at n$_{2comp,mean}$ = 2.0. 
    
\subsection{Kormendy relation}
\label{subsection:kormendy}

\begin{figure}[t]
    \centering
    \includegraphics[width=\linewidth]{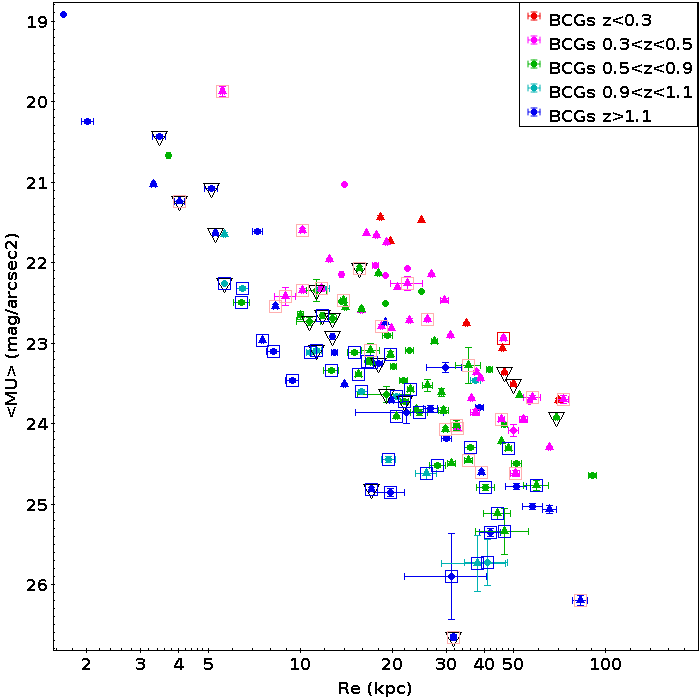}
    \caption{Kormendy relation \citep{kormendy1977brightness} using the parameters obtained with one S\'ersic component and the outer component of the two S\'ersic component model. Different colors represent different redshift bins. Symbols are the same as in \Cref{fig:evol_redshift_MAG_ABS_Re}.}
    \label{fig:kormendy_redshift}
\end{figure}

The Kormendy relation \citep{kormendy1977brightness} links the (mean) effective surface brightness of elliptical galaxies to their effective radius. This relation is plotted in \Cref{fig:kormendy_redshift}. The different colors represent different redshift bins: z $\leq$ 0.3, 0.3 $<$ z $\leq$ 0.5, 0.5 $<$ z $\leq$ 0.9, 0.9 $<$ z $\leq$ 1.3, and z $>$ 1.3. We show that all the BCGs seem to follow the Kormendy relation with the same slope, but the ordinate at the origin of the line decreases with increasing redshift.

While applying a linear regression to the relation obtained in each redshift bin (R $>$ 0.80, p $< \alpha$), we find that the slope remains quite constant at all redshift bins: m = 3.33 $\pm$ 0.73, whereas the ordinate at the origin varies as c = 2.15*z + 16.65.

\subsection{Inner component}

\begin{figure}[h]
    \centering
    \includegraphics[width=\linewidth]{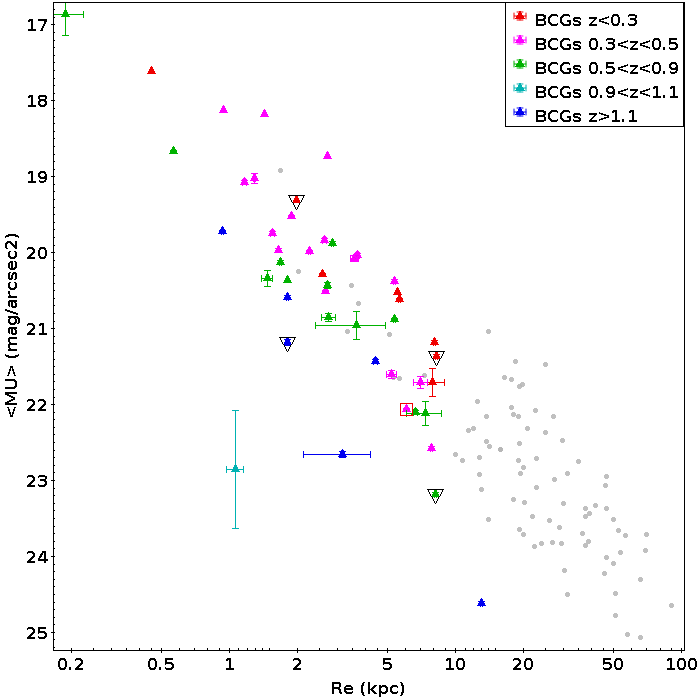}
    \caption{Same as \Cref{fig:kormendy_redshift}, but for the inner component of BCGs fitted with two components. Grey points are the same as in \Cref{fig:kormendy_redshift}.}
    \label{fig:kormendy_inner}
\end{figure}

The sample requiring an inner component consists of 46 BCGs. We find that the inner component follows a Kormendy relation (\Cref{fig:kormendy_inner}), and is a continuation of the Kormendy relation shown in \Cref{fig:kormendy_redshift} at brighter mean effective surface brightness and smaller effective radius (R$_{e, inn} \leq$ 20 kpc). 

We observe a very faint trend for the inner components to have brighter surface brightnesses with decreasing redshift, but the trend is not significant (R = 0.27, p = 0.06237 > $\alpha$). We don't find any clear correlation (correlation coefficient $\leq$ 0.2) between redshift and the absolute magnitude, effective radius or S\'ersic indices of the inner component of the BCGs.

\subsection{Alignment of the BCG with its host cluster}

\begin{figure}[h]
\centering
\includegraphics[width=\linewidth]{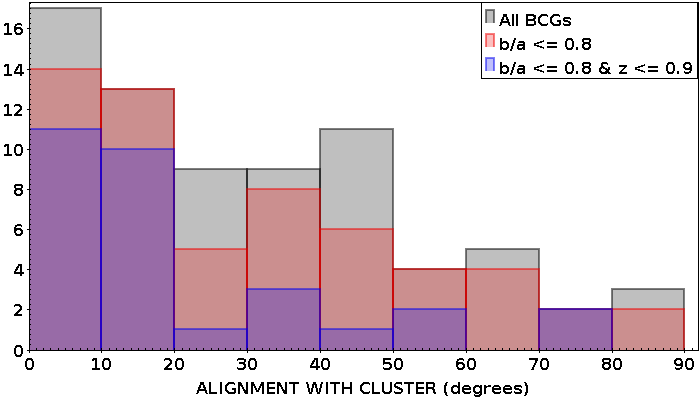}
\caption{Difference between the PA of the cluster \citep[see][]{west2017ten,Durret_2019} and that of the BCG as returned by GALFIT. Only clusters found in \citet{west2017ten} and \citet{Durret_2019} are included here. The histogram is to be compared to that shown in \citet{west2017ten}, as a way to check that our results agree with theirs. All BCGs are represented on the grey diagram, while only those with an axis ratio b/a $\leq$ 0.8 are included in the red diagram. We also exclude all BCGs at redshift z $>$ 0.9 on the blue histogram.}
\label{fig:hist_align}
\end{figure}

Some studies have shown that BCGs tend to have a similar orientation (hereafter PA) to that of their host cluster \citep{west2017ten,Durret_2019}. As a comparison, we reproduce this study and compare our results with those of these two papers. The PA of the host clusters are taken from \citet{west2017ten} (computed from the moments of inertia of the distribution of red sequence galaxies) and \citet{Durret_2019} (computed from density maps of red sequence galaxies), and the PAs of the BCGs are measured here with GALFIT. If measures are given in both papers, the PA$_{cluster}$ in \citet{Durret_2019} is taken, unless the  PA$_{cluster}$ is ill-defined (when the clusters are circular in shape), in which case the PA from  \citet{west2017ten} is taken. We did not measure the PA of the host clusters for the clusters that are not presented in the above quoted papers, as our images are not large enough to accurately measure the full extent and shape of the cluster.

We include all BCGs for which the PA of the cluster was measured (73 BCGs), including those observed with too blue filters, as the PA measured by GALFIT is the same regardless of the filter (see \Cref{fig:alignment_red_blue}). 
We show the histogram of the alignment between the BCGs and their host clusters (defined as the difference of PA between that of the cluster and the BCG) in \Cref{fig:hist_align}. 

We find that 39 BCGs (53\% of the BCGs) are aligned with their host cluster with a difference smaller than 30 degrees. This already shows a tendency for BCGs to align with their host clusters, as a random orientation of the BCGs would result in a flat distribution. BCGs with the highest PA difference tend to be circular in shape (elongation = b/a $\approx$ 1, for which it is more difficult to measure a PA, resulting in high uncertainties). We thus chose to exclude all BCGs with axis ratio b/a $\geq$ 0.8, in order to eliminate BCGs with ill-defined PAs, as shown in red on the histogram. We then find that 32 out of 58 of BCGs are aligned with their host cluster within 30 degrees, slightly rising the percentage to 55\%. There is a secondary peak between a PA difference of 30 to 40 degrees, mainly corresponding to BCGs at redshift z $\geq$ 0.9. At such high redshifts, galaxies appear smaller and therefore, the accuracy of the measured PA is probably worse. 
If we only consider galaxies at z $\leq$ 0.9 (blue histogram), we find that 22 BCGs out of 30 (73\%) align with their cluster within less than 30 degrees. This shows that in majority BCGs tend to align with their host cluster at least at z $\leq$ 0.9.

\subsection{BCG physical properties as a function of host cluster properties}
\label{section:mass_X_lum}

\begin{figure*}[ht!]
  \centering
  \begin{tabular}{@{}c@{}}
    \includegraphics[width=.49\linewidth]{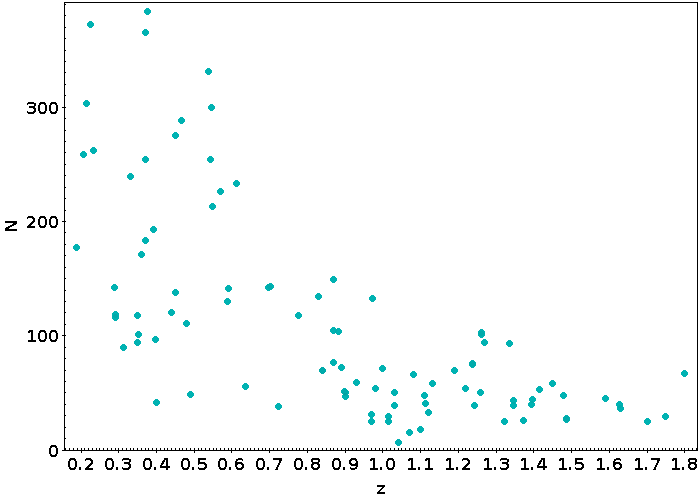}
  \end{tabular}
  \begin{tabular}{@{}c@{}}
    \includegraphics[width=.49\linewidth]{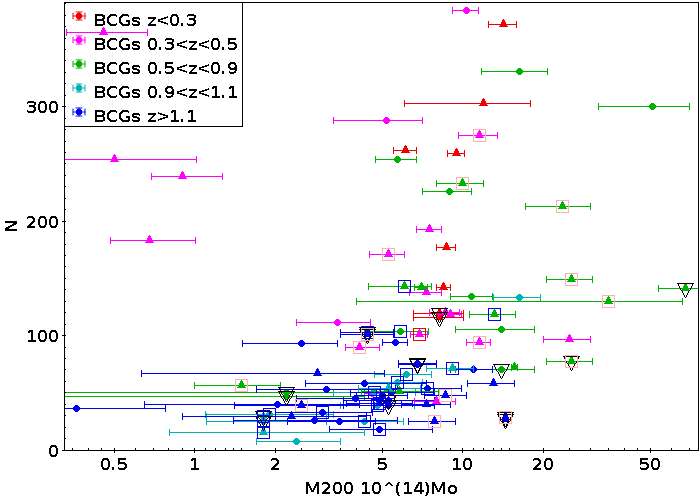}
  \end{tabular}
  \caption{(Left) Richness of the cluster (see text) as a function of redshift. (Right) Richness of the cluster as a function of its mass M$_{200,c}$.  The colors represent different redshift bins.}
\label{fig:mass_richness}
\end{figure*}

\begin{figure*}[ht!]
  \centering
  \begin{tabular}{@{}c@{}}
    \includegraphics[width=.49\linewidth]{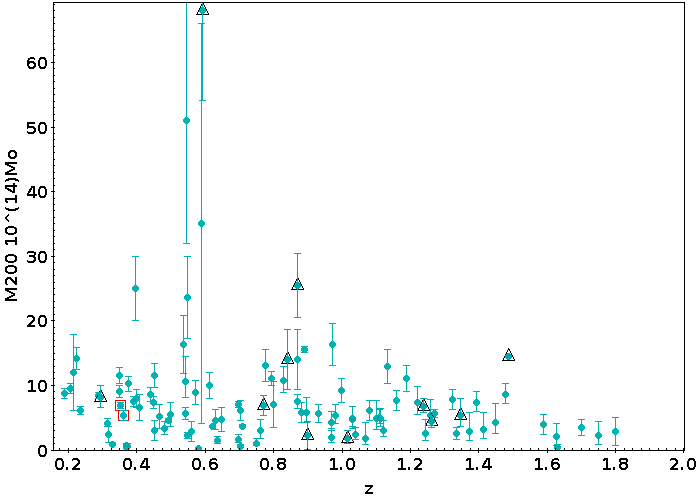}
  \end{tabular}
  \begin{tabular}{@{}c@{}}
    \includegraphics[width=.49\linewidth]{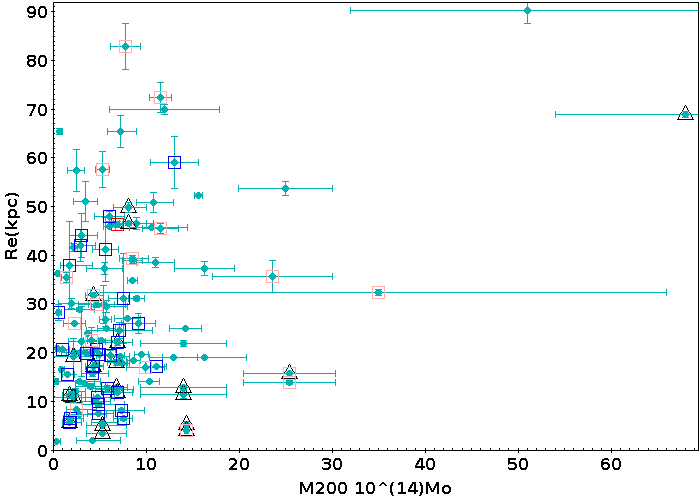}
  \end{tabular}
  \caption{ (Left) Cluster mass M200 (in units of 10$^{14} M_{\odot}$ as a function of redshift. (Right) BCG effective radius as a function of cluster mass.} 
    \label{fig:Re_mass}
\end{figure*}

We browsed the available bibliography to retrieve the cluster masses and X-ray center coordinates. The corresponding data can be found in \Cref{table:mass_coordX}. We prefer lensing based mass estimates if available. We bring all the masses to M$_{200}$, applying the conversion factor between M$_{500}$ and M$_{200}$: M$_{500}$ = 0.72 M$_{200}$ \citep{Pierpaoli_2003}.

We show the richness of the cluster as a function of redshift and cluster mass in \Cref{fig:mass_richness}. The richness N of the cluster is defined here as the number of red sequence galaxies (found in \Cref{section:detection}) in an aperture of 500 kpc radius around the BCG. We obtain different values of N for two different BCGs in the same cluster because the richness is computed in an aperture centered on each BCG. 

As can be seen on \Cref{fig:mass_richness} (left panel), clusters seem to become richer with decreasing redshift (correlation coefficient in logarithmic scale R = $-0.70$ and p-value of p $<$ 10$^{-5}$). Clusters at higher redshifts (z $\geq$ 1.0) have a lower richness, with a number of detected red sequence galaxies N $\leq$ 60. The right panel also shows that the most massive clusters are also the richest, and the high redshift clusters (blue points on the plot) with a low richness are also the less massive (M$_{200,c} \leq$ 5 $\times$ 10$^{14}$ M$_{\odot}$). This low value of N could in principle be due to the depth of our images, as we have a bias due to the distance: at higher redshifts, it is more difficult to detect objects and only the brightest ones can show up. 

However, when looking at \Cref{fig:Re_mass}, the left panel shows that we do not observe very massive clusters at high redshifts. So we have no bias due to the distance of the galaxies when measuring cluster masses: the masses are measured via lensing or derived from X-ray or SZ maps, which are independent of distance. Thus, we conclude that clusters become richer with time, and this result is not due to the depth of our images. However, although we only observe very massive clusters at lower redshifts (M$_{200,c} \geq$ 30 $\times$ 10$^{14}$ M$_{\odot}$ at z $\leq$ 0.8), the masses of the clusters do not vary much with time (R $<$ 0.20).
The right panel shows that the very massive clusters only host bigger BCGs: cluster with masses M$_{200,c} \geq$ 25 $\times$ 10$^{14}$ M$_{\odot}$ only have BCGs with effective radii R$_{e} \geq$ 30 kpc.

We find no correlations (R $\leq$ 0.2) between the BCG surface brightnesses or S\'ersic indices and the cluster masses. 

Using the relation given in \citet{bai2014inside}, we compute an estimate of the BCG masses from the cluster masses: M$_{BCG}^{*} \propto$ M$_{cluster}^{0.6}$. 
We find that the most massive BCGs are also the biggest in size (moderate correlation with R = 0.46 and p = 0.00007) and also tend to be brighter ($R = -0.32$, p = 007353). No correlation between the BCGs masses and redshift is seen (R $<$ 0.20).

\begin{figure*}[ht!]
  \centering
  \begin{tabular}{@{}c@{}}
    \includegraphics[width=.49\linewidth]{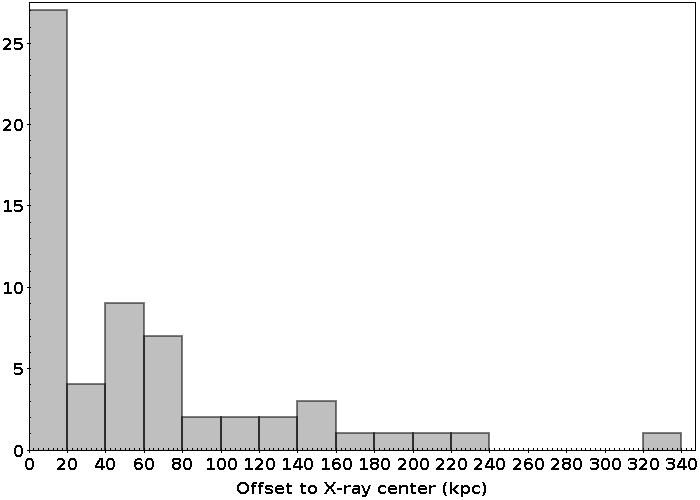}
  \end{tabular}
  \begin{tabular}{@{}c@{}}
    \includegraphics[width=.49\linewidth]{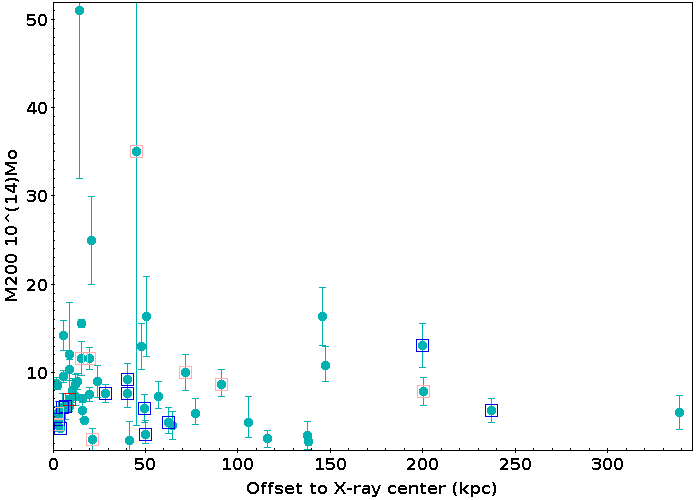}
  \end{tabular}
  \begin{tabular}{@{}c@{}}
    \includegraphics[width=.49\linewidth]{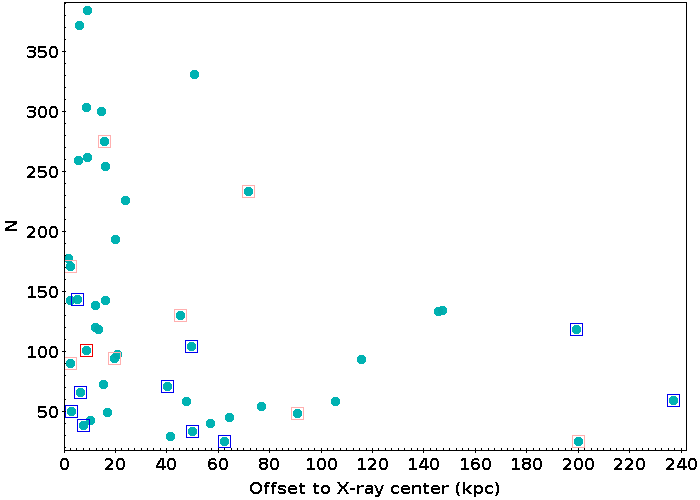}
  \end{tabular}
  \begin{tabular}{@{}c@{}}
    \includegraphics[width=.49\linewidth]{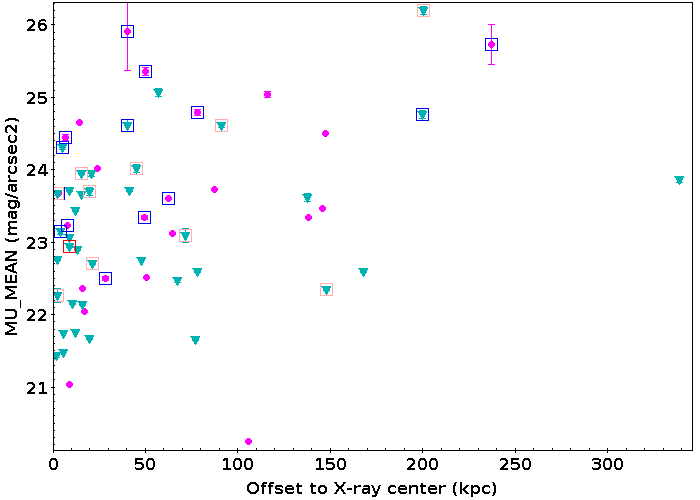}
  \end{tabular}
  \caption{ (Top left) Histogram of the offsets between the BCG and the X-ray center. As a function of the offsets between the BCG and the cluster X-ray center: (Top right) Masses M$_{200,c}$ of the clusters (Bottom left) Richness of the cluster (Bottom right) Mean effective surface brightness of the BCGs.} 
    \label{fig:offset_dX}
\end{figure*}

We also study how the BCGs behave depending on their offsets to the cluster X-ray center. We exclude superclusters and clusters which present several substructures and/or several BCGs. We show the histogram of the offets on \Cref{fig:offset_dX}, top left panel. We find that 31 out of 61 (51\%) are within a 30 kpc radius range from the X-ray center of the cluster, showing that BCGs tend to lie close to the cluster X-ray centers. The two star forming BCGs that have undergone recent mergers and are not at equilibrium are also located at the center of the cluster (D$_{X}$ $\leq$ 10 kpc). But, we confirm that there can be a significant offset between the two (12 out of 61 BCGs, 20\%, present an offset bigger than 100 kpc). Although the corresponding plots are not shown here, we find no correlation between the offset and the absolute magnitude, effective radius, S\'ersic index of the BCGs, or with the alignment previously computed. 

As can be seen on the top right and bottom left panels of \Cref{fig:offset_dX}, however, the more massive and the richer the cluster (or the BCG, as we converted the cluster masses to BCG masses), the closer the BCG is to the X-ray center of the cluster: the objects with the biggest offets ( $\geq$ 100 kpc) have masses M$_{cluster} \leq$ 10 $\times$ 10$^{14}$M$_{\odot}$ and richnesses N $\leq$ 100. We also find that there is a moderate correlation between the offset and the mean effective surface brightness of the BCG (see \Cref{fig:offset_dX}, bottom right): BCGs tend to have brighter mean effective surface brightnesses the closer they are to the X-ray center (in logarithmic scale, R = 0.34, p = 0.0395). 

We also analysed whether the most luminous BCGs are special \citep[see][]{Lin_2010, lauer2014brightest}. We studied the distribution of the difference in magnitude between the BCGs and the second ranked galaxies of the clusters. We found that the distribution was continue, with most BCGs having a difference smaller than one magnitude with the second ranked galaxy. By selecting BCGs which are at least brighter than 1 magnitude than the second ranked galaxy of the cluster (9 BCGs), we find that the most luminous BCGs do not occupy a specific place in the observed relations.


\section{Discussion and conclusions}
\label{section:conclusion}
 
Our work deals with the largest sample (to our knowledge) of BCGs with HST imaging, covering a broad redshift interval from z = 0.1 to z = 1.8 (see \Cref{fig:comp_study}), and thus enabling us to trace the evolution of BCGs through time. Our sample is larger than most studies found in the literature based on HST images, such as \citet{bai2014inside}, \citet{DeMaio_2019, Durret_2019}. We also study the luminosity profiles of these galaxies and how they evolve as a function of redshift. HST images allow us to perform profile fitting with precision, and GALFIT returns accurate parameters from model fitting. 

We developed a new tool to detect automatically red BCGs on optical images. We successfully detected all the red BCGs regardless of their peculiar characteristics (see \Cref{section:detection}). We did not manage to detect in a similar way the blue BCGs, which represent here only 2\% of our sample. 

We then proceeded to model the luminosity profiles of these automatically detected BCGs, as well as those which have only one filter available, bringing this sample to 149 BCGs. We removed all BCGs observed in too blue filters as well as BCGs better modeled with two components for which the inner component has an important contribution to the total luminosity of the galaxy. Our final sample consists then of 86 BCGs. 

We studied how the photometric properties of BCGs correlate with redshift, and we find that, although the correlation is weak -but significant, the absolute magnitude presents a faint trend to become brighter with time.

We show that there is a faint trend (see \Cref{fig:evol_redshift_MAG_ABS_Re}, right) for BCGs to become bigger with decreasing redshift. This is the behaviour we expect for galaxies that grow in size with time, by accreting gas and merging with other smaller galaxies. This trend was also observed in \citet{Durret_2019} up to redshift 0.9, and can be confirmed up to redshift z=1.8. Based on this relation, we find that BCGs grow by more than a factor 3 between redshifts 1.8 and 0.1. The dispersion can be linked with the Kormendy relation, that indicates that galaxies with higher surface brightnesses have smaller effective radii. 

We find no strong correlation between the other photometric properties (surface brightness or S\'ersic index) of the BCGs and redshift. This is in agreement with \citet{bai2014inside} who do not find any correlation between the magnitude or the mean surface brightness of the BCGs and redshift, up to redshift z = 0.9. We add that no evolution can be observed up to redshift z = 1.8 either. 

Although we do only observe massive clusters at lower redshifts (M$_{200,c} \geq$ 30 $\times$ 10$^{14}$ M$_{\odot}$ at z $\leq$ 0.8), overall, the masses of the clusters do not correlate with redshift. The growth of the cluster is mainly to be linked with the cluster richness (\Cref{fig:mass_richness}): clusters become richer with time, and we find that the number of red sequence galaxies in an aperture of 500 kpc centered on the BCG increases by almost a factor 10 between z = 1.8 and z = 0.1. We confirmed that the low richness we measured at higher redshift is not due to the depth of our images. This growth mainly seems to be happening at z $\leq$ 1.0, as we do not observe a significant variation of the richness of the clusters before that time.

We use the relation found in \citet{bai2014inside} to compute the BCG masses from the cluster masses, based on the relation found by \citet{bai2014inside}. We find that bigger BCGs are also more massive (see \Cref{fig:mass_richness}): R$_{e} \propto$ 4.42  $\times$ M$_{BCG}$, but the masses do not show a significant growth with redshift. 

We thus find that the sizes of the BCGs grow faster than their masses in the same redshift range. Although we do not find that the masses of the BCGs grow significantly with time, whereas \citet{bai2014inside} finds a factor 2 since z = 2, we agree that the sizes have grown significantly faster than the masses in the same redshift range. \citet{bai2014inside} find that the sizes grow more than twice as fast as the masses. We confirm that the masses and sizes of BCGs do not grow at the same rate. This is in favor of a scenario in which BCGs grow thanks to minor dry mergers at the later stages of their formation and evolution. Indeed, a growth mainly due to major dry mergers would make the sizes and masses grow at the same rate. 

To summarize, we can say that the sizes of the BCGs, as well as the richnesses of the clusters, evolve with redshift: clusters become richer with time and, at the same time, BCGs undergo dry mergers that increase their sizes.

Another interesting result is the distribution of S\'ersic indices (see \Cref{fig:redshift_n}) that shows two different populations with low S\'ersic indices, mainly at low redshift (z $\leq$ 0.8) and high S\'ersic indices. The limit is also to be linked to the fact that BCGs at lower redshifts often require a second component to correctly take into account the brighter core of the galaxy.
We find that BCGs better modeled with two components have a peak S\'ersic index n = 2, while those that were fit with a single component have a peak at n = 4. Those modeled with only one S\'ersic component are thus comparable to pure elliptical galaxies which can be well modeled with a deVaucouleurs profile. This slightly differs from the results shown in \citet{bai2014inside} who find a median value of n = 5.7. But, \citet{bai2014inside} only fit a single S\'ersic profile to all the BCGs in their sample. If we only look at the distribution we obtained for BCGs modeled with a single component, we find that this distribution is more comparable to that of ETGs shown in their paper. Another difference with that study is related to the filters chosen to model the luminosity profiles of the BCGs. While we consider the same spectral region of the SED for all clusters in order to only look at the same old red stellar population at all redshifts, \citet{bai2014inside} observe all BCGs with the ACS F814W filter, which we find is already too blue for clusters at redshifts z $\geq$ 0.57. We also showed, by studying the parameters obtained in two different filters for a sample of BCGs, that the parameters will vary depending on the part of the SED you look at (when looking at a bluer filter, the absolute magnitude and mean effective surface brightness become fainter, the effective radius becomes smaller, and the S\'ersic indices vary without any clear trend).

Finally, we find that the Kormendy relation \citep{kormendy1977brightness} is also a function of redshift, with the relation shifted towards fainter mean effective surface brightnesses at higher redshifts. This relation shows that, at the effective radius, smaller galaxies are brighter and denser than the bigger ones. The slope of 3.33 $\pm$ 0.73 measured with our sample remains constant with redshift. 
Our value is in good agreement with that given in \citet{bai2014inside}:

<$\mu$>=(3.50$\pm$ 0.18)logR$_{e}$+(18.01$\pm$ 0.23)

\noindent
and agrees within one $\sigma$ with the one given in \citet{Durret_2019}: 

<$\mu$>=(2.64$\pm$ 0.35)logR$_{e}$+(19.7$\pm$ 0.5).

We should note that cosmology or selection effects might be contributing to the results in \Cref{fig:evol_redshift_MAG_ABS_Re}, which shows a trend for BCG sizes and luminosities to increase with time. 
Despite its faintness, the contribution of the ICL should be taken into account. The ICL blends with the envelope of the BCG, making it difficult to differentiate the galaxy from the ICL, and this may affect our measurements (in particular those of the effective radii and S\'ersic indices). The ICL might contribute at some level to measured sizes and luminosities of  galaxies at low redshifts, yet might be missed in high-redshift clusters because of cosmological surface brightness dimming, or perhaps because the ICL has not yet developed in these young clusters. A concern comes from the value of the background, as GALFIT is sensitive to it, but its computation is limited by the sizes of the images (even without cropping).

We broaden the work by \citet{west2017ten} and \citet{Durret_2019} on the alignment of BCGs with their host clusters. We removed BCGs with ill measured position angles due to their circular shape, as well as BCGs at higher redshifts, z $\geq$ 0.9, as they will appear smaller on the CCD, and will thus be less resolved and have less accurate measured PAs.
This enables us to conclude that BCGs in majority tend to align with their host cluster at least at z $\leq$ 0.9, as after this selection 73\% of the remaining BCGs are aligned with their host cluster within 30 degrees. This is a tighter alignment than that of \citet{Durret_2019}, who found an alignment for 12 out of 21 BCGs (57\%) between redshifts 0.21 and 0.89. \citet{Okabe_2020} study the alignment of 45 dark matter (DM) haloes and their BCGs up to z = 0.97, and find that BCGs tend to be well aligned with their DM haloes, with a mean difference of 22.2 $\pm$ 3.9 degrees. A similar  study was done by \citet{Ragone_Figueroa_2020} on the alignment of BCGs both with the distribution of cluster galaxies and DM haloes, by analyzing cosmological hydro-simulations of 24 clusters. They find a strong alignment at z $<$ 2, and add that relaxed clusters tend to host BCGs that align with their major axis. Similar conclusions are made by \citet{De_Propris_2020}, who show that BCGs are generally aligned with their host cluster even when the offset between the BCG and the X-ray center is significant.

\citet{Cerulo_2019} found that 9\% the BCGs between 0.05 $\leq$ z $\leq$ 0.35 have colors bluer than $2 \sigma$ of the median color of the cluster red sequence. During this study, we found two peculiar blue BCGs in our sample. Apart from their colors and complex structures, these two peculiar BCGs do not have photometric properties different from the other BCGs. It would be interesting to continue this study by considering a larger sample of SF blue BCGs, to see where they lie in the previous plots. 

We plan to apply the method described in this paper to more than a thousand clusters from the CFHTLS, detected by \citet{Sarron_2018}, up to redshift z = 0.7. This will enable us to better evaluate the accuracy of our BCG detection method on ground-based-data, and although the resolution will not be as good, the sample will be significantly larger.
We also found two BCGs (2\%) with blue colors, and it would be interesting to estimate the fraction of blue BCGs in the Universe up to redshift z = 0.7. We can wonder if these BCGs evolve differently from the red BCGs that we detect.


\begin{acknowledgements}
We thank the referee, M. West, for his constructive comments and suggestions.
    F.D. acknowledges continuous support from CNES since 2002. 
   IM acknowledges  financial  support  from the State Agency for Research of the Spanish MCIU through the "Center of Excellence Severo Ochoa" award to the Instituto de Astrofísica de Andalucía (SEV-2017-0709), and through the programs AYA2016-76682C3-1-P and PID2019-106027GB-C41.
   Based on observations made with the NASA/ESA Hubble Space Telescope, and obtained from the Hubble Legacy Archive, which is a collaboration between the Space Telescope Science Institute (STScI/NASA), the Space Telescope European Coordinating Facility (ST-ECF/ESA) and the Canadian Astronomy Data Centre (CADC/NRC/CSA).
   
\end{acknowledgements}


\bibliographystyle{aa}


\begin{table*}[htbp]
\centering
\setcounter{table}{0}
  \caption{Sample of the 149 BCGs studied in this paper. The columns are: full cluster name, coordinates of the BCG, redshift, class of the BCG (if two BCGs are defined for a cluster, class 1 represents the brightest of the two), instrument, filter used to model the luminosity profile of the BCG (see \Cref{section:luminosity_profiles}), associated scale, color computed to extract the red sequence of the cluster (see \Cref{section:detection}). The BCGs with no values in the last column only had data in one filter, and their coordinates were taken from the literature.}
\label{tab:BCGs_coord}
\begin{tabular}{| l | l | l | l | l | l | l | l | l |}
\hline
Name & RA$_{BCG}$ & DEC$_{BCG}$ & Redshift & Class & Instrument &  Filter & Scale & Color \\    
 & (J2000) & (J2000) & & & & & (kpc/\arcsec) & \\    
\hline                                   
  SPT-CLJ0000-5748         &    0.2504   &    -57.8093 &    0.702    &    1    &     ACS\_WFC &      F814W  &     7.128 &      F606W-F814W  \\
  Cl0016+1609              &    4.64     &    16.4378  &    0.5455   &    1    &     ACS\_WFC &      F850LP &     2.83  &      F606W-F775W  \\
  SpARCS-J0335             &    8.9571   &    -43.2065 &    1.335    &    1    &     WFC3\_IR &      F140W  &     8.353 &      F105W-F140W  \\
  ACO2813                  &    10.8528  &    -20.6282 &    0.2924   &    1    &     ACS\_WFC &      F814W  &     4.368 &      F435W-F606W  \\
  ACO2813                  &    10.8548  &    -20.6169 &    0.2924   &    2    &     ACS\_WFC &      F814W  &     4.368 &      F435W-F606W  \\
  XDCPJ0044-2033           &    11.0236  &    -20.5651 &    1.59     &    1    &     WFC3\_IR &      F160W  &     8.42  &      F105W-F140W  \\
  RXJ0056-27               &    14.2374  &    -27.675  &    0.56     &    1    &     ACS\_WFC &      F814W  &     6.449 &                \\
  SPT-CLJ0102-4915         &    15.721   &    -49.2528 &    0.87     &    1    &     WFC3\_IR &      F105W  &     7.681 &      F625W-F775W  \\
  SPT-CLJ0102-4915         &    15.7409  &    -49.2719 &    0.87     &    2    &     ACS\_WFC &      F850LP &     7.681 &      F625W-F775W  \\
  RXJ0110+19               &    17.5758  &    19.6387  &    0.317    &    1    &     ACS\_WFC &      F814W  &     4.617 &                \\
  Abell209                 &    22.9689  &    -13.6112 &    0.206    &    1    &     ACS\_WFC &      F775W  &     3.373 &      F475W-F606W  \\
  CLJ0152-1357             &    28.1649  &    -13.9739 &    0.87     &    1    &     WFC3\_IR &      F105W  &     7.681 &      F775W-F850LP \\
  CLJ015244.18-135715.84   &    28.182   &    -13.9552 &    0.84     &    2    &     WFC3\_IR &      F105W  &     7.598 &      F625W-F850LP \\
  CLJ015244.18-135715.84   &    28.1824  &    -13.9555 &    0.84     &    1    &     WFC3\_IR &      F105W  &     7.598 &      F625W-F850LP \\
  SPT-CLJ0205-5829         &    31.451   &    -58.4802 &    1.322    &    1    &     WFC3\_IR &      F140W  &     8.346 &      F814W-F105W  \\
  XMMXCSJ022045.1-032555.0 &    35.1895  &    -3.4333  &    0.33     &    1    &     WFC3\_IR &      F105W  &     4.743 &      F105W-F160W  \\
  RCSJ0220-0333            &    35.2316  &    -3.5633  &    1.03     &    1    &     ACS\_WFC &      F850LP &     8.024 &      F775W-F850LP \\
  RCSJ0221-0321            &    35.4299  &    -3.3542  &    1.016    &    1    &     ACS\_WFC &      F850LP &     8.0   &      F775W-F850LP \\
  RCSJ0221-0321            &    35.4369  &    -3.3678  &    1.016    &    2    &     ACS\_WFC &      F850LP &     8.0   &      F775W-F850LP \\
  XLSSJ0223-0436           &    35.7636  &    -4.6043  &    1.22     &    1    &     ACS\_WFC &      F850LP &     8.269 &      F775W-F850LP \\
  RCS0224-02               &    36.014   &    -2.471   &    0.408    &    1    &     ACS\_WFC &      F814W  &     5.425 &                \\
  SpARCS-J0224             &    36.1092  &    -3.3929  &    1.63     &    1    &     WFC3\_IR &      F160W  &     8.419 &      F105W-F140W  \\
  JKCS041                  &    36.6815  &    -4.6894  &    1.8      &    1    &     WFC3\_IR &      F160W  &     8.39  &      F105W-F160W  \\
  Abell383                 &    42.0141  &    -3.5292  &    0.1871   &    1    &     ACS\_WFC &      F775W  &     3.127 &      F475W-F606W  \\
  MACS0329-0211            &    52.4232  &    -2.1962  &    0.45     &    1    &     ACS\_WFC &      F850LP &     5.745 &      F606W-F625W  \\
  SpARCS-J0330             &    52.733   &    -28.7166 &    1.626    &    1    &     WFC3\_IR &      F160W  &     8.42  &      F105W-F140W  \\
  RCS0337-2844             &    54.4539  &    -28.7517 &    1.1      &    1    &     ACS\_WFC &      F850LP &     8.132 &      F775W-F850LP \\
  RCS0350-08               &    57.6131  &    -8.9157  &    0.584    &    1    &     ACS\_WFC &      F814W  &     6.581 &                \\
  RCS0351-09               &    57.9159  &    -9.9406  &    0.304    &    1    &     ACS\_WFC &      F814W  &     4.487 &                \\
  MACSJ0416-2403           &    64.038   &    -24.0676 &    0.396    &    1    &     ACS\_WFC &      F850LP &     5.328 &      F606W-F625W  \\
  MACS0429-0253            &    67.4001  &    -2.8851  &    0.399    &    1    &     ACS\_WFC &      F850LP &     5.353 &      F606W-F625W  \\
  RCS0439-2904             &    69.9072  &    -29.0839 &    0.97     &    1    &     ACS\_WFC &      F850LP &     7.913 &      F814W-F850LP \\
  RCS0444-28               &    71.036   &    -28.3379 &    0.437    &    1    &     ACS\_WFC &      F814W  &     5.649 &                \\
  MACSJ0454.1-0300         &    73.5453  &    -3.0146  &    0.5377   &    1    &     ACS\_WFC &      F850LP &     6.321 &      F775W-F850LP \\
  RCS0511-42               &    77.8703  &    -42.5867 &    0.518    &    1    &     ACS\_WFC &      F814W  &     6.201 &                \\
  RCS0515-43               &    78.9047  &    -43.4207 &    0.44     &    1    &     ACS\_WFC &      F814W  &     5.672 &                \\
  RCS0518-43               &    79.6411  &    -43.4186 &    0.396    &    1    &     ACS\_WFC &      F814W  &     5.328 &                \\
  RCS0518-43               &    79.7313  &    -43.252  &    0.508    &    1    &     ACS\_WFC &      F814W  &     6.139 &                \\
  RCS0519-42               &    79.8316  &    -42.7978 &    0.603    &    1    &     ACS\_WFC &      F814W  &     6.679 &                \\
  RCS0519-44               &    79.9184  &    -44.0392 &    0.827    &    1    &     ACS\_WFC &      F814W  &     7.561 &                \\
  SPT-CLJ0533-5005         &    83.4035  &    -50.0957 &    0.881    &    1    &     ACS\_WFC &      F814W  &     7.709 &      F606W-F814W  \\
  SPT-CLJ0546-5345         &    86.657   &    -53.7586 &    1.16     &    1    &     ACS\_WFC &      F814W  &     8.207 &                \\
  SPT-CLJ0559-5249         &    89.9301  &    -52.8242 &    0.6112   &    1    &     ACS\_WFC &      F814W  &     6.721 &      F606W-F814W  \\
  SPT-CLJ0615-5746         &    93.9657  &    -57.7801 &    0.972    &    1    &     WFC3\_IR &      F125W  &     7.917 &      F814W-F105W  \\
  MACSJ0647+7015           &    101.9575 &    70.2488  &    0.5907   &    2    &     WFC3\_IR &      F105W  &     6.616 &      F625W-F775W  \\
  MACSJ0647+7015           &    101.961  &    70.2483  &    0.5907   &    1    &     WFC3\_IR &      F105W  &     6.616 &      F625W-F775W  \\
  MACSJ0717+3745           &    109.3985 &    37.7548  &    0.548    &    1    &     ACS\_WFC &      F850LP &     6.381 &      F625W-F775W  \\
  MACSJ0744+3927           &    116.22   &    39.4574  &    0.6976   &    1    &     WFC3\_IR &      F105W  &     7.11  &      F606W-F775W  \\
  Abell611                 &    120.2367 &    36.0566  &    0.288    &    1    &     ACS\_WFC &      F814W  &     4.322 &      F475W-F606W  \\
  RXJ0841+64               &    130.2818 &    64.3739  &    0.36     &    1    &     ACS\_WFC &      F814W  &     5.02  &                \\
  RXJ0847+34               &    131.7991 &    34.8144  &    0.56     &    1    &     ACS\_WFC &      F814W  &     6.449 &                \\
  RDCSJ0849+4452           &    132.2443 &    44.8659  &    1.261    &    2    &     ACS\_WFC &      F850LP &     8.304 &      F775W-F850LP \\
  RDCSJ0849+4452           &    132.2445 &    44.8658  &    1.261    &    1    &     ACS\_WFC &      F850LP &     8.304 &      F775W-F850LP \\
  RDCSJ0910+5422           &    137.6915 &    54.3687  &    1.11     &    1    &     ACS\_WFC &      F850LP &     8.146 &      F775W-F850LP \\
  RXJ0926+12               &    141.6529 &    12.7177  &    0.49186  &    1    &     ACS\_WFC &      F814W  &     6.034 &                \\
  RCS0928+36               &    142.0883 &    36.7744  &    0.393    &    1    &     ACS\_WFC &      F814W  &     5.303 &                \\
  Abell851                 &    145.7417 &    46.9867  &    0.4069   &    1    &     ACS\_WFC &      F814W  &     5.416 &                \\
\hline         
\end{tabular}
\end{table*}

\begin{table*}[htbp]
\centering
\setcounter{table}{0}
  \caption{Continued.}
\begin{tabular}{| l | l | l | l | l | l | l | l | l |}
  \hline
  Name & RA$_{BCG}$ & DEC$_{BCG}$ & Redshift & Class & Instrument &  Filter & Scale & Color \\    
 & (J2000) & (J2000) & & & & & (kpc/\arcsec) & \\    
\hline                                   
  MOOJ1014+0038            &    153.5304 &    0.6408   &    1.27     &    1    &     WFC3\_IR &      F140W  &     8.31  &      F105W-F125W  \\
  LCDCS0110                &    159.4681 &    -12.747  &    0.5789   &    1    &     ACS\_WFC &      F814W  &     6.553 &                \\
  LCDCS0130                &    160.1678 &    -11.9345 &    0.7043   &    1    &     ACS\_WFC &      F814W  &     7.137 &                \\
  SpARCS-J1049             &    162.3441 &    56.6754  &    1.7      &    1    &     WFC3\_IR &      F160W  &     8.412 &      F105W-F160W  \\
  LCDCS0172                &    163.6016 &    -11.7722 &    0.6972   &    1    &     ACS\_WFC &      F814W  &     7.109 &                \\
  LCDCS0173                &    163.6813 &    -12.7645 &    0.7498   &    1    &     ACS\_WFC &      F814W  &     7.309 &                \\
  MS1054-0321              &    164.2496 &    -3.6265  &    0.83     &    1    &     ACS\_WFC &      F850LP &     7.569 &      F775W-F850LP \\
  RCS1102-03               &    165.6373 &    -3.318   &    0.423    &    1    &     ACS\_WFC &      F814W  &     5.543 &                \\
  RCS1102-05               &    165.7463 &    -5.3527  &    0.395    &    1    &     ACS\_WFC &      F814W  &     5.32  &                \\
  CLJ1103.7-1245           &    165.8957 &    -12.7798 &    0.63     &    1    &     ACS\_WFC &      F814W  &     6.812 &                \\
  RCS1104-04               &    166.1669 &    -4.7495  &    0.637    &    1    &     ACS\_WFC &      F814W  &     6.845 &                \\
  RCS1107-05               &    166.8505 &    -5.3862  &    0.794    &    1    &     ACS\_WFC &      F814W  &     7.459 &                \\
  RCS1107-05               &    166.9754 &    -5.2778  &    0.579    &    1    &     ACS\_WFC &      F814W  &     6.554 &                \\
  RCS1108-04               &    167.0611 &    -4.5142  &    0.638    &    1    &     ACS\_WFC &      F814W  &     6.85  &                \\
  MACSJ1115+0129           &    168.9662 &    1.4986   &    0.349    &    1    &     ACS\_WFC &      F850LP &     4.921 &      F475W-F606W  \\
  SG1120-1202-2            &    170.0312 &    -12.0858 &    0.3704   &    1    &     WFC3\_IR &      F105W  &     5.112 &      F606W-F814W  \\
  SG1120-1202-1            &    170.0555 &    -11.9808 &    0.3707   &    1    &     WFC3\_IR &      F105W  &     5.114 &      F606W-F814W  \\
  SG1120-1202-3            &    170.0924 &    -12.0295 &    0.3713   &    1    &     WFC3\_IR &      F105W  &     5.119 &      F606W-F814W  \\
  RCS1122+24               &    170.6077 &    24.375   &    0.799    &    1    &     ACS\_WFC &      F814W  &     7.475 &                \\
  LCDCS0340                &    174.5423 &    -11.5607 &    0.4798   &    1    &     WFC3\_IR &      F105W  &     5.954 &      F814W-F105W  \\
  MOOJ1142+1527            &    175.6976 &    15.4531  &    1.19     &    1    &     WFC3\_IR &      F140W  &     8.24  &      F105W-F140W  \\
  MACSJ1149.5+2223         &    177.3987 &    22.3985  &    0.5444   &    1    &     ACS\_WFC &      F850LP &     6.36  &      F625W-F775W  \\
  Abell1423                &    179.3223 &    33.611   &    0.2142   &    1    &     ACS\_WFC &      F775W  &     3.476 &      F475W-F606W  \\
  MACS1206-0847            &    181.5506 &    -8.8009  &    0.44     &    1    &     ACS\_WFC &      F850LP &     5.672 &      F606W-F625W  \\
  LCDCS0504                &    184.1883 &    -12.0215 &    0.7943   &    1    &     ACS\_WFC &      F814W  &     7.46  &                \\
  CLJ1226+3332             &    186.7427 &    33.5468  &    0.89     &    1    &     WFC3\_IR &      F125W  &     7.732 &      F814W-F850LP \\
  LCDCS0531                &    186.9953 &    -11.5872 &    0.6355   &    1    &     WFC3\_IR &      F105W  &     6.838 &      F814W-F105W  \\
  XMMUJ1229+0151           &    187.3721 &    1.8561   &    0.98     &    1    &     WFC3\_IR &      F105W  &     7.933 &      F775W-F850LP \\
  LCDCS0541                &    188.1262 &    -12.8435 &    0.5414   &    1    &     ACS\_WFC &      F814W  &     6.342 &                \\
  RDCSJ1252-2927           &    193.2267 &    -29.4549 &    1.237    &    1    &     WFC3\_IR &      F125W  &     8.284 &      F775W-F850LP \\
  RDCSJ1252-2927           &    193.2273 &    -29.4548 &    1.237    &    2    &     WFC3\_IR &      F125W  &     8.284 &      F775W-F850LP \\
  MACSJ1311-0310           &    197.7577 &    -3.1777  &    0.49133  &    1    &     ACS\_WFC &      F850LP &     6.031 &      F606W-F625W  \\
  RCS1319-02               &    199.8005 &    -2.1197  &    0.354    &    1    &     ACS\_WFC &      F814W  &     4.966 &                \\
  RCS1323+30               &    200.892  &    30.3802  &    0.46163  &    1    &     ACS\_WFC &      F814W  &     5.828 &                \\
  ZWCl1332.8+5043          &    203.5857 &    50.5177  &    0.62     &    1    &     ACS\_WFC &      F775W  &     6.764 &                \\
  RXJ1347-1145             &    206.8776 &    -11.7527 &    0.451    &    1    &     ACS\_WFC &      F850LP &     5.752 &      F606W-F625W  \\
  LCDCS0853                &    208.5406 &    -12.5172 &    0.7627   &    1    &     ACS\_WFC &      F814W  &     7.355 &                \\
  RXJ1354-02               &    208.5716 &    -2.3664  &    0.54786  &    1    &     ACS\_WFC &      F814W  &     6.38  &                \\
  WARPSJ1415+3612          &    213.7962 &    36.2009  &    1.03     &    1    &     ACS\_WFC &      F850LP &     8.024 &      F775W-F850LP \\
  RCS1419+53               &    214.8006 &    53.4367  &    0.71     &    1    &     ACS\_WFC &      F814W  &     7.16  &                \\
  MACSJ1423+2404           &    215.9495 &    24.0784  &    0.5431   &    1    &     ACS\_WFC &      F850LP &     6.352 &      F606W-F775W  \\
  IDCSJ1426.5+3508         &    216.6373 &    35.1399  &    1.75     &    1    &     WFC3\_IR &      F160W  &     3.734 &      F105W-F140W  \\
  ISCS1429+3437            &    217.3237 &    34.6207  &    1.258    &    1    &     WFC3\_IR &      F160W  &     8.301 &      F775W-F850LP \\
  ISCSJ1432.3+3253         &    218.0757 &    32.8894  &    1.396    &    1    &     WFC3\_IR &      F140W  &     8.382 &      F105W-F140W  \\
  ISCSJ1432.4+3250         &    218.0947 &    32.8268  &    1.487    &    1    &     WFC3\_IR &      F160W  &     8.41  &      F814W-F160W  \\
  ISCSJ1432.4+3250         &    218.1038 &    32.8347  &    1.487    &    2    &     WFC3\_IR &      F160W  &     8.41  &      F814W-F160W  \\
  ISCSJ1432+3332           &    218.1137 &    33.5521  &    1.112    &    1    &     ACS\_WFC &      F850LP &     8.148 &      F775W-F850LP \\
  ISCSJ1432+3436           &    218.1574 &    34.6063  &    1.347    &    2    &     WFC3\_IR &      F160W  &     8.36  &      F850LP-F160W \\
  ISCSJ1432+3436           &    218.1578 &    34.6084  &    1.347    &    1    &     WFC3\_IR &      F160W  &     8.36  &      F850LP-F160W \\
  ISCSJ1434+3427           &    218.6255 &    34.4492  &    1.243    &    1    &     WFC3\_IR &      F125W  &     8.289 &      F850LP-F110W \\
  ISCSJ1434.5+3519         &    218.6913 &    35.3314  &    1.373    &    1    &     WFC3\_IR &      F160W  &     8.373 &      F775W-F850LP \\
  ISCSJ1438+3414           &    219.5352 &    34.2368  &    1.414    &    1    &     WFC3\_IR &      F160W  &     8.389 &      F850LP-F105W \\
  RCS1446+08               &    221.7281 &    8.4512   &    0.628    &    1    &     ACS\_WFC &      F814W  &     6.803 &                \\
  MACSJ1447.4+0827         &    221.8584 &    8.4737   &    0.3755   &    1    &     ACS\_WFC &      F814W  &     5.156 &      F606W-F814W  \\
  RCS1450+08               &    222.6692 &    8.679    &    0.769    &    1    &     ACS\_WFC &      F814W  &     7.376 &                \\
  RCS1452+08               &    223.1133 &    8.5821   &    0.395    &    1    &     ACS\_WFC &      F814W  &     5.32  &                \\
  RCS1511+09               &    227.7654 &    9.0533   &    0.97     &    1    &     ACS\_WFC &      F850LP &     7.913 &      F775W-F850LP \\
  RXJ1532+3020             &    233.2241 &    30.3498  &    0.3615   &    1    &     ACS\_WFC &      F850LP &     5.033 &      F606W-F625W  \\
  RXJ1540+14               &    235.2249 &    14.7655  &    0.44025  &    1    &     ACS\_WFC &      F814W  &     5.673 &                \\
  ClGJ1604+4304            &    241.1003 &    43.0772  &    0.9      &    2    &     ACS\_WFC &      F850LP &     7.757 &      F814W-F850LP \\
   \hline
\end{tabular}
\end{table*}

\begin{table*}[htbp]
\centering
\setcounter{table}{0}
  \caption{Continued.}
\begin{tabular}{| l | l | l | l | l | l | l | l | l |}
  \hline
  Name & RA$_{BCG}$ & DEC$_{BCG}$ & Redshift & Class & Instrument &  Filter & Scale & Color \\    
 & (J2000) & (J2000) & & & & & (kpc/\arcsec) & \\    
\hline                                   
  ClGJ1604+4304            &    241.1029 &    43.0741  &    0.9      &    1    &     ACS\_WFC &      F850LP &     7.757 &      F814W-F850LP \\
  RCS1620+29               &    245.0423 &    29.4901  &    0.8696   &    1    &     ACS\_WFC &      F814W  &     7.68  &                \\
  MACSJ1621.4+3810         &    245.3531 &    38.1691  &    0.465    &    1    &     WFC3\_IR &      F110W  &     5.852 &      F606W-F814W  \\
  OC02J1701+6412           &    255.348  &    64.2366  &    0.453    &    1    &     ACS\_WFC &      F814W  &     5.767 &                \\
  MACSJ1720+3536           &    260.0698 &    35.6073  &    0.3913   &    1    &     ACS\_WFC &      F850LP &     5.289 &      F606W-F625W  \\
  Abell2261                &    260.6133 &    32.1325  &    0.224    &    1    &     ACS\_WFC &      F814W  &     3.597 &      F475W-F606W  \\
  MACSJ1932-2635           &    292.9567 &    -26.5756 &    0.352    &    1    &     ACS\_WFC &      F850LP &     4.948 &      F606W-F625W  \\
  SPT-CLJ2040-5725         &    310.0551 &    -57.4208 &    0.93     &    1    &     ACS\_WFC &      F814W  &     7.827 &      F606W-F814W  \\
  SPT-CLJ2040-4451         &    310.2382 &    -44.8595 &    1.478    &    1    &     WFC3\_IR &      F160W  &     8.408 &      F105W-F140W  \\
  SPT-CLJ2043-5035         &    310.8233 &    -50.5923 &    0.723    &    1    &     ACS\_WFC &      F814W  &     7.21  &      F606W-F814W  \\
  SPT-CLJ2106-5844         &    316.5192 &    -58.7411 &    1.132    &    1    &     WFC3\_IR &      F140W  &     8.174 &      F814W-F105W  \\
  MACSJ2129-0741           &    322.3588 &    -7.691   &    0.5889   &    1    &     ACS\_WFC &      F850LP &     6.606 &      F625W-F775W  \\
  RXJ2129+0005             &    322.4164 &    0.0892   &    0.234    &    1    &     ACS\_WFC &      F775W  &     3.717 &      F475W-F606W  \\
  MS2137-2353              &    325.0633 &    -23.6611 &    0.313    &    1    &     ACS\_WFC &      F814W  &     4.578 &      F475W-F625W  \\
  RCS2152-06               &    328.2013 &    -6.1599  &    0.649    &    1    &     ACS\_WFC &      F814W  &     6.901 &                \\
  RCS2156-0448             &    329.1759 &    -4.8013  &    1.07     &    1    &     ACS\_WFC &      F850LP &     8.089 &      F775W-F850LP \\
  XMMUJ2205-0159           &    331.4611 &    -1.9917  &    1.12     &    1    &     ACS\_WFC &      F850LP &     8.159 &      F775W-F850LP \\
  XMMXCSJ2215.9-1738       &    333.9973 &    -17.634  &    1.45     &    1    &     WFC3\_IR &      F160W  &     8.401 &      F850LP-F125W \\
  XMMJ2235.3-2557          &    338.8369 &    -25.9611 &    1.393    &    1    &     WFC3\_IR &      F160W  &     8.381 &      F850LP-F105W \\
  RCS2239-60               &    339.9784 &    -60.7457 &    0.429    &    1    &     ACS\_WFC &      F814W  &     5.589 &                \\
  RXJ2248-4431             &    342.1832 &    -44.5306 &    0.349    &    1    &     ACS\_WFC &      F850LP &     4.921 &      F606W-F625W  \\
  RCS2316-00               &    349.2302 &    -0.1965  &    0.56     &    1    &     ACS\_WFC &      F814W  &     6.449 &                \\
  RCX2319+0038             &    349.9727 &    0.637    &    0.8972   &    1    &     ACS\_WFC &      F850LP &     7.75  &      F775W-F850LP \\
  RXJ2328+14               &    352.2178 &    14.8786  &    0.49885  &    1    &     ACS\_WFC &      F814W  &     6.08  &                \\
  SPT-CLJ2331-5051         &    352.9631 &    -50.865  &    0.5707   &    1    &     ACS\_WFC &      F814W  &     6.509 &      F606W-F814W  \\
  SPT-CLJ2337-5942         &    354.3653 &    -59.7014 &    0.775    &    1    &     ACS\_WFC &      F814W  &     7.397 &      F606W-F814W  \\
  SPT-CLJ2341-5119         &    355.3015 &    -51.3291 &    0.9983   &    1    &     ACS\_WFC &      F814W  &     7.968 &      F606W-F814W  \\
  RCS2342-35               &    355.5794 &    -35.5713 &    0.802    &    1    &     ACS\_WFC &      F814W  &     7.484 &                \\
  SPT-CLJ2342-5411         &    355.6913 &    -54.1848 &    1.08     &    1    &     ACS\_WFC &      F814W  &     8.104 &      F606W-F814W  \\
  RCS2345-3632             &    356.3729 &    -36.5461 &    1.04     &    1    &     WFC3\_IR &      F160W  &     8.041 &      F105W-F160W  \\
  SPT-CLJ2359-5009         &    359.9286 &    -50.1672 &    0.77     &    2    &     ACS\_WFC &      F814W  &     7.38  &                \\
  SPT-CLJ2359-5009         &    359.9326 &    -50.1723 &           &    1    &     ACS\_WFC &      F814W  &     7.38  &                \\
  \hline
\end{tabular}
\end{table*}


\begin{table*}[htbp]
\centering
\setcounter{table}{1}
  \caption{Parameters obtained from fitting the luminosity profiles of the BCGs with GALFIT. Only the parameters obtained for the chosen model are shown. If fitted by two S\'ersic profiles, the parameters of the outer component are given (the parameters for the inner component are then given in \Cref{tab:GALFIT_inn}). The columns are: full cluster name, class of the galaxy, best model (S\'ersic being a model with a single component, S\'ersic2 a model with two components, and S\'ersic* by fixing the S\'ersic index n = 4), absolute magnitude, mean effective surface brightness, effective radius, S\'ersic index, elongation (ratio of the major to minor axis), position angle, alignment of the BCG with its host cluster.}
\label{tab:GALFIT_out}
\begin{tabular}{| l | l | l | l | l | l | l | l | l | l |}   \hline
Name & Class & Model & m$_{ABS}$ & $<\mu_{e}>$ & R$_{e}$ & n & b/a & PA & Alignment  \\ 
 & &  & (mag) & (mag/arcsec$^{2}$) & (kpc) & & & (degrees) & (degrees) \\
\hline                                   
Cl0016+1609  &  1  &  S\'ersic  &    &    &    &    &    &    & \\  
SpARCS-J0335  &  1  &  S\'ersic  &  -26.088  &  25.034  &  57.488  &  8.7  &  0.88  & 158 & 4 \\
XDCPJ0044-2033  &  1  &  S\'ersic  &  -25.073  &  23.119  &  12.944  &  3.39  &  0.36  & 11 & 63 \\
CLJ0152-1357  &  1  &  S\'ersic  &  -24.859  &  23.471  &  21.802  &  3.59  &  0.7  & 49 & \\  
CLJ015244.18-135715.84  &  2  &  S\'ersic  &  -24.378  &  22.698  &  12.74  &  4.24  &  0.75  & 42 & \\  
CLJ015244.18-135715.84  &  1  &  S\'ersic  &  -24.478  &  22.343  &  11.331  &  7.96  &  1.0  & 10 & \\  
RCSJ0220-0333  &  1  &  S\'ersic  &  -24.511  &  23.111  &  10.804  &  4.08  &  0.75  & 24 & 64 \\
RCSJ0221-0321  &  1  &  S\'ersic  &  -23.916  &  22.257  &  5.667  &  0.88  &  0.61  & 111 & 63 \\
RCSJ0221-0321  &  2  &  S\'ersic  &  -24.566  &  23.098  &  11.261  &  5.43  &  0.74  & 21 & 27 \\
XLSSJ0223-0436  &  1  &  S\'ersic  &  -24.675  &  23.103  &  8.146  &  4.52  &  0.74  & 47 & 69 \\
SpARCS-J0224  &  1  &  S\'ersic  &  -24.905  &  18.918  &  1.682  &  2.87  &  0.64  & 128 & \\  
SpARCS-J0330  &  1  &  S\'ersic  &  -25.891  &  24.187  &  30.167  &  5.43  &  0.67  & 9 & 34 \\
RCS0337-2844  &  1  &  S\'ersic  &  -24.102  &  23.468  &  9.433  &  4.21  &  0.72  & 114 & \\  
RCS0350-08  &  1  &  S\'ersic  &  -24.849  &  23.089  &  22.851  &  7.63  &  0.71  & 159 & \\  
RCS0439-2904  &  1  &  S\'ersic  &  -24.625  &  23.598  &  15.81  &  5.81  &  0.77  & 112 & \\  
MACSJ0454.1-0300  &  1  &  S\'ersic*  &  -24.66  &  22.513  &  19.069  &  4.0  &  0.74  & 119 & 31 \\
RCS0515-43  &  1  &  S\'ersic  &  -24.928  &  24.088  &  49.921  &  9.58  &  0.74  & 9 & \\  
RCS0519-42  &  1  &  S\'ersic  &  -23.293  &  20.669  &  3.71  &  3.95  &  0.81  & 112 & \\  
RCS0519-44  &  1  &  S\'ersic  &  -24.83  &  23.117  &  15.098  &  5.56  &  0.81  & 66 & \\  
SPT-CLJ0533-5005  &  1  &  S\'ersic  &  -24.463  &  23.344  &  12.609  &  6.03  &  0.76  & 149 & 72 \\
SPT-CLJ0546-5345  &  1  &  S\'ersic  &  -25.154  &  25.904  &  31.178  &  8.1  &  0.78  & 105 & 27 \\
SPT-CLJ0615-5746  &  1  &  S\'ersic  &  -26.055  &  23.471  &  37.3  &  4.46  &  0.66  & 19 & 49 \\
RXJ0841+64  &  1  &  S\'ersic*  &  -25.353  &  23.728  &  56.139  &  4.0  &  0.6  & 48 & \\  
RDCSJ0910+5422  &  1  &  S\'ersic  &  -24.369  &  24.853  &  19.696  &  8.36  &  0.78  & 18 & 13 \\
MOOJ1014+0038  &  1  &  S\'ersic  &  -25.528  &  23.814  &  26.754  &  3.3  &  0.68  & 52 & 13 \\
LCDCS0130  &  1  &  S\'ersic  &  -24.316  &  24.52  &  28.215  &  7.59  &  0.92  & 169 & \\  
SpARCS-J1049  &  1  &  S\'ersic  &  -26.55  &  24.782  &  51.065  &  7.63  &  0.94  & 85 & \\  
MS1054-0321  &  1  &  S\'ersic  &  -25.802  &  24.494  &  50.904  &  6.37  &  0.62  & 43 & \\  
RCS1107-05  &  1  &  S\'ersic  &  -25.196  &  24.795  &  40.238  &  5.37  &  0.65  & 66 & \\  
RCS1107-05  &  1  &  S\'ersic  &  -24.28  &  22.484  &  13.684  &  5.94  &  0.79  & 22 & \\  
RCS1108-04  &  1  &  S\'ersic  &  -24.563  &  23.294  &  20.204  &  4.42  &  0.48  & 98 & \\  
RCS1122+24  &  1  &  S\'ersic  &  -25.0  &  23.861  &  24.589  &  6.03  &  0.7  & 147 & \\  
LCDCS0340  &  1  &  S\'ersic  &  -23.979  &  22.153  &  13.621  &  4.41  &  0.74  & 92 & \\  
MOOJ1142+1527  &  1  &  S\'ersic  &  -26.175  &  23.799  &  38.599  &  4.16  &  0.66  & 76 & 4 \\
MACSJ1149.5+2223  &  1  &  S\'ersic  &  -25.909  &  24.645  &  90.254  &  4.1  &  0.71  & 131 & 12 \\
RDCSJ1252-2927  &  1  &  S\'ersic  &  -25.353  &  23.253  &  18.022  &  4.97  &  0.77  & 73 & 30 \\
RDCSJ1252-2927  &  2  &  S\'ersic  &  -24.943  &  22.918  &  12.784  &  2.57  &  0.73  & 66 & 23 \\
MACSJ1311-0310  &  1  &  S\'ersic  &  -24.793  &  22.043  &  17.613  &  3.57  &  0.86  & 132 & 34 \\
RXJ1354-02  &  1  &  S\'ersic  &  -25.709  &  23.334  &  41.547  &  5.57  &  0.77  & 20 & \\  
WARPSJ1415+3612  &  1  &  S\'ersic  &  -25.351  &  23.665  &  20.719  &  2.73  &  0.72  & 23 & 45 \\
MACSJ1423+2404  &  1  &  S\'ersic  &  -25.399  &  22.363  &  24.917  &  2.16  &  0.6  & 37 & 5 \\
ISCS1429+3437  &  1  &  S\'ersic  &  -26.15  &  23.308  &  29.892  &  8.65  &  0.93  & 58 & 28 \\
ISCSJ1432+3436  &  2  &  S\'ersic  &  -24.524  &  20.43  &  3.476  &  4.14  &  0.84  & 53 & \\  
ISCSJ1432+3436  &  1  &  S\'ersic  &  -24.714  &  21.079  &  5.116  &  3.53  &  0.88  & 31 & \\  
ISCSJ1434.5+3519  &  1  &  S\'ersic  &  -24.979  &  21.619  &  7.272  &  2.84  &  0.65  & 77 & 56 \\
ISCSJ1438+3414  &  1  &  S\'ersic  &  -25.233  &  23.866  &  22.274  &  9.86  &  0.78  & 36 & 81 \\
MACSJ1447.4+0827  &  1  &  S\'ersic  &  -24.961  &  21.031  &  13.976  &  1.45  &  0.54  & 86 & \\  
RCS1450+08  &  1  &  S\'ersic  &  -25.294  &  24.299  &  36.181  &  6.01  &  0.61  & 140 & \\  
RCS1452+08  &  1  &  S\'ersic  &  -24.575  &  22.162  &  19.019  &  4.15  &  0.92  & 11 & \\  
RCS1511+09  &  1  &  S\'ersic  &  -23.975  &  22.327  &  6.495  &  4.35  &  0.79  & 137 & 33 \\
ClGJ1604+4304  &  2  &  S\'ersic  &  -24.364  &  22.737  &  10.716  &  4.87  &  0.74  & 121 & \\  
ClGJ1604+4304  &  1  &  S\'ersic  &  -24.724  &  23.643  &  19.195  &  7.29  &  0.76  & 22 & \\  
RCS1620+29  &  1  &  S\'ersic  &  -23.819  &  22.501  &  6.405  &  5.03  &  0.78  & 93 & \\  
MACSJ1621.4+3810  &  1  &  S\'ersic  &  -24.991  &  22.075  &  22.448  &  3.72  &  0.77  & 129 & 4 \\
SPT-CLJ2040-5725  &  1  &  S\'ersic  &  -24.933  &  25.728  &  41.119  &  6.96  &  0.83  & 15 & 6 \\
SPT-CLJ2043-5035  &  1  &  S\'ersic  &  -24.501  &  23.233  &  16.648  &  2.74  &  0.67  & 111 & \\  
\hline         
\end{tabular}
\end{table*}

\begin{table*}[htbp]
\centering
\setcounter{table}{1}
  \caption{Continued.}
\begin{tabular}{| l | l | l | l | l | l | l | l | l | l |}   \hline
Name & Class & Model & m$_{ABS}$ & $<\mu_{e}>$ & R$_{e}$ & n & b/a & PA & Alignment  \\ 
 & &  & (mag) & (mag/arcsec$^{2}$) & (kpc) & & & (degrees) & (degrees) \\
 \hline     
XMMUJ2205-0159  &  1  &  S\'ersic  &  -25.794  &  25.354  &  42.005  &  5.36  &  0.64  & 18 & \\
XMMXCSJ2215.9-1738  &  1  &  S\'ersic  &  -23.698  &  20.249  &  2.019  &  6.25  &  0.61  & 93 & 48 \\  
RCS2316-00  &  1  &  S\'ersic  &  -24.509  &  22.908  &  19.319  &  7.9  &  0.91  & 71 & \\ 
SPT-CLJ2331-5051  &  1  &  S\'ersic  &  -25.302  &  24.013  &  46.462  &  5.24  &  0.73  & 8 & 20 \\  
SPT-CLJ2342-5411  &  1  &  S\'ersic  &  -25.128  &  24.448  &  19.462  &  3.45  &  0.68  & 7 & 42 \\  
RCS2345-3632  &  1  &  S\'ersic  &  -24.659  &  22.317  &  11.956  &  8.04  &  0.75  & 28 & 56 \\  
SPT-CLJ2359-5009  &  2  &  S\'ersic  &  -24.486  &  22.655  &  11.804  &  3.19  &  0.75  & 123 & 6 \\  
SPT-CLJ2359-5009  &  1  &  S\'ersic  &  -24.766  &  23.726  &  21.984  &  4.15  &  0.69  & 123 & 6 \\  
\hline                                    
SPT-CLJ0000-5748  &  1  &  S\'ersic2  &  -25.601  &  24.306  &  47.94  &  1.85  &  0.53  & 162 & 8 \\
ACO2813  &  1  &  S\'ersic2  &  -24.891  &  23.516  &  49.81  &  4.7  &  0.72  & 175 & \\  
ACO2813  &  2  &  S\'ersic2  &  -24.894  &  23.362  &  46.464  &  1.3  &  0.53  & 148 & \\  
RXJ0056-27  &  1  &  S\'ersic2  &  -24.641  &  23.613  &  28.883  &  1.83  &  0.64  & 94 & \\  
SPT-CLJ0102-4915  &  1  &  S\'ersic2  &  -24.87  &  22.478  &  13.894  &  1.62  &  0.87  & 118 & 29 \\
SPT-CLJ0102-4915  &  2  &  S\'ersic2  &  -25.753  &  22.069  &  15.653  &  1.26  &  0.57  & 134 & 13 \\
RXJ0110+19  &  1  &  S\'ersic2  &  -24.411  &  22.709  &  26.043  &  1.1  &  0.72  & 44 & \\  
Abell209  &  1  &  S\'ersic2  &  -24.327  &  21.734  &  19.715  &  2.05  &  0.71  & 134 & 23 \\
SPT-CLJ0205-5829  &  1  &  S\'ersic2  &  -25.701  &  26.197  &  82.853  &  1.09  &  0.33  & 32 & 8 \\
XMMXCSJ022045.1-032555.0  &  1  &  S\'ersic2  &  -24.403  &  21.638  &  16.486  &  2.34  &  0.91  & 65 & \\  
RCS0224-02  &  1  &  S\'ersic2  &  -23.824  &  21.606  &  10.172  &  1.91  &  0.94  & 97 & \\  
JKCS041  &  1  &  S\'ersic2  &  -25.156  &  23.512  &  13.984  &  1.57  &  0.6  & 64 & 36 \\
Abell383  &  1  &  S\'ersic2  &  -24.417  &  21.437  &  18.34  &  1.82  &  0.84  & 17 & 1 \\
MACS0329-0211  &  1  &  S\'ersic2  &  -25.17  &  21.756  &  19.19  &  2.34  &  0.84  & 168 & 32 \\
RCS0351-09  &  1  &  S\'ersic2  &  -23.868  &  19.87  &  5.573  &  2.77  &  1.0  & 57 & \\  
MACSJ0416-2403  &  1  &  S\'ersic2  &  -24.997  &  23.951  &  53.747  &  0.92  &  0.4  & 50 & 15 \\
MACS0429-0253  &  1  &  S\'ersic2  &  -25.332  &  22.155  &  26.926  &  1.52  &  0.64  & 169 & 12 \\
RCS0444-28  &  1  &  S\'ersic2  &  -24.586  &  22.716  &  22.713  &  1.61  &  0.7  & 157 & \\  
RCS0511-42  &  1  &  S\'ersic2  &  -24.997  &  22.985  &  27.435  &  2.09  &  0.74  & 173 & \\  
RCS0518-43  &  1  &  S\'ersic2  &  -23.365  &  22.326  &  11.748  &  1.71  &  0.7  & 74 & \\  
RCS0518-43  &  1  &  S\'ersic2  &  -24.326  &  23.526  &  26.018  &  3.89  &  0.73  & 31 & \\  
SPT-CLJ0559-5249  &  1  &  S\'ersic2  &  -24.29  &  23.094  &  17.035  &  1.93  &  0.92  & 41 & 45 \\
MACSJ0647+7015  &  2  &  S\'ersic2  &  -26.169  &  23.921  &  68.842  &  2.7  &  0.46  & 109 & 24 \\
MACSJ0647+7015  &  1  &  S\'ersic2  &  -25.779  &  23.92  &  57.494  &  0.06  &  0.23  & 111 & 26 \\
MACSJ0717+3745  &  1  &  S\'ersic2  &  -25.353  &  23.278  &  35.644  &  1.58  &  0.66  & 66 & 4 \\
MACSJ0744+3927  &  1  &  S\'ersic2  &  -25.312  &  22.136  &  17.986  &  2.05  &  0.77  & 21 & 80 \\
Abell611  &  1  &  S\'ersic2  &  -24.919  &  22.756  &  34.895  &  1.4  &  0.61  & 43 & 14 \\
RXJ0847+34  &  1  &  S\'ersic2  &  -24.387  &  22.589  &  15.827  &  2.29  &  0.89  & 4 & \\  
RDCSJ0849+4452  &  2  &  S\'ersic2  &  -24.723  &  24.822  &  17.068  &  0.57  &  0.58  & 116 & 38 \\
RDCSJ0849+4452  &  1  &  S\'ersic2  &  -24.236  &  26.658  &  31.769  &  0.07  &  0.84  & 59 & 85 \\
RXJ0926+12  &  1  &  S\'ersic2  &  -24.127  &  22.596  &  15.816  &  1.56  &  0.59  & 66 & \\  
RCS0928+36  &  1  &  S\'ersic2  &  -23.84  &  22.791  &  18.376  &  0.73  &  0.8  & 43 & \\  
Abell851  &  1  &  S\'ersic2  &  -23.891  &  21.962  &  12.473  &  2.84  &  0.73  & 71 & \\  
LCDCS0110  &  1  &  S\'ersic2  &  -24.251  &  22.556  &  14.045  &  3.9  &  0.81  & 25 & \\  
LCDCS0172  &  1  &  S\'ersic2  &  -24.068  &  23.396  &  15.465  &  2.52  &  0.69  & 136 & 36 \\
LCDCS0173  &  1  &  S\'ersic2  &  -24.404  &  23.912  &  20.679  &  1.81  &  0.66  & 20 & \\  
RCS1102-03  &  1  &  S\'ersic2  &  -25.381  &  22.472  &  29.721  &  4.62  &  0.88  & 24 & \\  
RCS1102-05  &  1  &  S\'ersic2  &  -23.882  &  24.06  &  32.648  &  0.78  &  0.45  & 112 & \\  
CLJ1103.7-1245  &  1  &  S\'ersic2  &  -23.595  &  22.659  &  10.011  &  3.71  &  0.62  & 67 & \\  
RCS1104-04  &  1  &  S\'ersic2  &  -24.515  &  23.577  &  22.86  &  1.31  &  0.66  & 17 & \\  
MACSJ1115+0129  &  1  &  S\'ersic2  &  -24.69  &  22.903  &  31.105  &  1.79  &  0.62  & 150 & 60 \\
SG1120-1202-2  &  1  &  S\'ersic2  &  -24.237  &  23.693  &  36.234  &  3.11  &  0.58  & 164 & \\  
SG1120-1202-1  &  1  &  S\'ersic2  &  -24.918  &  24.301  &  65.446  &  1.76  &  0.55  & 156 & \\  
SG1120-1202-3  &  1  &  S\'ersic2  &  -24.419  &  22.31  &  20.783  &  2.44  &  0.61  & 27 & \\  
Abell1423  &  1  &  S\'ersic2  &  -25.131  &  23.716  &  70.008  &  1.97  &  0.51  & 57 & 5 \\
MACS1206-0847  &  1  &  S\'ersic2  &  -24.991  &  23.437  &  38.844  &  2.42  &  0.42  & 107 & 18 \\
LCDCS0504  &  1  &  S\'ersic2  &  -24.855  &  23.241  &  17.146  &  1.72  &  0.89  & 2 & \\  
CLJ1226+3332  &  1  &  S\'ersic2  &  -26.382  &  23.657  &  52.251  &  2.03  &  0.49  & 96 & 13 \\
LCDCS0531  &  1  &  S\'ersic2  &  -24.27  &  24.461  &  35.401  &  1.52  &  0.5  & 22 & \\  
XMMUJ1229+0151  &  1  &  S\'ersic2  &  -24.05  &  21.654  &  5.645  &  2.53  &  0.9  & 118 & 45 \\
LCDCS0541  &  1  &  S\'ersic2  &  -25.022  &  24.224  &  45.61  &  1.7  &  0.48  & 83 & \\  
RCS1319-02  &  1  &  S\'ersic2  &  -22.508  &  22.418  &  8.924  &  3.07  &  0.87  & 138 & \\  
\hline         
\end{tabular}
\end{table*}

\begin{table*}[htbp]
\centering
\setcounter{table}{1}
  \caption{Continued.}
\begin{tabular}{| l | l | l | l | l | l | l | l | l | l |}   \hline
Name & Class & Model & m$_{ABS}$ & $<\mu_{e}>$ & R$_{e}$ & n & b/a & PA & Alignment  \\ 
 & &  & (mag) & (mag/arcsec$^{2}$) & (kpc) & & & (degrees) & (degrees) \\
 \hline     
RCS1323+30  &  1  &  S\'ersic2  &  -25.091  &  23.366  &  37.632  &  1.56  &  0.74  & 128 & \\  
ZwCl1332.8+5043  &  1  &  S\'ersic2  &  -24.332  &  23.825  &  23.977  &  1.62  &  0.45  & 62 & \\  
RXJ1347-1145  &  1  &  S\'ersic2  &  -24.865  &  23.944  &  45.555  &  0.84  &  0.4  & 6 & 50 \\
LCDCS0853  &  1  &  S\'ersic2  &  -24.959  &  25.111  &  44.128  &  4.56  &  0.69  & 125 & \\  
RCS1419+53  &  1  &  S\'ersic2  &  -24.865  &  23.143  &  19.732  &  0.91  &  0.99  & 25 & \\  
IDCSJ1426.5+3508  &  1  &  S\'ersic2  &  -25.647  &  23.716  &  19.951  &  2.55  &  0.73  & 159 & 45 \\
ISCSJ1432.3+3253  &  1  &  S\'ersic2  &  -24.002  &  21.035  &  3.304  &  1.46  &  0.8  & 66 & \\  
ISCSJ1432.4+3250  &  1  &  S\'ersic2  &  -24.446  &  21.641  &  5.278  &  2.12  &  0.6  & 174 & \\  
ISCSJ1432.4+3250  &  2  &  S\'ersic2  &  -24.256  &  21.246  &  4.033  &  2.92  &  0.75  & 123 & \\  
ISCSJ1432+3332  &  1  &  S\'ersic2  &  -24.156  &  22.964  &  7.506  &  0.74  &  0.43  & 164 & \\  
ISCSJ1434+3427  &  1  &  S\'ersic2  &  -24.355  &  22.543  &  8.306  &  0.83  &  0.77  & 137 & 87 \\
RCS1446+08  &  1  &  S\'ersic2  &  -24.208  &  24.501  &  31.293  &  1.38  &  0.49  & 115 & \\  
RXJ1532+3020  &  1  &  S\'ersic2  &  -25.286  &  23.673  &  57.604  &  2.61  &  0.56  & 66 & 15 \\
RXJ1540+14  &  1  &  S\'ersic2  &  -23.255  &  22.342  &  10.161  &  0.22  &  0.53  & 48 & \\  
OC02J1701+6412  &  1  &  S\'ersic2  &  -24.245  &  22.823  &  19.929  &  1.79  &  0.67  & 126 & 36 \\
MACSJ1720+3536  &  1  &  S\'ersic2  &  -24.848  &  21.668  &  17.715  &  2.1  &  0.81  & 177 & 25 \\
Abell2261  &  1  &  S\'ersic2  &  -25.192  &  21.472  &  25.01  &  1.31  &  0.81  & 177 & 38 \\
MACSJ1932-2635  &  1  &  S\'ersic2  &  -25.628  &  22.946  &  46.235  &  4.72  &  0.58  & 154 & 30 \\
SPT-CLJ2040-4451  &  1  &  S\'ersic2  &  -25.832  &  24.611  &  39.304  &  0.55  &  0.31  & 99 & 1 \\
SPT-CLJ2106-5844  &  1  &  S\'ersic2  &  -25.568  &  22.74  &  18.973  &  2.07  &  0.6  & 170 & 33 \\
MACSJ2129-0741  &  1  &  S\'ersic2  &  -24.552  &  24.023  &  32.387  &  0.78  &  0.51  & 79 & 45 \\
RXJ2129+0005  &  1  &  S\'ersic2  &  -25.002  &  23.063  &  45.994  &  2.1  &  0.47  & 66 & 15 \\
MS2137-2353  &  1  &  S\'ersic2  &  -24.548  &  22.265  &  22.504  &  5.75  &  0.82  & 9 & 42 \\
RCS2152-06  &  1  &  S\'ersic2  &  -24.621  &  24.074  &  29.826  &  0.95  &  0.51  & 31 & \\  
RCS2156-0448  &  1  &  S\'ersic2  &  -24.766  &  25.742  &  37.946  &  1.75  &  0.45  & 149 & \\  
XMMJ2235.3-2557  &  1  &  S\'ersic2  &  -26.339  &  25.068  &  65.437  &  2.13  &  0.43  & 20 & 16 \\
RCS2239-60  &  1  &  S\'ersic2  &  -24.372  &  24.618  &  50.625  &  1.12  &  0.41  & 101 & \\  
RXJ2248-4431  &  1  &  S\'ersic2  &  -25.687  &  23.701  &  72.473  &  1.22  &  0.5  & 55 & 2 \\
RCX2319+0038  &  1  &  S\'ersic2  &  -25.504  &  23.833  &  29.438  &  2.58  &  0.7  & 100 & 14 \\
RXJ2328+14  &  1  &  S\'ersic2  &  -24.797  &  23.862  &  37.316  &  1.59  &  0.7  & 141 & \\  
SPT-CLJ2337-5942  &  1  &  S\'ersic2  &  -25.894  &  24.764  &  59.064  &  4.61  &  0.92  & 70 & 42 \\
SPT-CLJ2341-5119  &  1  &  S\'ersic2  &  -25.241  &  24.615  &  25.975  &  2.12  &  0.8  & 153 & 66 \\
RCS2342-35  &  1  &  S\'ersic2  &  -24.912  &  25.342  &  46.534  &  1.27  &  0.56  & 5 & \\  
\hline         
\end{tabular}
\end{table*}

\begin{table*}[htbp]
\centering
\setcounter{table}{2}
  \caption{Parameters obtained for the inner component, for BCGs fitted with two S\'ersic profiles. The columns are: full cluster name, class of the galaxy, absolute magnitude, mean effective surface brightness, effective radius, S\'ersic index, elongation (ratio of the major to minor axis), position angle.}
\label{tab:GALFIT_inn}
\begin{tabular}{| l | l | l | l | l | l | l | l | l |}   \hline
Name & Class & m$_{ABS, inn}$ & $<\mu_{e, inn}>$ & R$_{e, inn}$ & n$_{inn}$ & b/a$_{inn}$ & PA$_{inn}$  \\ 
 & &  & (mag) & (mag/arcsec$^{2}$) & (kpc) & & (degrees) \\
\hline                                   
SPT-CLJ0000-5748  &  1  &  -23.363  &  22.079  &  6.133  &  1.18  &  0.66  & -6 \\
ACO2813  &  1  &  -22.077  &  19.314  &  1.968  &  0.37  &  0.76  & -3 \\
ACO2813  &  2  &  -23.127  &  21.373  &  8.236  &  2.03  &  0.87  & -45 \\
RXJ0056-27  &  1  &  -23.153  &  22.119  &  7.314  &  4.47  &  0.89  & -32 \\
SPT-CLJ0102-4915  &  1  &  -23.625  &  19.466  &  1.956  &  1.6  &  0.73  & -12 \\
SPT-CLJ0102-4915  &  2  &  -24.621  &  19.962  &  3.521  &  1.54  &  0.47  & -48 \\
RXJ0110+19  &  1  &  -23.367  &  19.475  &  3.634  &  2.32  &  0.92  & 23 \\
Abell209  &  1  &  -20.251  &  17.611  &  0.452  &  1.28  &  0.93  & 68 \\
SPT-CLJ0205-5829  &  1  &  -24.978  &  23.76  &  19.333  &  7.67  &  0.72  & -19 \\
XMMXCSJ022045.1-032555.0  &  1  &  -21.69  &  18.129  &  0.939  &  1.51  &  0.72  & 67 \\
RCS0224-02  &  1  &  -22.619  &  18.586  &  1.454  &  2.43  &  0.81  & -27 \\
JKCS041  &  1  &  -23.623  &  20.588  &  1.795  &  1.06  &  0.9  & 25 \\
Abell383  &  1  &  -22.317  &  21.714  &  7.922  &  3.2  &  0.91  & -78 \\
MACS0329-0211  &  1  &  -22.402  &  22.575  &  7.82  &  1.11  &  0.38  & -67 \\
RCS0351-09  &  1  &  -22.581  &  19.864  &  3.074  &  6.34  &  0.85  & -57 \\
MACSJ0416-2403  &  1  &  -23.553  &  20.384  &  5.349  &  1.49  &  0.85  & 56 \\
MACS0429-0253  &  1  &  -22.303  &  21.6  &  5.171  &  2.21  &  0.55  & 79 \\
RCS0444-28  &  1  &  -23.215  &  20.076  &  3.582  &  2.88  &  0.88  & -53 \\
RCS0511-42  &  1  &  -23.198  &  19.874  &  2.859  &  1.48  &  0.76  & 77 \\
RCS0518-43  &  1  &  -22.388  &  19.456  &  1.998  &  3.06  &  0.77  & 44 \\
RCS0518-43  &  1  &  -22.63  &  20.958  &  3.652  &  9.01  &  0.72  & -22 \\
SPT-CLJ0559-5249  &  1  &  -23.059  &  19.4  &  1.763  &  1.72  &  0.78  & -90 \\
MACSJ0647+7015  &  2  &  -22.269  &  23.179  &  8.116  &  1.72  &  0.6  & 10 \\
MACSJ0647+7015  &  1  &  -25.779  &  23.09  &  39.227  &  2.79  &  0.55  & -71 \\
MACSJ0717+3745  &  1  &  -24.57  &  20.752  &  7.766  &  5.01  &  0.8  & 67 \\
MACSJ0744+3927  &  1  &  -22.898  &  20.437  &  2.705  &  1.08  &  0.97  & -31 \\
Abell611  &  1  &  -23.294  &  21.187  &  8.018  &  2.26  &  0.95  & 15 \\
RXJ0847+34  &  1  &  -21.982  &  20.124  &  1.68  &  0.89  &  0.8  & 49 \\
RDCSJ0849+4452  &  2  &  -23.235  &  20.429  &  1.138  &  1.6  &  0.76  & -89 \\
RDCSJ0849+4452  &  1  &  -23.205  &  19.765  &  0.826  &  4.96  &  0.63  & 18 \\
RXJ0926+12  &  1  &  -21.915  &  19.752  &  1.542  &  1.28  &  0.79  & 67 \\
RCS0928+36  &  1  &  -22.977  &  19.871  &  3.219  &  1.68  &  0.93  & 27 \\
Abell851  &  1  &  -21.477  &  19.977  &  1.644  &  1.08  &  0.63  & 65 \\
LCDCS0110  &  1  &  -21.156  &  18.668  &  0.564  &  1.27  &  0.85  & -81 \\
LCDCS0172  &  1  &  -22.451  &  20.336  &  1.794  &  2.61  &  0.77  & 34 \\
LCDCS0173  &  1  &  -22.193  &  20.628  &  1.646  &  1.75  &  0.97  & 59 \\
RCS1102-03  &  1  &  -22.021  &  19.018  &  1.289  &  2.0  &  0.61  & 44 \\
RCS1102-05  &  1  &  -23.407  &  19.859  &  3.79  &  4.84  &  0.91  & -59 \\
CLJ1103.7-1245  &  1  &  -20.752  &  16.862  &  0.187  &  0.72  &  0.48  & 80 \\
RCS1104-04  &  1  &  -22.928  &  20.564  &  2.748  &  1.44  &  0.95  & -30 \\
MACSJ1115+0129  &  1  &  -21.903  &  19.982  &  2.246  &  1.52  &  0.79  & 89 \\
SG1120-1202-2  &  1  &  -21.982  &  19.527  &  1.883  &  1.33  &  0.75  & -7 \\
SG1120-1202-1  &  1  &  -23.569  &  18.727  &  2.698  &  2.4  &  0.93  & -58 \\
SG1120-1202-3  &  1  &  -22.731  &  18.176  &  1.423  &  1.14  &  0.87  & 50 \\
Abell1423  &  1  &  -22.761  &  20.619  &  5.648  &  1.92  &  0.91  & 59 \\
MACS1206-0847  &  1  &  -22.985  &  21.713  &  6.969  &  3.28  &  0.8  & 88 \\
LCDCS0504  &  1  &  -21.873  &  22.417  &  2.972  &  0.76  &  0.8  & -30 \\
CLJ1226+3332  &  1  &  -24.205  &  20.878  &  5.332  &  1.64  &  0.77  & -89 \\
LCDCS0531  &  1  &  -23.086  &  20.342  &  3.077  &  1.64  &  0.7  & 29 \\
XMMUJ1229+0151  &  1  &  -19.227  &  22.851  &  1.063  &  0.06  &  0.24  & 81 \\
LCDCS0541  &  1  &  -22.957  &  22.1  &  6.624  &  3.28  &  0.94  & 57 \\
RCS1319-02  &  1  &  -21.208  &  17.99  &  0.638  &  2.36  &  0.78  & -57 \\
RCS1323+30  &  1  &  -23.371  &  20.032  &  3.672  &  2.27  &  0.95  & 5 \\
ZwCl1332.8+5043  &  1  &  -22.164  &  20.363  &  1.794  &  1.5  &  0.84  & 0 \\
RXJ1347-1145  &  1  &  -24.36  &  20.552  &  7.572  &  1.62  &  0.85  & -1 \\
LCDCS0853  &  1  &  -21.694  &  20.479  &  1.162  &  2.08  &  0.72  & 11 \\
RCS1419+53  &  1  &  -23.551  &  20.891  &  3.82  &  2.65  &  0.91  & -1 \\
IDCSJ1426.5+3508  &  1  &  -22.98  &  19.72  &  0.928  &  0.98  &  0.79  & 72 \\
ISCSJ1432.3+3253  &  1  &  -22.283  &  22.658  &  3.161  &  0.05  &  0.26  & 48 \\
\hline         
\end{tabular}
\end{table*}

\begin{table*}[htbp]
\centering
\setcounter{table}{2}
  \caption{Continued.}
\begin{tabular}{| l | l | l | l | l | l | l | l | l |}   \hline
Name & Class & m$_{ABS, inn}$ & $<\mu_{e, inn}>$ & R$_{e, inn}$ & n$_{inn}$ & b/a$_{inn}$ & PA$_{inn}$  \\ 
 & &  & (mag) & (mag/arcsec$^{2}$) & (kpc) & & (degrees) \\
\hline                  
ISCSJ1432.4+3250  &  1  &  -22.573  &  21.182  &  1.803  &  0.65  &  0.61  & -72 \\
ISCSJ1432.4+3250  &  2  &  -23.163  &  21.131  &  2.311  &  0.59  &  0.59  & -48 \\
ISCSJ1432+3332  &  1  &  -22.283  &  19.668  &  0.694  &  1.09  &  0.71  & 30 \\
ISCSJ1434+3427  &  1  &  -23.737  &  19.729  &  1.711  &  1.12  &  0.53  & -46 \\
RCS1446+08  &  1  &  -22.559  &  20.848  &  2.724  &  3.65  &  0.88  & -39 \\
RXJ1532+3020  &  1  &  -24.321  &  20.449  &  8.366  &  1.47  &  0.75  & 38 \\
RXJ1540+14  &  1  &  -22.786  &  19.669  &  2.391  &  2.31  &  0.93  & -73 \\
OC02J1701+6412  &  1  &  -22.18  &  20.509  &  2.653  &  0.98  &  0.97  & 40 \\
MACSJ1720+3536  &  1  &  -21.525  &  19.07  &  1.159  &  1.73  &  0.78  & 49 \\
Abell2261  &  1  &  -22.858  &  20.529  &  5.527  &  0.58  &  0.95  & -82 \\
MACSJ1932-2635  &  1  &  -22.085  &  22.067  &  6.037  &  0.11  &  0.86  & 50 \\
SPT-CLJ2040-4451  &  1  &  -25.29  &  22.836  &  13.515  &  6.35  &  0.75  & 89 \\
SPT-CLJ2106-5844  &  1  &  -22.864  &  24.62  &  12.985  &  0.28  &  0.28  & -54 \\
MACSJ2129-0741  &  1  &  -23.868  &  22.101  &  9.754  &  3.55  &  0.81  & -85 \\
RXJ2129+0005  &  1  &  -21.517  &  20.281  &  2.566  &  1.44  &  0.86  & -26 \\
MS2137-2353  &  1  &  -24.19  &  21.284  &  12.142  &  1.68  &  0.84  & 75 \\
RCS2152-06  &  1  &  -23.691  &  21.225  &  5.234  &  4.63  &  0.75  & -40 \\
RCS2156-0448  &  1  &  -22.666  &  21.32  &  1.883  &  3.21  &  0.99  & -32 \\
XMMJ2235.3-2557  &  1  &  -24.119  &  21.432  &  4.412  &  1.65  &  0.8  & 5 \\
RCS2239-60  &  1  &  -23.61  &  21.723  &  9.392  &  3.69  &  0.89  & -88 \\
RXJ2248-4431  &  1  &  -24.692  &  21.271  &  14.969  &  2.06  &  0.83  & 50 \\
RCX2319+0038  &  1  &  -22.494  &  20.339  &  1.473  &  1.91  &  0.93  & 55 \\
RXJ2328+14  &  1  &  -23.054  &  19.838  &  2.62  &  2.49  &  0.91  & 68 \\
SPT-CLJ2337-5942  &  1  &  -21.691  &  19.118  &  0.633  &  0.95  &  0.71  & -89 \\
SPT-CLJ2341-5119  &  1  &  -22.005  &  21.059  &  1.138  &  1.39  &  0.96  & -70 \\
RCS2342-35  &  1  &  -23.591  &  21.78  &  4.91  &  4.98  &  0.91  & -42 \\

\hline         
\end{tabular}
\end{table*}


\include{tableMassX}

\begin{table*}[htbp]
\centering
\setcounter{table}{3}
  \caption{Cluster properties. The columns are: full cluster name, X-ray coordinates of the cluster, mass of the cluster, method used to measure the mass of the cluster, PA of the cluster.}
\label{table:mass_coordX}      
\begin{tabular}{| l | l | l | l | l | l | l | l | l |}   \hline
Name & RA$_{cluster, X}$ & DEC$_{cluster, X}$ & Ref & M$_{200,c}$ & Method & Ref & PA$_{cluster}$ & Ref \\ 
 & (J200) & (J2000) & & 10$^{14}$ M$\odot$ & & & (degrees) & \\
\hline       
SPT-CLJ0000-5748 & 0.25 & -57.8093 & 1 & 6.04$^{+1.61}_{-1.61}$ & SZ & 1 & 170$\pm$22 & 2 \\  
Cl0016+1609 &  &  &  & 25.76$^{+6.66}_{-6.66}$ & WL & 3 & 35$\pm$0 & 4 \\  
SpARCS-J0035 & 8.9588 & -43.2029 & 5 & 2.5$^{+0.9}_{-1.0}$ & $\sigma$-M$_{200}$ & 6 & 154$\pm$8 & 2 \\  
ACO2813 & 10.8519 & -20.6229 & 7 & 8.17$^{+1.91}_{-1.61}$ & WL & 8 &  &  \\  
XDCPJ0044-2033 & 11.022 & -20.5665 & 9 & 3.98$^{+1.58}_{-1.58}$ & X & 10 & 128$\pm$49 & 2 \\  
RXJ0056−27 & 14.2338 & -27.67 & 11 & 2.84$^{+1.59}_{-1.59}$ & WL & 3 &  &  \\  
SPT-CLJ0102-4915 & 15.734 & -49.2656 & 1 & 25.4$^{+4.9}_{-4.9}$ & WL & 3 & 147$\pm$2 & 2 \\  
RXJ0110+19 & 17.575 & 19.6397 & 11 & 2.36$^{+1.22}_{-1.22}$ & WL & 3 &  &  \\  
Abell209 & 22.969 & -13.6108 & 12 & 9.5$^{+0.7}_{-0.7}$ & Lens & 13 & 131$\pm$0 & 4 \\  
CLJ0152−1357 & 28.1712 & -13.9686 & 14 & 14.0$^{+4.6}_{-4.6}$ & WL & 15 &  &  \\  
CLJ015244.18-135715.84 & 28.1712 & -13.9686 & 14 & 14.0$^{+4.6}_{-4.6}$ & WL & 15 &  &  \\  
SPT-CLJ0205-5829 & 31.4437 & -58.4856 & 1 & 7.85$^{+1.58}_{-1.58}$ & SZ & 1 & 40$\pm$20 & 2 \\  
XMMXCSJ022045.1−032555.0 &  &  &  & 0.9$^{+0.38}_{-0.21}$ & X & 16 &  &  \\  
RCSJ0220-0333 &  &  &  & 4.8$^{+1.8}_{-1.3}$ & Lens & 17 & 88$\pm$13 & 2 \\  
RCSJ0221-0321 &  &  &  & 1.8$^{+1.3}_{-0.7}$ & Lens & 17 & 174$\pm$9 & 2 \\  
XLSSJ0223-0436 &  &  &  & 7.4$^{+2.5}_{-1.8}$ & Lens & 17 & 116$\pm$17 & 2 \\  
SpARCS-J0224 &  &  &  & 0.36$^{+0.42}_{-0.14}$ & X & 16 &  &  \\  
JKCS041 &  &  &  & 2.86$^{+2.23}_{-2.23}$ & X &  18 & 100$\pm$18 & 2 \\  
Abell383 & 42.014 & -3.5291 & 12 & 8.7$^{+0.7}_{-0.7}$ & Lens & 13 & 16$\pm$6 & 2 \\  
MACS0329-0211 & 52.4237 & -2.1966 & 12 & 7.3$^{+1.0}_{-1.0}$ & Lens & 13 & 144$\pm$0 & 4 \\  
SpARCS-J0330 &  &  &  & 2.04$^{+2.08}_{-1.39}$ & X & 16 & 43$\pm$12 & 2 \\  
RCS0337−2844 &  &  &  & 4.9$^{+2.8}_{-1.7}$ & Lens & 17 &  &  \\  
MACSJ0416-2403 & 64.0391 & -24.0678 & 12 & 25.0$^{+5.0}_{-5.0}$ & Lens &  19 & 52$\pm$0 & 4 \\  
MACS0429-0253 & 67.4004 & -2.8856 & 12 & 8.0$^{+1.4}_{-1.4}$ & Lens & 13 & 1$\pm$16 & 2 \\  
RCS0439−2904 & 69.9083 & -29.0819 & 20 & 4.3$^{+1.7}_{-1.2}$ & Lens & 17 &  &  \\  
MACSJ0454.1-0300 & 73.5475 & -3.0142 & 21 & 16.3$^{+4.5}_{-4.5}$ & Lens & 22 & 150$\pm$0 & 4 \\  
SPT-CLJ0533-5005 & 83.406 & -50.0965 & 1 & 5.88$^{+1.57}_{-1.57}$ & SZ & 1 & 41$\pm$7 & 2 \\  
SPT-CLJ0546-5345 & 86.6548 & -53.759 & 1 & 7.61$^{+1.61}_{-1.61}$ & SZ & 1 & 78$\pm$26 & 2 \\  
SPT-CLJ0559-5249 & 89.9329 & -52.8266 & 1 & 9.94$^{+2.0}_{-2.0}$ & SZ & 1 & 176$\pm$11 & 2 \\  
SPT-CLJ0615-5746 & 93.957 & -57.778 & 1 & 16.32$^{+3.26}_{-3.26}$ & SZ & 1 & 68$\pm$47 & 2 \\  
MACSJ0647+7015 & 101.9585 & 70.2471 & 12 & 68.0$^{+14.0}_{-14.0}$ & Lens & 23 & 90$\pm$0 & 4 \\  
MACSJ0717+3745 & 109.3838 & 37.7558 & 14 & 23.6$^{+6.4}_{-6.4}$ & WL & 15 & 122$\pm$0 & 4 \\  
MACSJ0744+3927 & 116.22 & 39.4568 & 12 & 7.0$^{+0.4}_{-0.4}$ & Lens & 13 & 96$\pm$0 & 4 \\  
Abell611 & 120.2368 & 36.0567 & 12 & 8.5$^{+0.5}_{-0.5}$ & Lens & 13 & 29$\pm$8 & 2 \\  
RXJ0841+64 & 130.2792 & 64.3786 & 11 &  &  &  &  &   \\  
RXJ0847+34 & 131.7958 & 34.8211 & 11 &  &  &  &  &   \\  
RDCSJ0849+4452 & 132.2346 & 44.8711 & 20 & 4.4$^{+1.1}_{-0.9}$ & Lens & 17 & 154$\pm$38 & 2 \\  
RDCSJ0910+5422 &  &  &  & 5.0$^{+1.2}_{-1.0}$ & Lens & 17 & 31$\pm$36 & 2 \\  
RXJ0926+12 & 141.65 & 12.7156 & 11 &  &  &  &   &  \\  
Abell851 &  &  &  & 6.6$^{+2.0}_{-2.0}$ & WL & 15 &  &  \\  
MOOJ1014+0038 &  &  &  & 5.6$^{+0.6}_{-0.6}$ & SZ & 24 & 65$\pm$20 & 2 \\  
LCDCS0110 &  &  &  & 0.3$^{+0.2}_{-0.1}$ & M$_{200}$-$\sigma$ & 25 &  &  \\  
\hline         
\end{tabular}
\end{table*}

\begin{table*}[htbp]
\centering
\setcounter{table}{3}
  \caption{Continued.}
\begin{tabular}{| l | l | l | l | l | l | l | l | l |}   \hline
Name & RA$_{cluster, X}$ & DEC$_{cluster, X}$ & Ref & M$_{200,c}$ & Method & Ref & PA$_{cluster}$ & Ref \\ 
 & (J200) & (J2000) & & 10$^{14}$ M$\odot$ & & & (degrees) & \\
 \hline     
 LCDCS0130 &  &  &  & 0.6$^{+0.3}_{-0.2}$ & M$_{200}$-$\sigma$ & 25 &  &  \\  
SpARCS-J1049 &  &  &  & 3.47$^{+1.25}_{-1.25}$ & X & 16 &  &  \\  
LCDCS0172 &  &  &  & 1.6$^{+0.7}_{-0.5}$ & M$_{200}$-$\sigma$ & 25 & 100$\pm$0 & 4 \\  
LCDCS0173 &  &  &  & 1.0$^{+0.8}_{-0.3}$ & M$_{200}$-$\sigma$ & 25 &  &  \\  
MS1054−0321 & 164.2442 & -3.6269 & 26 & 10.8$^{+2.1}_{-1.8}$ & Lens & 17 &  &  \\  
RCS1102−03 & 165.64 & -3.3201 & 27 &  &  &  &  &  \\  
 CLJ1103.7-1245 &  &  &  & 4.61$^{+1.65}_{-1.46}$ & M$_{200}$-$\sigma$ & 28 &  &  \\  
RCS1107−05 & 166.85 & -5.3891 & 29 &  &  &  &  &  \\  
MACSJ1115+0129 & 168.9669 & 1.4991 & 12 & 9.0$^{+0.9}_{-0.9}$ & Lens & 13 & 147$\pm$0 & 4 \\  
SG1120−1202−2 &  &  &  & 0.46$^{+0.21}_{-0.13}$ & X & 16 &  &  \\  
SG1120−1202−1 &  &  &  & 0.68$^{+0.33}_{-0.19}$ & X & 16 &  &  \\  
SG1120−1202−3 &  &  &  & 0.5$^{+0.51}_{-0.21}$ & X & 16 &  &  \\  
RCS1122+24 &  &  &  & 7.1$^{+3.5}_{-3.5}$ & Velocity & 30 &  &  \\  
 & & & & &  dispersion  & & & \\
LCDCS0340 &  &  &  & 3.4$^{+1.1}_{-1.0}$ & M$_{200}$-$\sigma$ & 25 &  &  \\  
MOOJ1142+1527 &  &  &  & 11.0$^{+2.0}_{-2.0}$ & SZ & 31 & 80$\pm$12 & 2 \\  
MACSJ1149.5+2223 & 177.3994 & 22.3986 & 12 & 51.0$^{+19.0}_{-19.0}$ & Lens & 23 & 140$\pm$0 & 4 \\  
Abell1423 & 179.3219 & 33.6104 & 12 & 12.0$^{+5.9}_{-5.9}$ & Lens & 32 & 52$\pm$6 & 2 \\  
MACS1206-0847 & 181.5512 & -8.8007 & 12 & 8.6$^{+1.1}_{-1.1}$ & Lens & 13 & 89$\pm$12 & 2 \\  
LCDCS0504 &  &  &  & 11.14$^{+1.07}_{-1.07}$ & X & 33 &  &  \\  
CLJ1226+3332 & 186.7432 & 33.5465 & 12 & 15.6$^{+0.4}_{-0.4}$ & Lens & 13 & 83$\pm$11 & 2 \\  
LCDCS0531 &  &  &  & 1.5$^{+0.6}_{-0.5}$ & M$_{200}$-$\sigma$ & 25 &  &  \\  
XMMUJ1229+0151 & 187.3717 & 1.8588 & 5 & 5.3$^{+1.7}_{-1.2}$ & Lens & 17 & 73$\pm$42 & 2 \\  
LCDCS0541 &  &  &  & 10.6$^{+3.9}_{-2.4}$ & M$_{200}$-$\sigma$ & 25 &  &  \\  
RDCSJ1252-2927 & 193.2271 & -29.455 & 5 & 6.8$^{+1.2}_{-1.0}$ & Lens & 17 & 43$\pm$15 & 2 \\  
MACSJ1311-0310 & 197.757 & -3.1776 & 12 & 4.6$^{+0.3}_{-0.3}$ & Lens & 13 & 174$\pm$0 & 4 \\  
ZwCl1332.8+5043 &  &  &  & 3.69$^{+0.0}_{-0.0}$ & X & 34 &  &  \\  
RXJ1347-1145 & 206.8783 & -11.7525 & 14 & 11.6$^{+1.9}_{-1.9}$ & Lens & 13 & 56$\pm$32 & 2 \\  
LCDCS0853 &  &  &  & 3.05$^{+1.74}_{-1.3}$ & M$_{200}$-$\sigma$ & 28 &  &  \\  
RXJ1354−02 & 208.5667 & -2.3631 & 11 & 2.19$^{+0.4}_{-0.4}$ & X & 33 &  &  \\  
WARPSJ1415+3612 & 213.7962 & 36.2008 & 35 & 4.7$^{+2.0}_{-1.4}$ & Lens & 17 & 158$\pm$44 & 2 \\  
RCS1419+53 & 214.8004 & 53.4366 & 29 & 3.61$^{+0.31}_{-0.21}$ & X & 29 &  &  \\  
MACSJ1423+2404 & 215.949 & 24.0779 & 12 & 5.7$^{+1.0}_{-1.0}$ & Lens & 13 & 42$\pm$14 & 2 \\  
IDCSJ1426.5+3508 & 216.6361 & 35.1408 & 36 & 2.3$^{+2.1}_{-1.4}$ & Lens & 37 & 114$\pm$11 & 2 \\  
ISCS1429+3437 &  &  &  & 5.4$^{+2.4}_{-1.6}$ & Lens & 17 & 85$\pm$21 & 2 \\  
ISCSJ1432.4+3250 &  &  &  & 14.4$^{+0.2}_{-0.2}$ & X & 38 &  &  \\  
ISCSJ1432+3332 &  &  &  & 4.9$^{+1.6}_{-1.2}$ & Lens & 17 &  &  \\  
ISCSJ1432+3436 &  &  &  & 5.3$^{+2.6}_{-1.7}$ & Lens & 17 &  &  \\  
ISCSJ1434+3427 &  &  &  & 2.5$^{+2.2}_{-1.1}$ & Lens & 17 & 44$\pm$30 & 2 \\  
ISCSJ1434.5+3519 &  &  &  & 2.8$^{+2.9}_{-1.4}$ & Lens & 17 & 21$\pm$9 & 2 \\  
ISCSJ1438+3414 &  &  &  & 3.1$^{+2.6}_{-1.4}$ & Lens & 17 & 135$\pm$11 & 2 \\  
MACSJ1447.4+0827 & 221.8579 & 8.4736 & 28& 10.36$^{+1.11}_{-1.19}$ & X & 39 &  &  \\  
 \hline         
\end{tabular}
\end{table*}

\begin{table*}[htbp]
\centering
\setcounter{table}{3}
  \caption{Continued.}
\begin{tabular}{| l | l | l | l | l | l | l | l | l |}   \hline
Name & RA$_{cluster, X}$ & DEC$_{cluster, X}$ & Ref & M$_{200,c}$ & Method & Ref & PA$_{cluster}$ & Ref \\ 
 & (J200) & (J2000) & & 10$^{14}$ M$\odot$ & & & (degrees) & \\
\hline    
RCS1511+09 &  &  &  & 1.9$^{+1.4}_{-0.8}$ & Lens & 17 & 170$\pm$44 & 2 \\  
RXJ1532+3020 & 233.2241 & 30.3496 & 12 & 5.3$^{+0.8}_{-0.8}$ & Lens & 13 & 51$\pm$11 & 2 \\  
RXJ1540+14 & 235.2208 & 14.7594 & 11 &  &  &  &  &  \\  
ClGJ1604+4304 &  &  &  & 2.2$^{+2.1}_{-2.1}$ & WL & 40 &  &  \\  
RCS1620+29 & 245.0421 & 29.4891 & 29 & 7.57$^{+1.02}_{-1.04}$ & X & 41 &  &  \\  
MACSJ1621.4+3810 &  &  &  & 5.2$^{+1.9}_{-1.9}$ & WL & 15 & 125$\pm$0 & 4 \\  
OC02J1701+6412 &  &  &  & 3.0$^{+1.5}_{-1.5}$ & Lens & 15 & 90$\pm$0 & 4 \\  
MACSJ1720+3536 & 260.0706 & 35.6066 & 12 & 7.5$^{+0.8}_{-0.8}$ & Lens & 13 & 22$\pm$2 & 2 \\  
Abell2261 & 260.6135 & 32.1329 & 12 & 14.2$^{+1.7}_{-1.7}$ & Lens & 13 & 35$\pm$5 & 2 \\  
MACSJ1932-2635 & 292.9569 & -26.5761 & 12 & 6.9$^{+0.5}_{-0.5}$ & Lens & 13 & 4$\pm$15 & 2 \\  
SPT-CLJ2040-5725 & 310.0631 & -57.428 & 1 & 5.69$^{+1.35}_{-1.35}$ & SZ & 1 & 9$\pm$14 & 2 \\  
SPT-CLJ2040-4451 & 310.241 & -44.8618 & 36 & 8.6$^{+1.7}_{-1.4}$ & WL & 36 & 98$\pm$16 & 2 \\  
SPT-CLJ2043−5035 & 310.8238 & -50.5923 & 42 &  &  &  &  &  \\  
SPT-CLJ2106-5844 & 316.5179 & -58.7426 & 1 & 12.99$^{+2.56}_{-2.56}$ & SZ & 1 & 23$\pm$25 & 2 \\  
MACSJ2129-0741 & 322.3571 & -7.6919 & 14 & 35.0$^{+31.0}_{-31.0}$ & Lens & 23 & 34$\pm$3 & 2 \\  
RXJ2129+0005 & 322.4164 & 0.0886 & 12 & 6.1$^{+0.6}_{-0.6}$ & Lens & 13 & 78$\pm$0 & 4 \\  
MS2137-2353 & 325.0632 & -23.6613 & 12 & 4.11$^{+0.75}_{-0.46}$ & SL & 43 & 51$\pm$25 & 2 \\  
RCS2152−06 &  &  &  & 4.7$^{+1.9}_{-1.9}$ & Velocity & 30 &  &  \\  
 & & & & &  dispersion  & & & \\
RCS2156−0448 &  &  &  & 1.8$^{+2.5}_{-1.0}$ & Lens & 17 &  &  \\  
XMMUJ2205−0159 & 331.4596 & -1.9909 & 5 & 3.0$^{+1.6}_{-1.0}$ & Lens & 17 &  &  \\  
XMMXCSJ2215.9-1738 & 333.9938 & -17.6349 & 5 & 4.3$^{+3.0}_{-1.7}$ & Lens & 17 & 45$\pm$35 & 2 \\  
XMMJ2235.3-2557 & 338.835 & -25.962 & 5 & 7.3$^{+1.7}_{-1.4}$ & Lens & 17 & 36$\pm$14 & 2 \\  
RXJ2248-4431 & 342.1845 & -44.5301 & 12 & 11.6$^{+1.2}_{-1.2}$ & Lens & 13 & 57$\pm$12 & 2 \\  
RCX2319+0038 &  &  &  & 5.8$^{+2.3}_{-1.6}$ & Lens & 17 & 86$\pm$10 & 2 \\  
RXJ2328+14 & 352.2042 & 14.8867 & 11 & 5.5$^{+1.9}_{-1.9}$ & WL & 15 &  &  \\  
SPT-CLJ2331-5051 & 352.9634 & -50.864 & 1 & 8.96$^{+1.86}_{-1.86}$ & SZ & 1 & 168$\pm$13 & 2 \\  
SPT-CLJ2337-5942 & 354.3523 & -59.705 & 1 & 13.11$^{+2.54}_{-2.54}$ & SZ & 1 & 28$\pm$7 & 2 \\  
SPT-CLJ2341-5119 & 355.3 & -51.328 & 1 & 9.15$^{+1.82}_{-1.82}$ & SZ & 1 & 87$\pm$15 & 2 \\  
SPT-CLJ2342-5411 & 355.6916 & -54.1849 & 1 & 6.15$^{+1.49}_{-1.49}$ & SZ & 1 & 49$\pm$19 & 2 \\  
RCS2345-3632 &  &  &  & 2.4$^{+1.1}_{-0.7}$ & WL/SL & 17 & 152$\pm$19 & 2 \\  
SPT-CLJ2359-5009 & 359.9327 & -50.169 & 1 & 6.92$^{+1.61}_{-1.61}$ & SZ & 1 & 129$\pm$8 & 2 \\

  \hline         
\end{tabular}

\tablebib{
(1)~\citet{Chiu_2015}; (2) \citet{west2017ten}; (3) \citet{10.1093/mnras/stu2505}; (4) \citet{Durret_2019};
(5) \citet{Fassbender_2011}; (6) \citet{vanderburg2014}; (7) \citet{Bartalucci_2019}; (8) \citet{Okabe_2016};
(9) \citet{Tozzi_2015}; (10) \citet{Cooke_2019submil}; (11) \citet{Hoekstra_2010}; 
(12) \citet{Postman_2012_lens}; (13) \citet{Merten_2015}; (14) \citet{Sayers_2019}; (15) \citet{martinet2016};
(16) \citet{DeMaio_2019}; (17) \citet{jee2011scaling}; (18) \citet{Andreon2014}; (19) \citet{Zitrin_2012};
(20) \citet{Stott_2010}; (21) \citet{Czakon_2015}; (22) \citet{Herbonnet_2020}; (23) \citet{Sereno_2011};
(24) \citet{Brodwin_2015}; (25) \citet{Chan_2019}; (26) \citet{Jee_2005}; (27) \citet{10.1093/mnras/stt348};
(28) \citet{Just_2019}; (29) \citet{Hicks_2008}; (30) \citet{Noble_2011}; (31) \citet{Gonzalez_2015};
(32) \citet{Dahle_2006}; (33) \citet{guennou2014}; (34) \citet{J_rgensen_2018}; (35) \citet{santos2012};
(36) \citet{Jee_2017}; (37) \citet{Mo_2016}; (38) \citet{Martini_2013}; (39) \citet{Richard_Laferri_re_2020};
(40) \citet{Sereno_2013}; (41) \citet{Lidman_2012}; (42) \citet{Sanders_2017}; (43) \citet{Donnarumma_2009}.
}

\end{table*}

\end{document}